\documentclass[aps,superscriptaddress,showpacs,nofootinbib,eqsecnum]{revtex4}
\usepackage{graphics,amssymb,epsfig,float,psfrag}
\usepackage[usenames,dvips]{color}
\usepackage{times}
\begin{document}

\newcommand{\lsim}{\raisebox{-0.13cm}{~\shortstack{$<$ \\[-0.07cm] $\sim$}}~}
\newcommand{\gsim}{\raisebox{-0.13cm}{~\shortstack{$>$ \\[-0.07cm] $\sim$}}~}
\newcommand{\dx}{\mbox{\rm d}}
\newcommand{\ra}{\rightarrow}
\newcommand{\lra}{\longrightarrow}
\newcommand{\gam}{\gamma \gamma}
\newcommand{\tb}{\tan \beta}
\newcommand{\mlsp}{m_{\tilde \chi_1^0}}
\newcommand{\s}{\smallskip}
\newcommand{\nni}{\noindent}
\newcommand{\nn}{\nonumber}
\newcommand{\be}{\begin{equation}}
\newcommand{\ee}{\end{equation}}
\newcommand{\beq}{\begin{eqnarray}}
\newcommand{\eeq}{\end{eqnarray}}
\newcommand{\pslash}{\not\hspace*{-1.6mm}p}
\newcommand{\kslash}{\not\hspace*{-1.6mm}k}
\newcommand{\lslash}{\not\hspace*{-1.6mm}l}
\newcommand{\eslash}{\hspace*{-1.4mm}\not\hspace*{-1.6mm}E}
\def\ds{\displaystyle}
\def\t{\tilde}
\baselineskip=14pt
\thispagestyle{empty}
\hfill  PTA--07--054\\
\vspace{2cm}
\begin{center}
{\sc \Large \bf Bottom-Up Reconstruction Scenarios for}\\
\vspace{.5cm}
{\Large\sc\bf (un)constrained MSSM Parameters at the LHC}\\
\vspace {.7cm} 
{\large  J.-L. Kneur$^1$ and N. Sahoury$^{1,2}$}\\
\vskip 1cm
$^1$ Laboratoire de Physique Th\'eorique et Astroparticules, UMR5207--CNRS,\\
Univ. Montpellier 2, FR-34095 Montpellier Cedex 5, France. 

$^2$ Laboratoire de Physique Nucléaire et Hautes Energies, UMR7585--CNRS,
Univ. Paris VI, FR-75252 Paris Cedex 5, France.
\vspace{1cm}

{\bf Abstract:}
\end{center}

We consider some specific inverse problem or ``bottom-up"
reconstruction strategies at the LHC for both general and 
constrained MSSM parameters, 
starting from a plausibly limited set of 
sparticle identification and mass measurements, using 
mainly gluino/squark cascade decays, plus  
eventually the lightest Higgs boson mass. For the 
three naturally separated sectors of: gaugino/Higgsino, squark/slepton, 
and Higgs
parameters, we examine different step-by-step 
algorithms based on rather simple, entirely analytical, inverted 
relations between masses and basic 
MSSM parameters. This includes also reasonably good approximations of some of the 
relevant radiative correction calculations. 
We distinguish the constraints obtained for a general MSSM from those   
obtained with universality assumptions in the three different sectors.   
Our results are compared at different stages 
with the determination from more standard  
``top-down" fit of models to data, 
and finally combined into a global determination of all the 
relevant parameters. Our approach gives  
complementary information to more conventional analysis, 
and is not restricted to the specific LHC measurement specificities.    
In addition, the bottom-up renormalization group evolution of 
general MSSM parameters, being an important ingredient in this framework, 
is illustrated as a new publicly available option of 
the MSSM spectrum calculation code ``SuSpect". 
\newpage
\section{Introduction}
If supersymmetry shows up at the LHC, 
it may be that only a limited part of the predicted Minimal Supersymmetric
Standard Model (MSSM)\cite{MSSM}  
sparticles will be discovered and some of their 
properties measured. Hopefully, the lightest Higgs scalar $h$ could be 
discovered, and some of the squarks 
and the gluino could be copiously produced (if not too heavy) at the LHC 
due to their strong interactions. In addition some of the 
neutralinos, including the lightest supersymmetric sparticle (LSP), could be identified and have their 
masses extracted indirectly from a detailed study of squark and 
gluino cascade decays\cite{cascade,cascade1,cascade2}. Beyond that, the discovery
and measurement of the full set of MSSM sparticles may be very model 
dependent, and anyway challenging in many scenarios at the LHC. 
Various analyses have been conducted
(see e.g. \cite{LHCstudy,SPA}) to determine the basic MSSM
parameter space from the above assumed experimental measurements. 
A widely illustrated strategy, in a so-called ``top-down" approach,
is to start from a given supersymmetry-breaking model at a very high 
grand unification scale; predict for given input parameter values
the superpartner spectrum at experimentally accessible energy scales; 
then fit this spectrum (together with possibly other observables like 
cross-sections etc) to the data in order to extract constraints on the
basic model parameters. 
Constraints from past and present collider and non-collider data, with
consequent prospects for the LHC and future linear collider (ILC), 
have been analyzed typically from systematic scanning of MSSM 
parameter space\cite{msugrascan,mssmbayes} (though mostly in the 
constrained minimal supergravity (mSUGRA)~\cite{mSUGRA} case). 
In addition, more elaborated $\chi^2$ fitting procedure (or some 
generalizations\cite{pdg,minuit}) have been also used in many such 
studies, together  
with Monte-Carlo or other process simulation tools\cite{Pythia,Herwig,Comphep}, 
as well as other specific codes for parameter determination\cite{Sfitter,Fittino}.  
On general grounds, fitting and minimization procedures are efficient when 
the number of independent measurements is (much) greater 
than the number of fitted parameters of the underlying model, and  
provided that data are reasonably accurate. But clearly the minimization  
becomes less controllable\footnote{See however ref.\cite{dzerwas} for a
recent elaborated treatment of many-parameter cases.} 
in a general MSSM with more than $\sim$ 20 relevant basic 
parameters (even when neglecting flavor mixing in the sfermion sector). 
Alternatively, so-called inverse or bottom-up reconstruction 
approaches are 
often motivated\cite{inv1,zerwasetal,bpzm123,bpzm_next,SPA,kane,otherinv}. 
Also, a growing number of analysis for the LHC or the ILC appeared recently,  attempting to go beyond
conventional top-down fitting techniques\cite{otherinv} or supplementing these 
with more elaborated 
frequentist or Bayesian methods, with Markov chain Monte-Carlo (MCMC) 
techniques\cite{markov} in particular\cite{mssmbayes,ben_bayes,dzerwas}. 
Yet it has been stressed (for instance
in refs. \cite{otherinv}) that the mapping
from LHC data to the underlying basic MSSM parameters may be far from 
unique. However, most works still rely essentially 
on simulation tools fed with top-down MSSM Lagrangian-to-spectrum 
relations, while to our knowledge  reconstruction scenarios based on explicitly  
inverted relations (see e.g. \cite{inv1,zerwasetal,bpzm123,bpzm_next})  
appear not so widely explored in the literature. Moreover,    
many studies on MSSM parameter reconstruction\cite{Fittino,LHCstudy,SPA} often considered rather optimistic LHC or ILC scenarios, in the
sense that the results presented were obtained by assuming that most, if not all, 
MSSM sparticles masses and other relevant observables are measured 
with the best expected accuracy. At the same time it is often 
assumed that the more constrained
mSUGRA model\cite{mSUGRA} 
(with four continuous plus one discrete
parameters) is to be determined. While such
studies are certainly very useful guidelines for LHC and ILC 
analysis, these assumptions may be considered quite optimistic 
for the supersymmetry discovery prospects in general, especially at the LHC. 
It is thus worth to develop alternative (or rather complementary) 
strategies to foresee  more pessimistic scenarios, still trying to extract 
as much as possible informations on the nature of the  underlying 
supersymmetry-breaking model in case where only a handful of the 
predicted sparticles would be identified. 

  To start with, in this paper we explore specific bottom-up reconstructions,  
  which are more restricted and  
  certainly far from being fully realistic as concerns data simulations, 
  but that we expect 
  to be useful and complementary to the more standard simulation tools.
  Our approach is based essentially on analytical   
  inverse relations between the measured masses and basic parameters. 
   This ``inverse mapping'' for the MSSM spectrum has been
  investigated to some extent in the past years\cite{inv1,zerwasetal}  
  but mainly at the tree-level approximation, and moreover much often in 
  the context of the ILC data essentially.  
  It is generally expected that simple
  analytic expressions between observables and parameters are 
  more transparent or insightful than purely
  numerical results, providing e.g. explicit correlations among parameters.  
  Particularly in the MSSM, even at tree-level, this connexion
  is already quite involved so that it is difficult to grasp with   
  a good intuition the sensitivity of the different observables
  to MSSM parameters, unless having spend much time in doing 
  fits and related calculations. But more concretely than a
  useful insight, we also hope that such an approach could suggest new
  strategies for reconstruction of parameters, as 
  will be illustrated here. For example,  
  by exploiting well-known relations between the (first two generation)  
  squark and slepton soft mass parameters and physical masses, including the 
  renormalization group evolution (RGE) dependence\cite{rge,RGE2},
  we construct appropriate combinations of observables in this squark/slepton sector
  which appear to provide interesting and almost model-independent constraints.\\
  Deriving analytic inversion relations in the MSSM 
  may appear at first a rather academic exercise, quite remote
  from the complexity of the actual experimental situation especially at
  the LHC. This is because such inverted mapping remains relatively 
  simple only if restricted to the tree-level approximation. But it
  becomes a priori inextricable if including radiative corrections, 
  which are certainly necessary at the accuracy level expected for realistic
  LHC and ILC data analysis. More precisely through
  loop contributions almost all sparticle masses have
  a cumbersome dependence 
  on almost all MSSM parameters. Still, we
  will see how radiative corrections can be incorporated into
  our framework rather simply,
  essentially by (numerical) iterative
  procedure in a manageable way,
  in reasonable but often realistic approximations.  
  We emphasize that this procedure is very similar  
  to the way in which radiative
  corrections are included in more conventional top-down MSSM spectrum
  calculations\cite{isasugra,suspect,softsusy,spheno}, 
  and it allows to keep most advantages of the bottom-up approach. 
    
  Even if one can incorporate a fair amount of presently know radiative
  corrections into this framework, we stress that  
  our motivation is not to compete with the state of the art in  
  present analysis of MSSM constraints at LHC, merely by replacing 
  elaborated simulations
  tools with a bunch of rather simple analytic relations (and 
  simple combinations of data uncertainties as we will see).  
  Accordingly our analysis at this stage is still essentially     
  a theoretical exercise, not incorporating important ingredients  
  of the complexity of LHC measurements (such as detailed event 
  selections, detector properties etc) that are ultimately necessary and
  that pave the non-trivial steps in going from basic LHC data to sparticle
  mass measurements. Yet  
   our aim is to consider as much as possible realistic and minimal LHC 
   sparticle identifications, using a limited set of sparticle mass
   measurements. We then gradually
  consider different SUSY-discovery scenarios, going from 
  minimal input assumptions to more optimistic ones, defining 
  corresponding algorithms with definite input/output parameters.
  This step-by-step analysis may turn out to be closer to the actual  
  experimental situation, in which one will certainly not identify all
  sparticles at once, even for the most optimistic expectations. 
  However our approach is only one step in the
  vastly more ambitious program of so-called ``blind'' analysis of LHC data: 
  in particular when we consider a general MSSM case (i.e. departing from e.g.
  mSUGRA universality relations), we nevertheless assume a spectrum 
 pattern still allowing gluino/squark (long) cascade decays (i.e. 
with some neutralinos sufficiently heavy to decay in the LSP plus  
sfermions, but sufficiently light to be decay product of heavier squarks/gluinos). 
This pattern may admittedly be considered a not so general scenario 
within the MSSM. \\       
  Though our analysis essentially 
  concentrates on sparticles expected to be accessible at the LHC, 
  it will appear that some of the reconstruction
  algorithms used here could apply more or less directly to ILC measurements,  
  upon appropriate changes in data accuracies. We thus occasionally make 
  some comments on ILC prospects, but refrain
  to enter a detailed ILC analysis which is beyond the scope of the present paper, 
  since the other expected sparticle mass measurements at the ILC would need 
  rather different algorithms (though quite similar in spirit).\\ 
     
  The paper is organized as follows: in section 2 we briefly define and
  review   a plausible set of sparticle mass measurements at the LHC, with 
  accuracy on which is based our analysis. We  
  consider different levels of assumptions on the nature and number
  of identified sparticles, defining several 
  scenarios. We also gradually introduce universality assumptions
  for the soft-SUSY breaking parameters of the different sectors. 
  In sections 3--6 we expand 
  results of analytic inversion analysis, with some of these already 
  presented in ref. \cite{inv1}, 
  for different parameter sectors of the MSSM, recasting results  
  in the context of gluino/squark cascade
  decay mass measurements at the LHC, and incorporating radiative corrections.  
  We consider separately four different
  sectors: gaugino/Higgsino parameters (section 3); 
  squarks and sleptons (first and second
  generation) (section 4); third generation squarks (section 5); and finally the
  Higgs parameter sector in section 6. 
  These distinctions are quite natural 
  when considering both the interdependence between observables and  
  parameters and the experimental
  signatures which are expected at LHC from a given sector.
  We will delineate which relations and results
  are valid in a general (unconstrained) or a more constrained 
  MSSM (with universality relations at the GUT scale). 
  We also compare in some detail at different stages
  our reconstruction results with more standard
  top-down fitting procedure using MINUIT $\chi^2$ minimization\cite{minuit},
  with data and fitted parameters  
  directly set by the above step-by-step scenarios, rather than by performing
  ``all at once'' global fits. Conclusion and outlook are given in section 7.
  
  Finally we develop in Appendix A the explicit inverse solutions in the
  gaugino/Higgsino sector for different input/output assumptions and
  related issues, and in Appendix B the properties  
  of the bottom-up renormalization group 
  evolution, a necessary ingredient in this approach,
  implemented as an option of the SuSpect 
  code\cite{suspect}. 
  Important features, like the error propagation from
  low to high energy parameters that is implied by 
  RGE, are illustrated there.           
\section{Bottom-up strategy from plausible LHC measurements}
At the LHC, the  dominant production of pairs of gluinos or squarks (or 
gluinos associated with squarks) is expected due to their strong 
interaction.
The corresponding cross-sections are large for moderate masses but decrease
rapidly for large gluino and/or squark masses. Discovery prospects for
gluinos and squarks with masses up to a few TeVs are reported\cite{ATLAS,CMS,LHCstudy}, 
depending on the luminosity (and depending of course 
on the details of the supersymmetric models and spectra).
\subsection{Mass measurements from gluino/squark cascade decays}
From detailed studies of gluino/squark cascade decay products at the LHC,
the masses of the sparticles involved can be determined with a quite good
accuracy (a few percent)
using the so-called kinematic endpoints method\cite{cascade1,cascade2}. 
For a typical mSUGRA benchmark point like SPS1a \cite{benchmark}, 
which has been intensively studied 
in simulations, 
the masses of the sparticle involved in the gluino and squark
decays are obtained from  analysis of exclusive chain of (two-body) cascade decays,  typically\cite{cascade2}: 
\be
\tilde g \to \tilde q_L (\t b) q (b) \to \t \chi^0_2 \,q_f q \to \tilde l_R l q_f q
\to \t \chi^0_1 \, l_f l q_f q\;.
\label{casc}
\ee
Note however that we will subsequently use these data
as a blind input, with the aim to go beyond the SPS1a benchmark
(or even beyond a mSUGRA model)
as concerns the basic MSSM parameter reconstruction.

Actually the 
four masses of $\t q_L$, $\t e_R$ and $\t \chi^0_{1,2}$ (designated in what follows
as $m_{\t N_1}, m_{\t N_2}$)
can be determined from the cascade decay starting from
a $\t q_L$. The gluino mass $m_{\t g}$ can then be determined from the decay to
$\t b$ (see B.~Gjelsten et al p.~213 in \cite{LHCstudy}). 
Alternatively in ref. \cite{cascade2} the gluino mass as well as the four other
masses are determined from the full cascade Eq.~(\ref{casc}).
The sparticle mass determinations and accuracies assumed here 
are based on the results of 
ref. \cite{cascade2} together with ref. \cite{LHCstudy}, summarized 
in Tables \ref{tabinput} and \ref{tabexp}. 
These accuracies in Table \ref{tabexp} may be subject to some adjustments or
updates due to eventually more refined analysis, and should 
be considered here as illustrative, without drastically changing our 
procedure and results.
(For instance very recently even better prospects on mass accuracies 
have been reported\cite{NoPoTo,gunion_cas} by exploiting correlated decays 
of two such cascades.)

Among the different selection criteria, an important characteristic one 
is to look for two isolated, opposite sign, same flavour $e,\mu$ leptons\cite{LHCstudy,cascade2}. We will not be involved here with
a concrete analysis of these cascade events, and rather use 
directly the expected mass measurements extracted from such studies.
We refer to these references for more details and shall only briefly mention some of their main features.
For typical benchmark points, like SPS1a or other cases\cite{benchmark},  
more sparticles than those present in Eq.~(\ref{casc}) are in principle 
accessible from independent processes.
(Indeed the slepton $l_R$ can also be measured independently of the cascade
(\ref{casc}) from slepton pair production).  
Some of the other sparticles may be more difficult to identify, 
due e.g. to the fact that neutralinos, 
and charginos decay predominantly into
$\tilde \tau$ and $\tau$, experimentally
more challenging to detect than a dilepton signal typically\cite{LHCstudy}. 
(These effects are more pronounced for large $\tan\beta$ due to a larger mixing). 
Though gluinos decay predominantly in $\tilde q_L$ squark, and $\tilde \chi_2$ 
predominantly in $\tilde l_R$ slepton, the mass
measurements of $\tilde q_R$ (and $\tilde l_L$)
are also possible\cite{LHCstudy}, but may be 
less favored by the small $\t g \to \t q_R \,q$ branching ratio (B.R.) 
with respect to other
channels, though final statistics can be sufficient\cite{LHCstudy}. 
Moreover $\t q_R$ decays directly into the lightest neutralino $\t \chi_1$, 
since $\t \chi_1$ is mainly Bino for SPS1a (and this is more or less so
in most mSUGRA cases as well, similarly the next-to-lightest
neutralino $\t \chi_2$ is essentially Wino). 
Thus, $\tilde q_R$ being a $SU(2)$ singlet, it decays into the corresponding 
$q_R$ quark together with $\t \chi_1$ with a B.R. of almost 100\%.  

\begin{table}[h!]
\begin{center}
\caption{\label{tabinput} different plausible gradually optimistic 
assumptions on the amount of sparticle mass
measurements at the LHC, from gluino cascade and other decays, 
defining our input for various reconstruction  
scenarios.} 
\vspace{.5cm}
\begin{tabular}{|c||c||c|}
\hline\hline
Input scenarios & mass    & decay or
process \\
(+theory assumptions)          &           &         \\
\hline
 (minimal): & $m_{\t g}$,   
   &  $\t g$ cascade decay \\
$S_1$(MSSM),    & $m_{\t N_1}$,   & " " \\
$S_2$(universality)  & $m_{\t N_2}$.   & " " \\
$S_4$,  & $m_{\t q_L}$,     & " "  \\
$S^\prime_4$ (universality) & $m_{\t l_R}$    & " "   \\
\hline
$S_3 =S_1$ plus:  & $m_{\t N_4}$  &  $\t q_L \to \t \chi^0_4+..$ cascade \\
\hline
$S_5$, & $m_{\t b_1}$,     & $\t g$ cascade decay   \\
 $S^\prime_5$ (universality)      & $m_{\t b_2}$      & " "   \\
\hline
$S_6$ $=S_2+S^\prime_4+S^\prime_5$ plus:  & $m_h$       & $h \to 
\gamma\gamma$ (mainly)   \\  
\hline\hline
\end{tabular}
\end{center}
\end{table}
More generally the nature of sparticles involved in such cascades or 
other considered processes strongly depends on MSSM parameters, i.e. 
the specific masses, branching ratios and other observables, 
and also on some properties inherent to the MSSM structure.
Thus at present it is hard to guess which process may be actually
favoured at the LHC. Indeed the parameter space where decay chains such as 
in Eq.~(\ref{casc}) can occur
may be considered already quite specific, as it requires (non-LSP) neutralinos 
heavier than the first two generation sleptons, but  
light enough to be produced by gluinos and squarks. 
Accordingly we insist that considering in this work only
the sparticle masses accessible from the decay (\ref{casc}) (plus the lightest Higgs) is not a strong prejudice against the possibilities of other processes and
extra sparticle identification. As motivated in introduction, it is simply to consider what this approach can do from
a well-defined ``minimal'' input set, and indeed most of our inversion 
algorithms could be easily extended if more (or different) 
sparticles will be available. 
  
Another important feature of the decay in Eq.~(\ref{casc}) 
is that there is no way to
distinguish the different squarks $\t q$ from each others: 
this is not so much a property of this specific decay but rather
due to the fact that 
there is no realistic mean at present of tagging light quark charge and/or flavor 
(moreover they all have almost indistinguishable B.R.). 
Accordingly the first squark entering the decay chain, 
resulting from the decaying 
gluino, can be either $\t u, \t d, \t c, \t s$ or $\t b_1, \t b_2$
(in general it could also be $\t t_1, \t t_2$ but this is not kinematically 
allowed for the SPS1a input parameters\cite{cascade2,LHCstudy}). 
One can identify the $\t b$ to some extent: 
the decay of a gluino into $\tilde b_1$ is 
dominant over the $\tilde b_2$ one due to the smaller mass,  
and the $\tilde b_1$ decay leads to a b-quark that can be tagged. 
In addition, one may be able to extract a signal even for $\tilde b_2$
(i.e. distinguish it from $\tilde b_1$), but with less statistics 
(correspondingly with a larger mass error), and only for the
large luminosity prospect of 300 fb$^{-1}$\cite{cascade1}.
We will thus consider in addition to our minimal input scenario a 
next scenario where either  
$\t b_1$ alone or both $\t b_1$ and $\t b_2$ masses can be extracted. \\

Finally, on top of the sparticle masses measured via the gluino cascade,
we will consider in section 6 what additional constraints
are obtained within our approach if the lightest Higgs mass is 
assumed to be measured
via its $\gamma \gamma $ decay modes\cite{LHCstudy}, which is mainly responsible
of the expected accuracy as quoted in Table \ref{tabexp}.   
\subsection{Outline of bottom-up reconstruction algorithms}
According to the previous experimental possibilities, 
we define in Table \ref{tabinput}
successive scenarios to be studied and differing on the amount of sparticle
masses measured at the LHC,
from $S_1$ to $S_6$: Scenarios $S_1$-$S_3$  may be considered to range 
from a minimal
input scenario to gradually more optimistic ones, while some scenarios
differ by model assumptions (general MSSM, or with universality relations in the sfermion and/or gaugino sectors typically).  

In our study we shall first generate ``data" with central values, e.g. for
the SPS1a benchmark point, by running the code SuSpect\cite{suspect} 
for the (constrained MSSM) input:
\be
m_0 = -A_0= 100\: \mbox{GeV},\;\;m_{1/2}= 250 \:\mbox{GeV},\;\;\tb(m_Z)=10,\;\;
\mu>0.
\label{defsps1a}  
\ee
\begin{table}[h!]
\caption{\label{susoutput} Soft and other basic parameters, plus 
sparticle pole masses (relevant to our study) 
for SPS1a input (with $m_{top}=175$ GeV), calculated with the latest  
SuSpect ver 2.41, for two illustrative optional choices: 1) full two-loop
in RGE and full radiative corrections to sparticle masses (second and fifth columns); 2) one-loop
RGE, no radiative corrections to squarks, gluino, neutralinos,
charginos masses, simple approximation for $m_h$ radiative corrections
(third and sixth columns).} 
\vspace{.2cm}
\begin{tabular}{|c||c|c||c||c|c|}
\hline
basic par. & 2-loop RGE & 1-loop RGE & relevant  & 
2-loop RGE & 1-loop RGE \\ 
       & +full R.C. & +approx. R.C. & pole masses & +full R.C. & +approx. R.C.\\
\hline\hline
$Q_{EWSB}$&  465.5 & 468.2 &   &    &      \\ 
\hline
 $M_1$ &      101.5    & 108.8  & $m_{\t N_1}$ &   97.2 &  105.1      \\
 $M_2$ &      191.6    &  208.9 & $m_{\t N_2}$   & 180.8  & 189.9    \\
 $M_3$ &     586.6   &   603.8  & $m_{\t g}$ & 606.1 &  603.8 \\
\hline
$\mu$        &   356.9  &   340.6     &  $m_{\t N_4}$   & 381.8     &    369.6          \\
$\tan\beta$ &   9.74 &  9.75  &            &  &              \\
\hline
$m^2_{H_d}$ &$(179.9)^2$ & $(187.3)^2$ &  $m_h$  & 110.85 &  111.28  \\
$m^2_{H_u}$ & $-(358.1)^2$ & $-(341.7)^2$ &   & &  \\
\hline
$m_{e_L}$ &     195.5  & 201.5       & &  &  \\
$m_{\tau_L}$&      194.7 &  200.6    & &  & \\
$m_{e_R}$ &    136   &  138.6         &$m_{\t e_2}$ &  142.8  & 145.4 \\
$m_{\tau_R}$&      133.5 &  136.2    & &    &   \\
 $m_{Q_L1,2}$&     545.8   & 554.1     & $m_{\t u_1}$ & 562.3 &  551.6  \\
$m_{Q_L3}$&      497  &  502.9        & $m_{\t b_1}$ & 516.2 & 502.1     \\
$m_{u_R}$&     527.8   & 531.6       & &   &  \\
$m_{t_R}$&         421.5  &  421.6  & & &     \\
$m_{d_R}$&    525.7   &  528.7         & & &     \\
$m_{b_R}$&        522.4 & 525.4      &$m_{\t b_2}$ & 546.3 & 530.1    \\
\hline
$-A_t$  &  494.5   & 501.0            & &  & \\
 $-A_b$ &   795.2   & 791.3                & &   & \\
 $-A_\tau$ &   251.7 & 255.0               &  &  & \\
 $-A_u$ &  677.3   & 686.6                  & &  &  \\
 $-A_d$ &  859.4  & 857.2                  &  &  &  \\
 $-A_e$ &   253.4  & 256.7                 & &  &  \\
\hline\hline
\end{tabular}
\end{table}
The resulting spectrum in Table \ref{susoutput} is calculated from the latest
version 2.41 of SuSpect. We used two available options on RGE and 
sparticle mass radiative corrections, both for illustration and the  
need of our analysis, as will be developed later on. 
Note that we use everywhere a value of the top mass 
$m_{top} =175$ GeV rather than the latest experimental 
top mass values: $m_{top}= 172.6 \pm 1.4$ GeV \cite{lastmt}, in order to be
more consistent with the analysis performed in ref.\cite{cascade1,cascade2}.
We assume that this shift in the central value of the top mass should not
affect qualitatively our analysis (although what could be important is
the impact of the top mass uncertainties).  

We then use the sparticle masses contributing to the gluino 
cascade as blind input, within different
reconstruction scenarios, without necessarily assuming a constrained MSSM
with universality relations. The aim is to examine
what can be reconstructed under different gradually constrained 
assumptions on the MSSM parameters. We illustrate the determination 
uncertainties from the SPS1a
sparticle mass error as reference, since it is one of the most simulated  
benchmark in the literature.    

Apart from distinguishing different scenarios
as indicated in Table \ref{tabinput}, 
most of our study is based on specific bottom-up algorithms
depending on the assumed input sparticle masses and output basic parameters.
In defining these algorithms it is convenient to consider
separately and gradually the three different sectors of gaugino/Higgsino,
squark/slepton, and Higgs sector respectively (distinguishing also the
third from the first two generations in the squark sector, since those 
necessitate different treatments due to the mixing in the third generation). 
We will also carefully distinguish different scenarios depending 
on the amount of theoretical assumptions, eventually reducing the number 
of basic MSSM parameters, like universality
of gaugino and/or scalar mass terms typically. 
Our different algorithms obviously depend on specific assumptions, 
since input and output parameters may be completely different depending on
these.    
We describe in detail in the next sections 
these particular algorithms depending on the parameter 
sectors and theoretical assumptions considered. The starting point 
is always the use of tree-level relations giving some specific Lagrangian
parameters in terms of appropriate input sparticle masses. For example 
in the gaugino/higgsino sector, one of the inverted
relations we shall consider has the form 
\be
f(m_{\t N_1}, m_{\t N2}; M_1, M_2) +\Delta f_{rad.corr.} =0, 
\label{definv}
\ee
where f gives the output parameters $M_1, M_2$ in terms
of two neutralino mass input: 
$m_{\t N_1}, m_{\t N_2}$, or other such 
relations for different input/output choices.  Whenever possible, 
the relations defining $f$ in Eq.~(\ref{definv}) are entirely 
analytical and often giving a linear (unique) 
or at most quadratic solution (with eventually corresponding 
twofold solutions). In addition some input/output parameter choices
need extra numerical calculations, typically iterations. These are
needed anyway to take
into account the radiative corrections, symbolized by the 
term $\Delta f_{rad.corr.}$ in Eq.~(\ref{definv}), which generally
depend on extra MSSM parameters or masses. As already mentioned,
it is clear that such approach cannot be very realistic if not including
at least some part of these radiative corrections, 
as is discussed in next sub-section.   

Once having  reconstructed from a relation like (\ref{definv}) the  
relevant MSSM parameters
at the ``physical" scale (generally identified as the electroweak symmetry 
breaking (EWSB) scale), 
another important step in this bottom-up approach is the
possibility of evolving these parameters consistently 
from low to high (GUT) scale,
with implications concerning the
propagation of parameter uncertainties from low to high scales. 
Such bottom-up RGE evolution of soft parameters 
had been considered in the past\cite{inv1,bpzm123} (see \cite{bpzm123}
notably for mass measurement error propagation), but meanwhile many
refinements e.g. on radiative corrections have been included in
public MSSM codes. Accordingly we have 
implemented an up-to-date option 
in the code SuSpect to perform this bottom-up RGE, which  
is used at different stages in our analysis and illustrated in more detail  
in Appendix B.
 
\subsection{Including radiative corrections in bottom-up reconstruction}
\begin{figure}[h!]
\begin{center}
\mbox{
\epsfig{figure=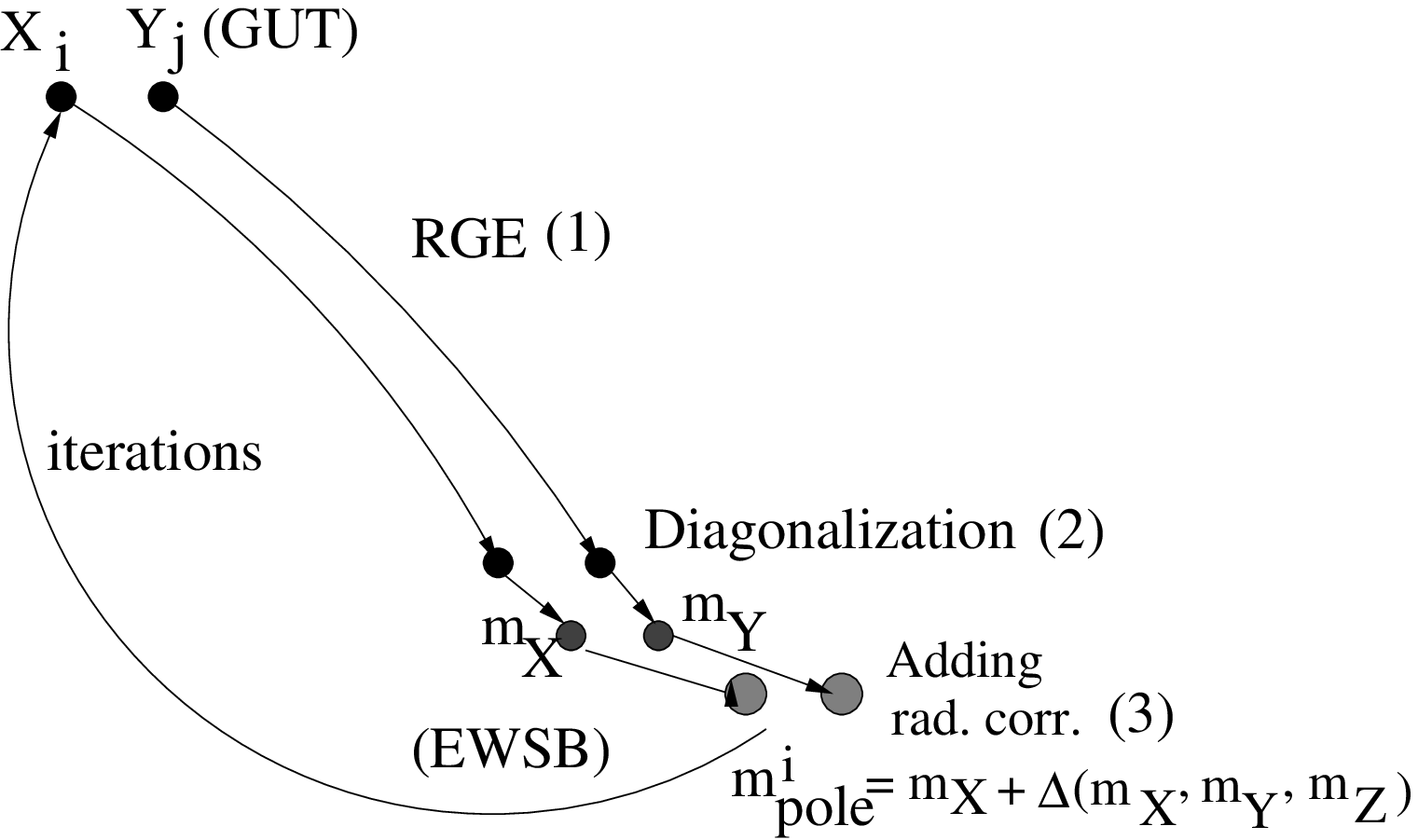,width=8.5cm}
\vspace{-0.5cm}
\epsfig{figure=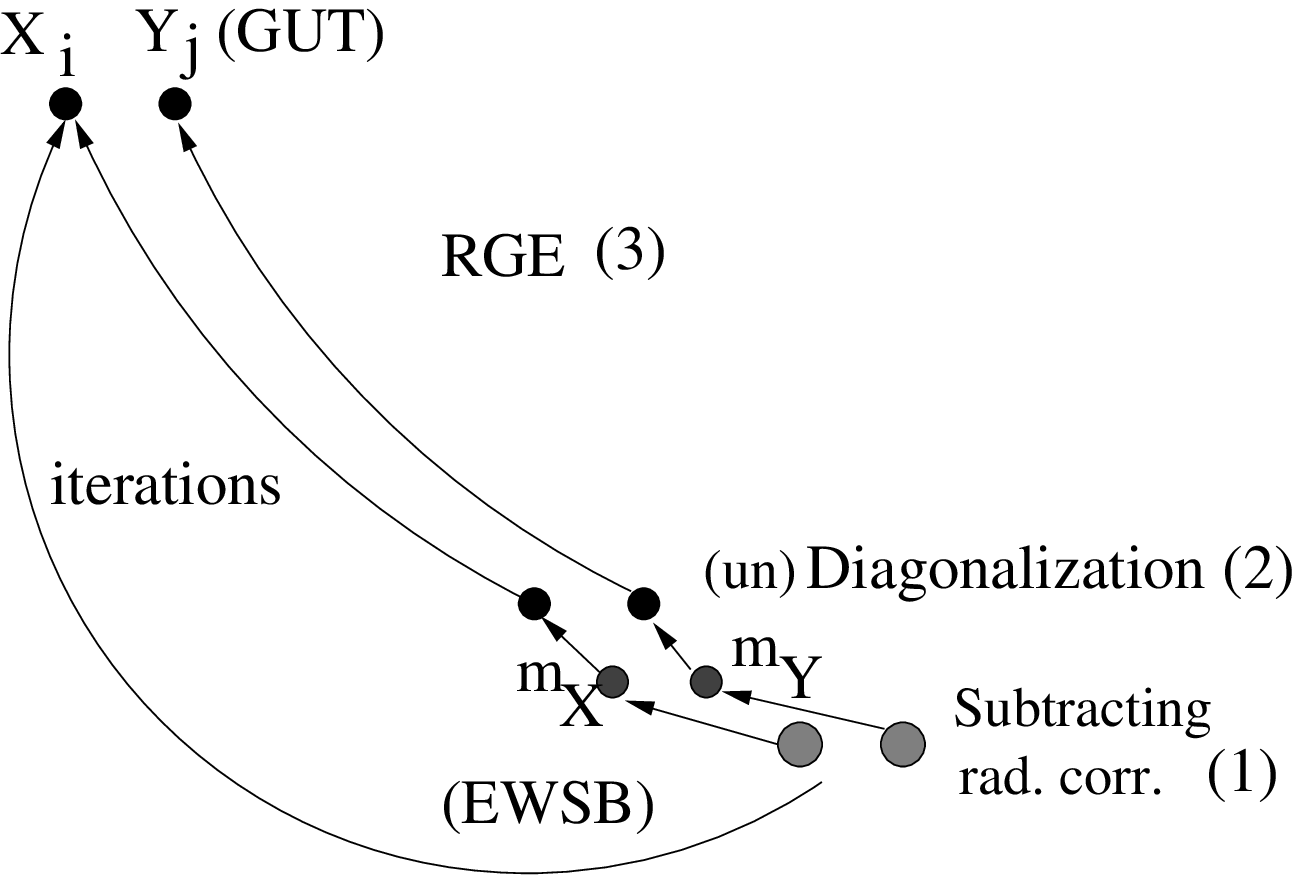,width=7.5cm}}
\caption[long]{Schematic algorithms of the top-down (left)  
and bottom-up (right) mappings illustrating their similarities in practical calculations. $X_i$ and $Y_j$ are a
set of (soft or SUSY) running parameters, RG-evolved between GUT and EWSB
scales (steps (1) or (3) respectively). $X_i$ and $Y_j$ may mix, and diagonalization
gives (running) mass eigenvalues $m_X, m_Y$ (step (2)). 
Then, additional radiative corrections linking 
running to pole masses $m^i_{pole}$  are added (resp. subtracted)
and may depend on extra unknown parameters or masses $Z_k, m_k$.
Iterations are needed at 
several steps in each approaches, as well as specific 
assumptions on some a priori unknown parameters.}
\label{figtdbu}
  \end{center}
 \end{figure}
We explain here on general grounds how we
incorporate radiative corrections into our algorithms, 
with specific mass/parameter relations   
to be given later, once having defined algorithms for the different
sectors more precisely. Prior to the bottom-up RGE comes  
the question of incorporating radiative corrections
linking the running parameters to the physical (pole) masses, as generically
indicated by the terms $\Delta f_{rad.corr.}$ in Eq.~(\ref{definv}).  
Clearly, incorporating the full radiative corrections to {\em all} 
sparticle pole
masses irremediably spoils
such simple analytic inversions, since the complete 
radiative corrections would introduce, already at one-loop level, 
highly non-linear dependence 
upon (almost) all parameters of the MSSM. However, 
upon assuming a certain level of (reasonably good) approximations 
for these radiative corrections, it turns out to be relatively
easy to incorporate these at a realistic level. This is especially 
the case for the sparticle masses entering
the relevant cascade decays: typically the (first two generation)
squarks receive radiative corrections that are largely dominated\cite{bpmz} 
(at one-loop) by gluon/squark and gluino/quarks QCD corrections, 
involving precisely the same sparticles entering the cascade. 
Other corrections, like the electroweak ones, are
fairly negligible\cite{bpmz} in comparison. Similarly, corrections
to the gluino mass are essentially dominated by gluon/squark QCD corrections.
Consequently it is rather simple to
``subtract out"  those corrections, starting from the experimentally
measured pole masses, and applying next tree-level algorithms
to the running masses. It eventually needs to apply this procedure
in several steps using numerical iterations. 
Although this procedure may appear rather involved, 
we emphasize 
that it is very similar to the manipulations which 
are performed in a standard top-down approach, where iterations
are anyway necessary in spectrum calculations\cite{suspect,softsusy,spheno,isasugra} 
once including radiative corrections. The similarities between standard
top-down and our bottom-up practical calculations are illustrated sketchily
in Fig.~\ref{figtdbu}: $X_i$ and $Y_j$ represent some of the relevant 
(soft or SUSY) running parameters, to be RG-evolved between GUT and EWSB
scales (step (1) in the top-down approach or (3) in the bottom-up approach respectively). Then these parameters can have  mixing so that diagonalization
in step (2) gives (running) mass eigenvalues $m_X, m_Y$ in the top-down case.
In the bottom-up case, rather than performing a brute force 
inverse diagonalization at step (2),
it is more convenient to use appropriate 
relations\cite{inv1} such that the required output 
MSSM parameters are in one-to-one relations
with the few accessible input masses. (We will see a specific example
of such relations for Eq.~(\ref{definv}) in the neutralino sector in section 3 below).      
Next, the necessary radiative corrections which link the  
running to the pole masses $m^i_{pole}$  are added (respectively subtracted
in the bottom-up case).
These may depend on extra unknown parameters or masses $Z_k, m_k$
(in which case definite assumptions on these unknown parameters are needed). 
Iterations are performed 
in each approaches for the radiative correction steps, since these
depend on final sparticle masses. (Iterations are also 
needed for the RGE since the EWSB
scale and other relevant parameters depend on the sparticle spectrum).  
In practice the kind of subtractions and other 
numerical manipulations that are needed specifically here are 
made easier by a number of possibilities
included in the latest version of the SuSpect\cite{suspect} code~\footnote{like 
e.g. 
the option to ``switch off" gradually some of the radiative corrections to the
sparticle masses, as illustrated in Table \ref{susoutput}.}. Concerning the neutralino masses, radiative corrections
are know to be reasonably small, and moreover to a very good approximation
one can incorporate the leading ones in the form of tree-level
deviations on the parameter $\mu$, $M_1$, $M_2$, allowing again 
subtractions and iteration procedures when applying
tree-level reconstruction algorithms. Moreover, in some cases we also incorporate
the extra unknown radiative corrections by assuming typically universality
relations {\em within} the loop-level calculations. This may induce
a little bias, but we consider (and have explicitly checked for the SPS1a case)
this to be a reasonably good approximation,
even when considering a general MSSM reconstruction. 

Concerning the Higgs sector, radiative
corrections to the light Higgs mass $m_h$ and pseudoscalar mass $m_A$
are known to be of primary importance. But, as is well known, the leading 
contributions essentially come from the stop sector, and more generally 
there exist approximations\cite{Svenetal} that are excellent 
to 1-2 GeV level~\footnote{The latter approximations are also incorporated 
as one option in
the SuSpect code, alternatively to the full one-loop, or full one-loop plus 
leading two-loop
calculations options.}, i.e. of the order of higher order uncertainties\cite{mh1-2loop}.   
 
On general grounds, even within the present
state of the art, the known radiative corrections to sparticle
and Higgs mass still suffer from uncertainties due to unknown higher
orders. Moreover at the LHC experimental errors are generally larger
than the latter theoretical errors (except for the lightest Higgs mass $m_h$). These features evidently affect our
analysis, but in the same way as any other more standard top-down approach
to the reconstruction of MSSM parameters at the LHC. 
Clearly the real limitation 
in incorporating radiative corrections does not come from the eventual
complexity of incorporating these numerically within a particular procedure,
but rather on the uncertainties resulting from unknown sparticle masses
contributing at the loop level to a given observable. 
In some cases where the latter uncertainties may particularly
affect our results, we take these into account 
as theoretical uncertainties, as will be specified. 
Overall we consider that
our treatment of radiative corrections as described here should be sufficient
for our rather limited purpose.   
\subsection{Treatment of mass uncertainties and interpretation}
\begin{table}[h!]
\begin{center}
\caption{\label{tabexp} 
Experimental accuracies on mass determinations 
from LHC gluino cascade and other decays (taken from refs.\cite{cascade}
and \cite{LHCstudy}), corresponding to the different input mass
scenarios defined in Table 
\ref{tabinput}.} 
\vspace{.5cm}
\begin{tabular}{|c|c||c|}
\hline\hline
 mass & expected LHC   & decay or
process \\
                 &  accuracy (GeV) &         \\
\hline
  $m_{\t g}$  &  7.2   &$\t g$ cascade decay \\
 $m_{\t N_1}$ & 3.7  & " " \\
 $m_{\t  N_2}$  & 3.6 & " " \\
 $m_{\t q_L}$  &  3.7    & " "  \\
 $m_{\t l_R}$ & 6.0    & " "   \\
\hline
 $m_{\t N_4}$ & 5.1  &  $\t q_L \to \t \chi^0_4+..$ cascade \\
\hline
 $m_{\t b_1}$ & 7.5      & $\t g$ cascade decay   \\
 $m_{\t b_2}$ & 7.9      & " "   \\
\hline
 $m_h$         & 0.25 (exp)--2 (th) & $h \to 
\gamma\gamma$ (mainly)   \\  
\hline\hline
\end{tabular}
\end{center}
\end{table}
For the different scenarios considered we will illustrate the expected accuracy on
the reconstructed parameters for given mass measurement accuracies assumed according
to Table \ref{tabexp} 
(eventually considering also theoretical errors). To delineate this
error propagation we have performed various scanning over the 
input mass values within errors, or over relevant MSSM parameters, either
with uniformly distributed random numbers, or alternatively also using random 
numbers with a Gaussian distribution (in which case
we can define confidence level intervals)~\footnote{There are a few cases 
in our analysis where uniform 
``flat prior'' distributions may give misleading ``density'' regions for the 
resulting constraints on some of the parameters (typically for $\tb$, see sec.
4.2 for illustration and discussion). In such cases we made 
obvious changes for more appropriate non-flat distributions, but  
have not made any attempt to define much refined priors in a Bayesian 
approach such as is done notably in \cite{ben_bayes}.}.
The sparticle mass errors as quoted in Table \ref{tabexp}
are, however, known to be not purely statistical: there
is a large part which comes from the systematic errors on jet
resolution\cite{cascade1,cascade2}, and moreover these errors are also
strongly correlated. 
We stress however that a more involved treatment of uncertainties, properly
combining the 
statistic and systematic ones, taking into account correlations etc, 
appears quite non-trivial\cite{dzerwas} 
and is beyond the scope of the present paper.
One could in principle make substantial improvement
in the final determination of parameters by using directly
the endpoint measurements\cite{cascade1,cascade2} of the
gluino cascade rather than the naive mass errors obtained from the latter.  
Consequently, one should keep in mind  that the interpretation
of the various domains and contours in parameter space that we shall
obtain are lacking a very precise statistical significance. 
(We plan to perform a more refined statistical analysis in the 
future\cite{prepa}). 
Despite these limitations, we will illustrate detailed      
comparisons for most considered scenarios of bottom-up
determination results with those obtained from
more standard statistical treatment with $\chi^2$ minimization 
in a top-down approach using MINUIT\cite{minuit}.   
\section{Gaugino/Higgsino parameter determination from gluino cascade}
We start by recalling some analytic inversion algorithms
at the tree-level, adapted to the LHC input scenarios (i.e. corresponding
to the extractable masses in the gluino decay chain as discussed above).
Note that our algorithms
may be valid more generally, e.g. at the ILC, provided that 
the same input masses would be available.  
\subsection{Gaugino/Higgsino parameter: general case inversion}
We thus consider the parameters relevant to the gaugino/Higgsino sector, 
starting from the neutralino mass matrix:
\beq
M_N = \left(
\begin{array}{cccc} {M_1} & 0 & -m_Z s_W \cos\beta & m_Z s_W \sin\beta  \\
 0 & { M_2} &  m_Z c_W \cos\beta & -m_Z c_W \sin\beta  \\
 -m_Z s_W \cos\beta & m_Z c_W \cos\beta & 0 & -{ \mu }\\
  m_Z s_W \sin\beta & -m_Z c_W \sin\beta & -{ \mu} & 0 
\end{array} \right)\;.
\label{mino}
\eeq
Rather than performing an involved 
inverse diagonalization, which would moreover need to know all the four
neutralino masses, 
it is much more convenient to use appropriate 
relations among parameters involving fewer input masses.
The four invariants (under diagonalization transformation):
\beq
  && { Tr M_N} ,\;\; 
 { \frac{(Tr M_N)^2}{2} - \frac{Tr (M^2_N)}{2}} \nn \\ 
  && { \frac{(Tr M_N)^3}{6} - \frac{Tr M  \,\,\, Tr (M^2_N)}{2} + 
    \frac{Tr (M^3_N)}{3}}\;,\;\; { Det M_N}  
\label{4inv}
\eeq
provide a system of equations\cite{inv1} which can be used in  
different ways depending on the choice of input and output parameters. 
Two equations are actually expressing necessary and sufficient conditions
for the existence of solutions to this system (see also Appendix B of ref.
\cite{inv1} for more details):
\be
P^2_{ij}+(\mu^2+m_Z^2-M_1 M_2+(M_1+M_2)S_{ij}-S_{ij}^2)P_{ij}+
\mu m_Z^2(c_W^2 M_1+s_W^2 M_2) \sin 2\beta-\mu^2 M_1 M_2=0 
\label{b4}
\ee
and
\beq
&(M_1+M_2-S_{ij})P^2_{ij}+(\mu^2(M_1+M_2)+m_Z^2(c_W^2 M_1+s_W^2 
M_2-\mu \sin 2\beta))P_{ij}\nn \\
&+\mu( m_Z^2(c_W^2 M_1+s_W^2 M_2) 
\sin 2\beta-\mu M_1 M_2)S_{ij}=0
\label{b5}
\eeq
where we define for short    
    $S_{ij} \equiv m_{\t N_i} + m_{\t N_j} $, 
     $P_{ij} \equiv m_{\t N_i} m_{\t N_j} $ 
     where $i,j=1,...4$~\footnote{Note that $m_{\t N_i}, m_{\t N_j}$ 
can be any two neutralinos, all these equations being symmetrical
under any neutralino mass permutations.}, and 
     $s_W=\sin\theta_W$, $c_W=\cos\theta_W$.\\ 
Note that Eqs.~(\ref{b4}), (\ref{b5}) involve only two neutralino masses,
which corresponds to our minimal input assumptions in Table
\ref{tabinput}.  
These are originally tree-level relations but,
as explained in sub-section 2.3, in our analysis we shall incorporate as much as
possible of realistic radiative corrections. To begin, the values of  
$s^2_W$ and $m_Z$ in expressions (\ref{b4}),(\ref{b5}) are understood
as the properly defined
$\overline{DR}$ scheme parameters: $\bar s^2_W$ and $\bar m_Z$.

If chargino masses were known at this stage  
Eqs.~(\ref{b4}), (\ref{b5}) would lead rather simply to a unique solution for 
$M_1$  for given $\mu$, $M_2$ and $\tan\beta$\cite{inv1}. This 
had been studied in the past for chargino and neutralino 
mass measurement prospects at the ILC. 
Precise determinations of the chargino/neutralino parameters at the 
ILC, partly based on analytic (or semi-analytic) inverted 
relations in the neutralino and chargino sector, have been largely
analysed in ref.\cite{zerwasetal}. But  
since we do not assume chargino masses to be measured in our scenarios (which
appears anyway more challenging at LHC),  and given the parameters 
entering the relevant gluino/squark decay,
it is more appropriate to use Eqs.~(\ref{b4}), (\ref{b5}) differently as 
we examine next~\footnote{NB another recent analysis of the neutralino 
system in the LHC context of 
gluino/squark cascade decays has been performed in ref. \cite{polesetal}, 
also partly based on semi-analytic relations, though very different from ours
and not relying on exactly the same input.}.  
\subsubsection{Scenario S1: determining $M_1$, $M_2$
from $m_{\t N_1}$, $m_{\t N_2}$ in general MSSM}\label{s31}
We first consider a general (unconstrained)
MSSM scenario S1, assuming non-universality of gaugino masses. 
We then use Eqs.(\ref{b4}), (\ref{b5})
to determine $M_1$ and $M_2$ from (any) two
neutralino mass input: we thus take
$m_{\t N_1}$, $m_{\t N_2}$ input extracted from the
cascade decay, for given $\mu$ and $\tb$ parameter input. It is 
straightforward after some algebra to work out from Eqs.(\ref{b4}), (\ref{b5}) 
these $M_1, M_2$ solutions (e.g. eliminating first $M_1$
which depends linearly on $M_2$ from one of the two equations, and obtaining
a quadratic equation for $M_2$). 
For completeness the explicit 
solutions and related issues are worked out in some detail  
in Appendix A (see Eqs.~(\ref{M1sol})--(\ref{abcm1m2})). We note here that
the solution for $M_1$, $M_2$ 
has actually a twofold ambiguity, being obtained from a quadratic equation
e.g. for $M_2$. More basically   
Eqs.~(\ref{b4}), (\ref{b5}) as well as all other relations
(\ref{4inv}) only use information on mass
eigenvalues, and are
invariant under any neutralino mass permutations, 
e.g. $m_{\t N_1} \leftrightarrow m_{\t N_2}$. 
Accordingly, without further theoretical assumptions on gaugino 
mass terms, one cannot establish 
the hierarchy between the two gaugino (and the Higgsino) mass parameters
from the sole knowledge of those two neutralino masses,
unless extra information on the diagonalizing matrix
elements is available (which amounts to have information on some of 
the neutralino couplings to other particles). 
Thus in a general gaugino mass scenario 
there are two cases to consider, depending on the  
relative values of the Bino and Wino soft mass terms: either 
$M_1 < M_2$, as
in most mSUGRA scenarios, or a reverse hierarchy  $M_2 < M_1$ (as in
the case of e.g. AMSB models).

When assuming a
well-defined Bino/Wino mass hierarchy, the $(M_1, M_2)$ solution is then  
unique~\footnote{We   
assume $M_2>0$ in addition, which one always has the freedom to choose in MSSM\cite{kane}.}.   
Now taking central values of the masses
$m_{\t N_1}$, $m_{\t N_2}$ plus the reference SPS1a values of $\mu$ and $\tb$ we
recover the correct SPS1a values of $M_1$, $M_2$ if assuming $M_1 < M_2$,  
or another possible solution in general MSSM with $M_2 < M_1$ as is 
examined further below (see also Appendix A for more details).   
More interesting 
than this explicit solution for fixed input values is to determine the expected 
accuracy on output parameters, 
given the experimental uncertainties on neutralino masses, 
and the sensitivity of $M_1, M_2$ to the presumably limited 
knowledge on the two other basic 
parameters $\mu$ and $\tb$. This error propagation and other issues in
the reconstruction of $M_1, M_2$ for
the SPS1a test case will be illustrated below in subsection
\ref{s34}.
\subsubsection{Scenario S2: determining $\mu$, $\tb$ with
gaugino mass universality}\label{s32}
In a different scenario we consider  
the very same basic Eqs.~(\ref{b4}), (\ref{b5}) but changing input/output: 
adding now the gaugino universality assumption: $M_1=M_2=M_3$ at the
GUT scale, we first determine $ M_1, M_2$ from $M_3$, at the EWSB scale. 
(This does not necessarily imply a mSUGRA model, since non-universal relations
could still hold for all other MSSM parameters apart gaugino mass terms. At this
stage one could also start from any other well-defined relation between
the $M_i$'s at some given scale, like is the case for AMSB and GMSB models).
As a consequence of the related RGE structure of gaugino masses and gauge couplings
at one-loop level, the relation in the universality case reads: 
\be
               \frac{M_1}{g_1^2}=\frac{M_2}{g_2^2}=\frac{M_3}{g_3^2} 
\label{guniv}
\ee                 
(where $g_i$ are the properly normalized gauge couplings) 
to be valid at any scale. Then, Eqs.~(\ref{b4}), (\ref{b5}) are now used
to determine $\mu$ and $\tb$ for (universal) $M_1, M_2$ input, 
as a linear system for $\sin 2 \beta$ and  $\mu^2$. 
It is simple after some algebra to work out 
 those explicit solutions, e.g. first eliminating $\sin 2\beta$
 to get an expression for $\mu^2$ that only depends on $M_1, M_2$, 
and the two input neutralino masses. 
For completeness explicit solutions are given in Appendix A 
(see Eqs.~(\ref{mu2N}), (\ref{tb2N})). 
Our conventions are the usual ones such that $0 <\beta <\pi/2$, so that $\tb>0$
(and real). This is not a restriction on parameter space, 
since an eventual phase of $\tb$ can be absorbed by a consistent redefinition
of the Higgs doublet fields\cite{kane}.
The sign of $\mu$, however, is not determined by these equations, so we 
have to consider the two possible solutions for $\mu >0$ and $\mu<0$ a priori. 
As previously, as a cross-check we can plug in 
these expressions the central SPS1a values
for the masses $M_{\tilde g}$,  $\tilde{M}_{N_1}$, $\tilde{M}_{N_2}$
as obtained e.g from SuSpect, obtaining 
the correct values of $\tb$ and $\mu$.
In the numerical
applications for SPS1a reconstruction, illustrated in subsection
\ref{s34} below, we shall thus consider both 
$\mu>0$ or $\mu <0$ case (examining whether the latter may be eventually
eliminated when taking into account input mass accuracies.)  
\subsubsection{Scenario S3: three neutralino mass input (with and without
gaugino universality)}\label{s33}
What could be more constraining is the (more optimistic) scenario 
where three neutralino masses could be determined at the LHC, 
involving another squark decay measurement
(independent from the first gluino cascade decay)\cite{LHCstudy}, 
according to the input $S_3$ in Table \ref{tabinput} above.
In this case one can use very simply 
an extra relation originating from Eqs.~(\ref{4inv})
to get a determination of either $\mu$ or $\tb$. 
More precisely from the trace of the matrix (\ref{mino}) and the
second invariant in Eqs.~(\ref{4inv}), one obtains
a simple expression for $\mu^2$:
\beq      
      \mu^2 = M_1 M_2-m^2_Z-\left(P_{124}+S_{124}( M_1 +M_2-S_{124})\right)
\label{mu3N}
\eeq
where $S_{124}\equiv  m_{\t N_1}+m_{\t N_2}+m_{\t N_4}$
and $P_{124} \equiv  m_{\t N_1} m_{\t N_2} + m_{\t N_1} m_{\t N_4}+
 m_{\t N_2} m_{\t N_4}$.
Eq.~(\ref{mu3N}) can be first used in the non-universal scenario $S_1$ above,
thus determining $M_1, M_2$ and $|\mu|$ from three neutralino masses
(plus $\tb$) input. 
(Alternatively one may also solve this
system for $\tb$ instead of $\mu$, but since all expressions  
only depend on $\sin 2\beta$, it becomes rapidly insensitive for 
large enough $\tb$. Accordingly
we anticipate without calculations that it is unlikely 
to get any interesting $\tb$ (upper) bounds given the input mass
LHC accuracies, irrespectively of the amount of
neutralino masses measured.)       
Solving Eq.~(\ref{mu3N}) together with Eqs.~(\ref{b4}), (\ref{b5})
gives in fact a high (sixth) order polynomial equation for 
$M_1, M_2, \mu$ which thus cannot be solved fully analytically. It is however
easy to solve iteratively using e.g. Eq.~(\ref{mu3N}) on the solutions 
(\ref{M2sol}), (\ref{M1sol}) (upon having chosen a definite $M_1, M_2$ 
hierarchy). This iterative solution converges
very quickly (see Appendix A for more details).

When applied to the reconstruction for SPS1a test example, with 
corresponding input mass error propagation, 
this will result in a much more precise determination of $\mu$, 
as will be illustrated in subsection \ref{s36}.   
Note however that the sign of $\mu$ remains 
undetermined from this additional information. 
\subsection{Reconstructing $M_1$, $M_2$
in MSSM without universality assumptions: SPS1a test case}\label{s34}
\begin{figure}[h!]
\begin{center}
\includegraphics*[width=12cm,angle=-90]{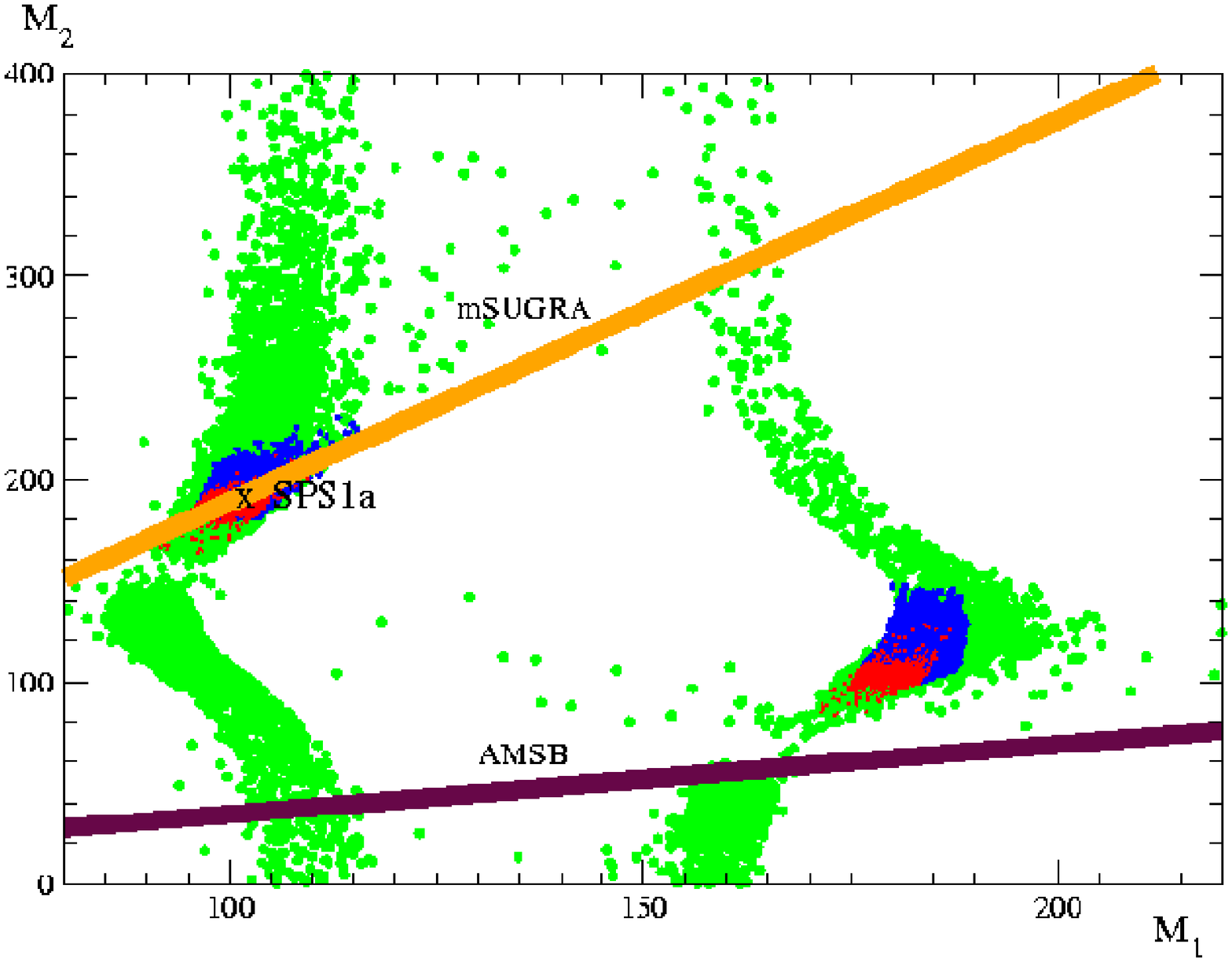}
\caption[long]{$M_1$, $M_2$ (in GeV units and at $Q_{EWSB}$ scale)
determination from two (resp. three) neutralino
masses $m_{\t N_1}$, $m_{\t N_2}$ ($m_{\t N_4}$) in unconstrained 
MSSM for different assumptions: 
1) green regions: $m_{\t N_1}$, $m_{\t N_2}$ input, 
$\mu =\mu(\mbox{SPS1a}) \pm 1$ TeV, $1 \lsim \tb \lsim 50$;
2) blue regions:   $m_{\t N_1}$, $m_{\t N_2}$ input, $1 < \tb <50$, $\Delta\mu=100$ GeV;
3) red regions:  $m_{\t N_1}$, $m_{\t N_2}$ + $m_{\t N_4}$ input, $1 \lsim \tb \lsim 50$
($\mu$ is thus more constrained due to the third neutralino mass, see main text).
Also shown are the mSUGRA or GMSB  (resp. AMSB) $M_1/M_2$ relations in orange 
(respectively in maroon)
including experimental errors on the masses.}
\label{figm1m2}
  \end{center}
 \end{figure}

We now apply the general solutions obtained for $M_1$, $M_2$ in the non-universal gaugino mass
scenario S1 (as described in subsection \ref{s31} and detailed in Appendix A), 
to the
actual reconstruction of those parameters for the SPS1a test case, taking into account  input mass error propagation~\footnote{Strictly speaking 
one should include here the theoretical
uncertainties on $\bar s^2_W$ as well, but the latter are rather small
in comparison to the experimental uncertainties on the other parameters. Yet,
this is related to the consistent inclusion of radiative corrections,
which contribute to $\bar s^2_W$. In fact this induces a small shift of the
central values but will affect very little the variation range of 
the output parameters $M_1$, $M_2$ here.}.
This is shown in Fig.~\ref{figm1m2}, where domains 
in the $M_1$, $M_2$ plane are obtained for accuracies on the two 
neutralino masses taken from Table \ref{tabexp}, resulting from a scan
with uniformly distributed random numbers. This illustrates in particular 
the twofold ambiguity in reconstructing $M_1, M_2$ from the sole
knowledge of two neutralino masses, as discussed in subsection \ref{s31}. 
We also consider 
different assumptions on the $\mu$ or $\tb$ range of variation, anticipating the difficulty 
in determining $\tb$ at the LHC solely from this cascade decay information, 
as will be confirmed more quantitatively in next sections. In practice 
we vary widely $\tb$, $1 < \tb < 50$.

We thus illustrate the
cases where both $\mu$ and $\tb$ would be largely undetermined, and 
how the $M_1, M_2$ determination is improving if a
more precise determination of $\mu$ can be available (anticipating
the better accuracies that may be obtained from more theoretical
assumptions, or other LHC processes, 
 or alternatively if supplementing our analysis with 
 ILC determination of parameters). 
One observes from
Fig.~\ref{figm1m2} that for largely unknown $\mu$ and $\tb$ 
the constraint obtained on $M_1$ from using solely the two neutralino mass input is fairly
reasonable: 80 GeV $\lsim M_1 \lsim $ 120 GeV (for the $M_1 < M_2$ mSUGRA-like pattern), and a similar accuracy for the $M_1 >M_2$ case. 
In contrast $M_2$ appears more poorly constrained in both cases.
Moreover note that only the region $M_2 \lsim 400$ GeV is shown on this plot,
while actually there are a few isolated points obtained from the scan with 
higher $M_2$ values, for reasons to be explained below.   
We have checked that using a regular grid scan, instead of random numbers, 
does not significantly alter the contours in Figs.~\ref{figm1m2}-\ref{figm1m2mu} or similar other figures as will be presented below. 
One should indeed be careful in the interpretation  
of the density levels of various regions, since our
scan was performed here with uniformly distributed random numbers.
Accordingly the density levels of points as appearing
in Fig.~\ref{figm1m2} principally reflect that  
the determination of $M_1$, $M_2$ from Eqs.~(\ref{b4})-(\ref{b5})  
is very non-linear with respect to $\mu$ (and with respect
to $\tb$ to some extent), see Eqs.~(\ref{M1sol}), (\ref{M2sol}) in Appendix A, 
and have thus no direct meaning of statistical confidence levels. In next
sections we often make explicit comparisons between uniform ``flat prior'' and Gaussian scanning of parameters: in the latter case, statistical confidence    levels can be more properly defined (with the cautions however mentioned in 
sub-section 2.4, regarding the fact that the data used in the present work  
are not purely statistical anyway). In some cases the differences are
significant and deserve a more careful analysis, as we will see.   
\begin{figure}[h!]
\begin{center}
\mbox{
\epsfig{figure=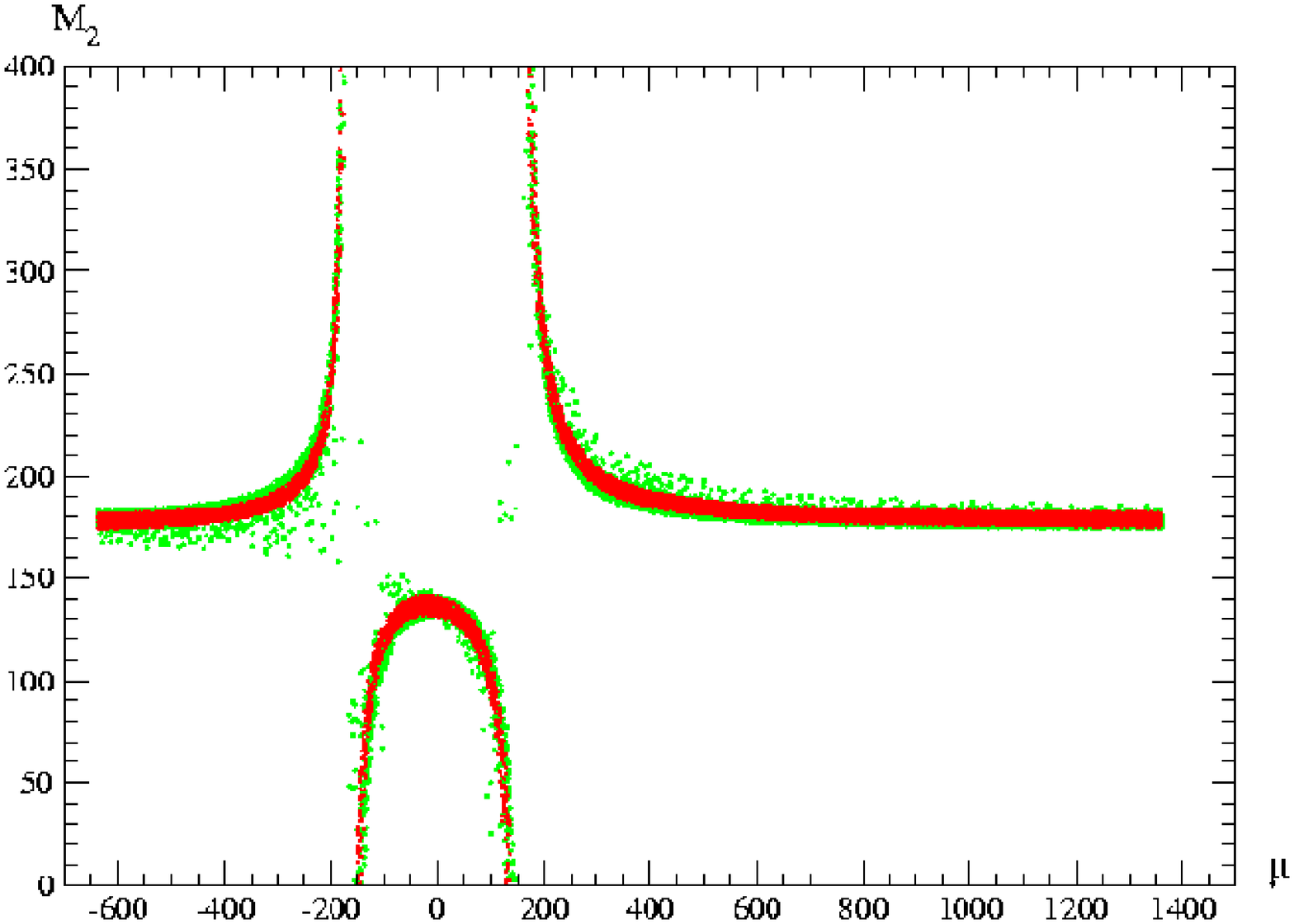,angle=-90,width=8cm}
\epsfig{figure=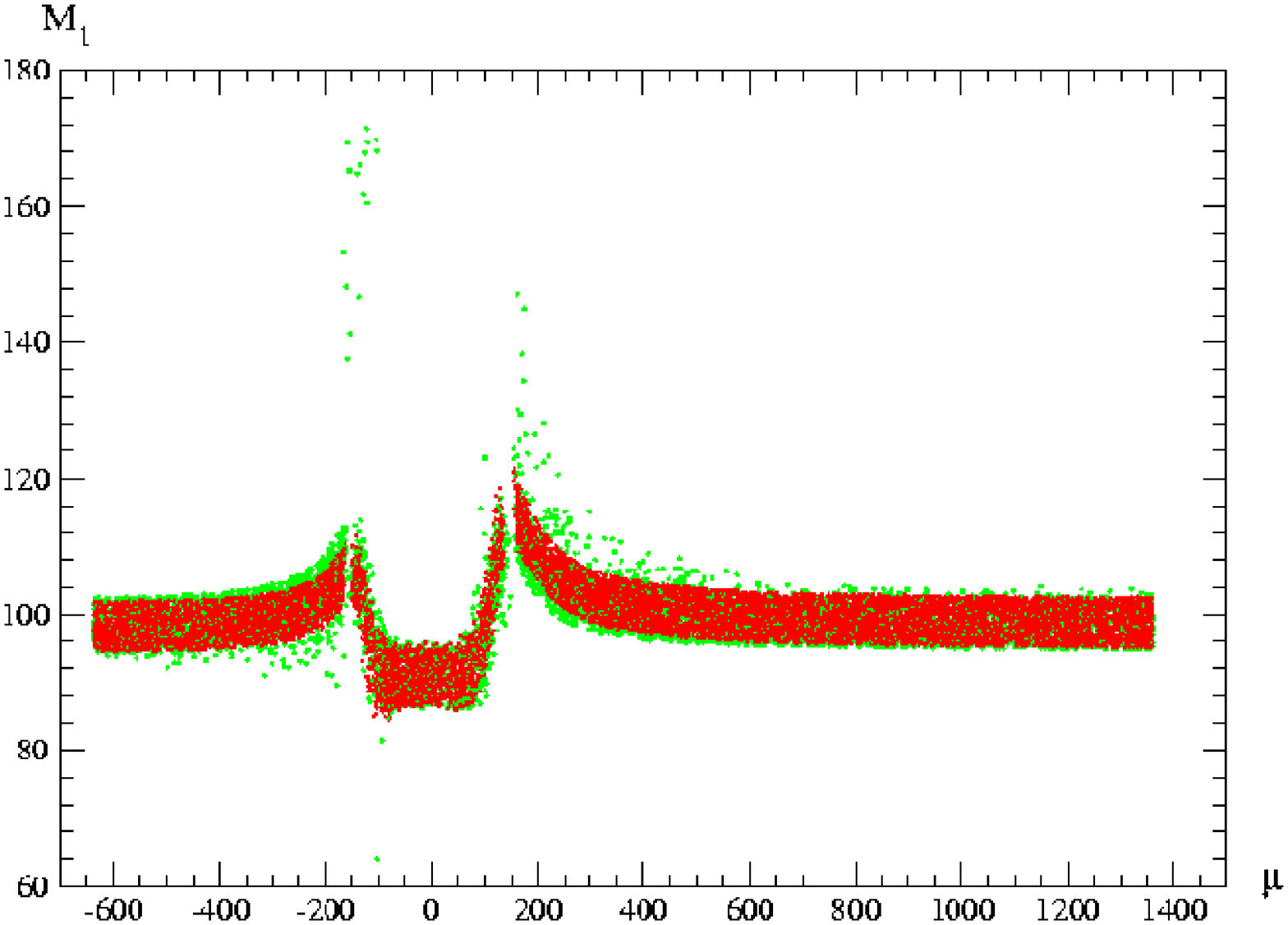,angle=-90,width=8cm}}
\caption{Correlated values of $M_2$ (left) and $M_1$ (right) for the 
$M_1 <M_2$ case
at $Q_{EWSB}$ scale as functions of $\mu$, for $m_{\t N_1}, m_{\t N_2}$ input. 
The spreading of points is
due to the variation $1 <\tb<50$ (in green) as well as the variation of $m_{\t N_1}$, $m_{\t N_2}$ within accuracy (in red).} 
\label{figm1m2mu}
  \end{center}
 \end{figure}

The cases of
moderate (blue region) and accurate (red region) determination of $\mu$
is giving much more interesting constraints. The red regions are anticipating 
the resulting  
accuracy on $\mu, M_1, M_2$ when a third neutralino $m_{\t N_4}$ can be measured,
as will be analyzed in a next subsection \ref{s36}.
According to Eq.~(\ref{mu3N}) in this case $\mu$ is determined (independently
of $\tb$) and can be combined with the previous Eqs.~(\ref{b4}), (\ref{b5})
to obtain much improved $M_1, M_2$ determination. 
(Alternatively another rather good 
determination of $\mu$ is also obtained when the latter is 
not arbitrary, as is assumed in a general MSSM, but calculated
from EWSB consistency conditions from universal Higgs and sfermion mass terms
at the GUT scale, as will be analyzed in section 6.)\footnote{For given $\mu$ input, the measurement of a third neutralino mass could in principle resolve
the twofold $M_1, M_2$ ambiguity, since the two different solutions give 
different $m_{\t N_3}$, $m_{\t N_4}$ values. But the latter masses
being essentially determined by $|\mu|$ (at least when $|\mu| > M_1, M_2$), the differences corresponding to those two solutions are often small (e.g. only
$\sim 4$ GeV for $m_{\t N_4}$, for fixed SPS1a
values of $\mu$, which is smaller than the expected $m_{\t N_4}$ LHC accuracy.) 
So one would need to determine $m_{\t N_4}$ 
(and $\mu$) accurately to really disentangle the two $M_1, M_2$ solutions.}    
Now, contour plots like those in Fig.~\ref{figm1m2} are not very
informative as concerns the improvement in $M_1$ or $M_2$
determination to be expected
when increasing $\mu$ (or eventually $\tb$) accuracies respectively. 
To trace more clearly this behaviour, we plot in Fig.~\ref{figm1m2mu}
the equivalent of the green contour of Fig.~\ref{figm1m2} but in the
$(\mu,M_2)$ and $(\mu,M_1)$ planes respectively (and for the $M_1 <M_2$ case). 
(Note that the values
of $M_1$ and $M_2$ are entirely correlated since both are obtained
from Eqs.~(\ref{M1sol}), {\ref{M2sol})). The spreading
of points in these plots is due to the variation of $\tb$ and
$m_{\t N_1}$, $m_{\t N_2}$. More precisely, what is shown 
in red in the effect of $m_{\t N_1}$, $m_{\t N_2}$ experimental errors only, for fixed
SPS1a value of $\tb=9.74$, while the additional green points 
correspond to $1< \tb < 50$. 
These plots are thus essentially the solutions $M_{1,2}(\mu)$ from
Eqs.~(\ref{M1sol})-(\ref{M2sol}), that would reduce to simple curves for fixed
$\tb$, $m_{\t N_1}$, $m_{\t N_2}$. 
One can see the structure of solutions for  
$(M_2,\mu)$ (and correspondingly $(M_1,\mu)$) with different
domains, originating from
the $\mu$ dependence in Eqs.~(\ref{M2sol}), 
with strong correlations. In fact $M_2(\mu)$ becomes arbitrarily large for
two $\mu$ values,  
for $\mu <0$ and $\mu >0$ (which 
are not exactly symmetrical, see Appendix A): for instance for SPS1a values of 
$\tb$, $m_{\t N_1}$, $m_{\t N_2}$, the positive ``pole'' is at $\mu\sim 118$ GeV.  
This explains the loose determination
of $M_2$ for large variations of $\mu$, and also explains the density
levels of scanned points in Fig.~\ref{figm1m2}. In contrast, $M_1$ always
remains finite when $M_2$ becomes arbitrary large, according to Eq.~(\ref{M1sol}).
This also explains the much better constraints on $M_1$ in Fig.~\ref{figm1m2} irrespectively of the $M_2$ behaviour.        

Next, one can see that both $M_2$  
and $M_1$ can be much better constrained, irrespectively
of $\tb$ values, as soon as the $\mu$ determination is slightly better 
(such that $\mu$ remains sufficiently far from these poles). 
This explains the much improved constraints on $M_2$ and $M_1$ for the 
blue contour in Fig.~\ref{figm1m2} (and a fortiori for the red contours
where $\mu$ is tightly constrained from the third neutralino mass as will be
discussed more in sub-sec.~\ref{s36}).   
All these properties are rather simple consequences 
of the basic Eqs.~(\ref{b4})-(\ref{b5}), 
and illustrate useful informations that would be very difficult
to delineate from a more standard top-down fit of parameters.    
Actually the poles for specific $\mu$ values are artifacts of our inversion equations, but more physically it simply means that to obtain precisely   
the $m_{\t N_1}$, $m_{\t N_2}$ SPS1a values for those particular $\mu$ values,  
 $M_2$ would have to be unreasonably large. 
Going back to the standard top-down approach, it also means that 
 performing e.g. a $\chi^2$ fit of the neutralino masses is likely to give  
 a very flat behaviour of the $\chi^2$ near this $\mu$ region: 
 more precisely, since $M_2$ varies widely around these $\mu$ values, 
 no clear ``best fit'' $M_2$ value will be
 found, or with a very large error, and/or that the $\chi^2_{min}$ 
 value will be bad. This is fully
 confirmed by the results of a MINUIT fit: 
 if $\mu$ is fixed to $\sim 120$ GeV the minimization does not give
useful constraints, 
MINUIT finds typically errors like:
 \be
M_1 = 2000 \pm 258\;\mbox{GeV}, \;\; M_2 = 2000 \pm 88\;\mbox{GeV}\;.
\ee
with even many more minima and errors found for very large $M_2$ value. \\
In contrast, fixing  $\mu$ (and $\tb$) to their SPS1a values and fitting
only $m_{\t N_1}$, $m_{\t N_2}$, gives very good accuracy on $M_1, M_2$, as 
will be discussed in a next sub-section below where other MINUIT 
fit results are given (see Table~\ref{minuit_m3mutb}):
 \be
M_1 = 108.8 \pm 5.8\;\mbox{GeV}, \;\; M_2 = 208.9 \pm 6.4\;\mbox{GeV}\;.
\ee

Next, if gaugino mass universality at the GUT scale is assumed, 
as expected one obtains stronger constraints. 
This is illustrated on Fig.~\ref{figm1m2} by the (orange)
band resulting from the ``mSUGRA" relation: $M_1 \simeq 0.53 M_2$ 
at the low energy scale $Q\simeq Q_{EWSB}$, from $M_1(Q_{GUT})=
M_2(Q_{GUT})$ at GUT scale. 
The width of this band results from the error on $M_3$, i.e. $m_{\t g}$  determination. 
We will see in next subsection how to make this study 
more precise when the (mSUGRA) gaugino mass universality is assumed. 
We anticipate, however, that for the given two neutralino and gluino mass 
accuracies, constraints on $\mu$ 
will be mild, even with gaugino mass universality assumptions, while 
those on $\tb$ almost absent. Indeed, one can see on Fig.~\ref{figm1m2} 
that the ``mSUGRA" band is 
compatible with a part of the green region 
where $\tb$ (and $\mu$) are essentially undetermined.\\   
Next, since the contours in Fig.~\ref{figm1m2} 
are valid for arbitrary gaugino
masses, it is straightforward to superpose different gaugino mass relations,
for instance in AMSB\cite{AMSB} models where the $M_1/M_2$ relation
is also fixed from high scale boundary conditions and RG evolution, but 
is very different: $M_1(Q_{EWSB}) \simeq 2.9 M_2(Q_{EWSB})$.
We show similarly on Fig.~\ref{figm1m2} this `AMSB" band (in maroon) 
including its width originating from the $M_3$ uncertainty. 
In this way one may possibly
distinguish, depending on the accuracy, between e.g. mSUGRA/GMSB and AMSB 
models from the neutralino mass measurements. 
(Note however that the relation between
$M_1$ and $M_2$ in GMSB models\cite{GMSB} is completely indistinguishable 
from the mSUGRA relation at this accuracy level).     
More precisely one can see here how AMSB would be excluded if moderate
(in blue) or accurate (in red) $\mu$ measurements could be achieved (even 
when considering the second solution with $M_1 > M_2$ which has
an AMSB-like hierarchy pattern).    
This is also a consistency cross-check, in the present analysis, since 
we started from a mSUGRA model SPS1a ``data".
\subsection{Reconstructing $\mu$, $\tb$ with
gaugino mass universality for SPS1a case}\label{s35}
Let us now consider 
scenario S2 as discussed above, assuming gaugino mass universality 
Eq.~(\ref{guniv}) to determine $M_1$ and $M_2$, and next
using basic Eqs.~(\ref{b4}), (\ref{b5}) to determine $\mu^2$ and $\tb$ for the
SPS1a test case (see explicit solutions Eqs.~(\ref{mu2N}), (\ref{a2b22N}) 
in Appendix A).   
Numerically, for SPS1a, gaugino mass universality at GUT scale 
gives at
the relevant low energy EWSB scale approximately:
\be
      M_1\simeq 0.17 M_3 \;,  \;\;\;\;
      M_2 \simeq 0.33  M_3\;,\;\;\;M_1 \simeq 0.53 M_2\;.
\ee 
To determine $M_1$, $M_2$ from $M_3$, we first extract $M_3$ from the  
the gluino pole mass $m_{\tilde g}$ as 
\be
m^{pole}_{\tilde g} = M_3(Q)+\Sigma_{\tilde g}(Q;m_{\t q},\cdots)
\ee 
by subtracting out the leading radiative corrections $\Sigma_{\tilde g}$ 
to the gluino
mass: those are dominantly due to squarks, and thus largely predictable in our
framework, as discussed above in sub-section 2.3. This induces a non-negligible
shift, since for SPS1a the correction  $\Sigma_{\tilde g} \sim 20$ GeV,
with $M_3(Q_{EWSB})\sim 600$ GeV.\\   
As mentioned in sub-section \ref{s32}, 
The solutions of (\ref{b4}), (\ref{b5}) for 
$\tb$ and  $\mu$ 
do not determine the sign of $\mu$,  so in the reconstruction with error
propagation we 
have to consider the two possible solutions for $\mu >0$ and $\mu<0$. 
We first vary $M_{\tilde g}$,  $m_{\t N_1}$, $m_{\t N_2}$ 
within accuracy according to Table \ref{tabexp}. 
Scanning the values with (uniformly distributed) random numbers, with the 
conditions:  $\mu$ real and $\tb>0$, is illustrated in Figs. 
\ref{fign1n2}--\ref{figtbmu-}.    
       \begin{figure}[h!]
  \begin{center}  
   \includegraphics*[width=8cm,angle=-90]{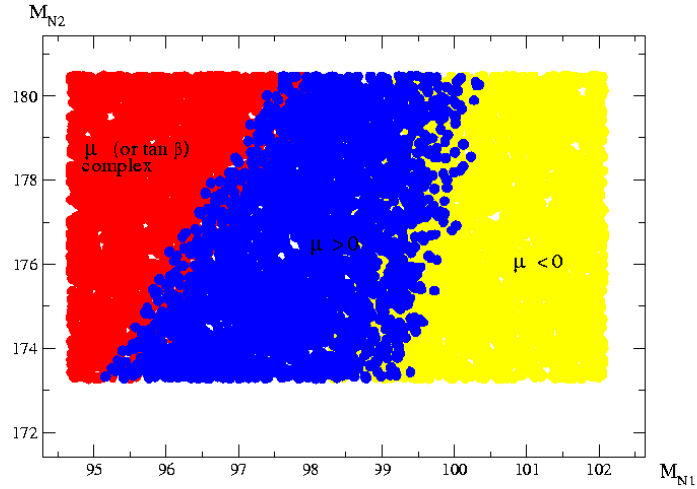}
\caption{Allowed domain for $m_{\t N_1}$,
$m_{\t N_2}$ corresponding to complex or real $\mu$.} 
\label{fign1n2}
  \end{center}
 \end{figure}
       \begin{figure}[h!]
  \begin{center}
  \includegraphics*[width=8cm,angle=-90]{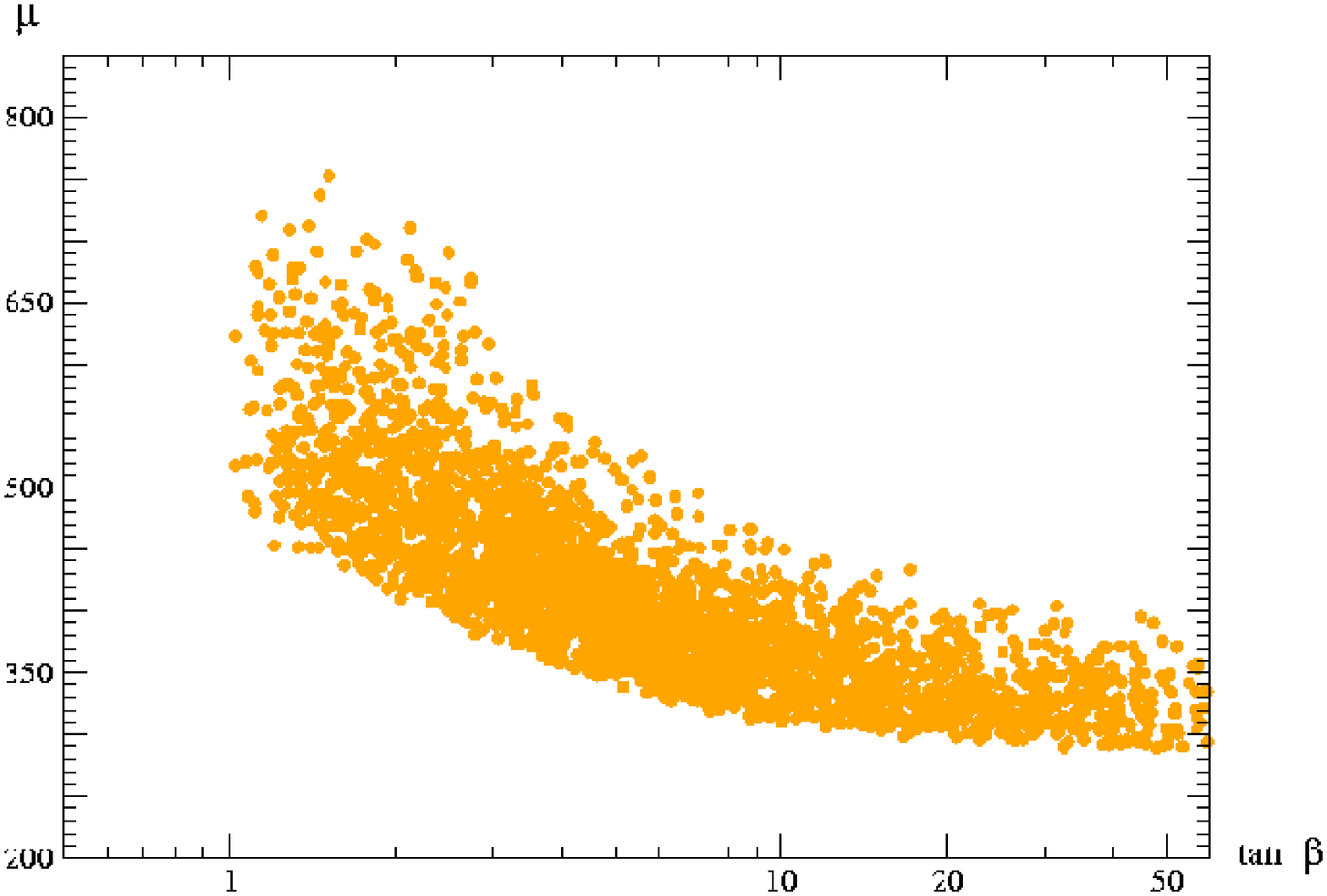}
\caption{Constraints on $\mu>0$, $\tb$ from gaugino universality and
gluino plus two neutralino mass measurements in gluino cascade decay. } 
\label{figtbmu+}
  \end{center}
 \end{figure}
        \begin{figure}[h!]
  \begin{center}
  \includegraphics*[width=8cm,angle=-90]{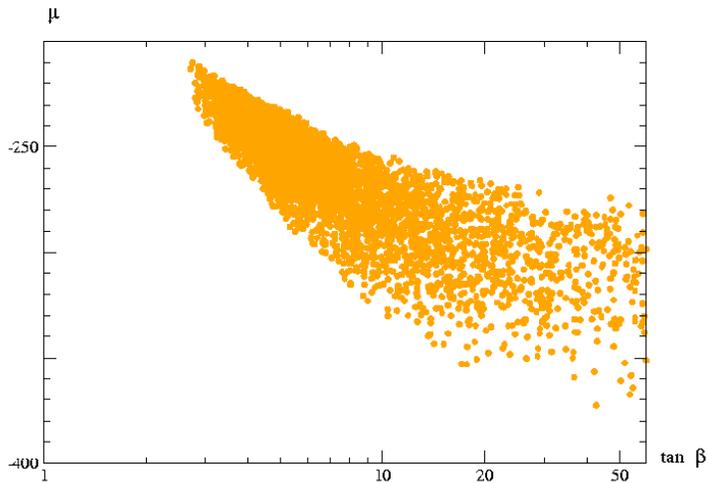}
\caption{Constraints on $\mu<0$, $\tb$ (at $Q_{EWSB}$ scale) from gaugino universality and
gluino plus two neutralino mass measurements in gluino cascade decay. } 
\label{figtbmu-}
  \end{center}
 \end{figure}
A number of remarks are worth here: first, requiring
$\mu$ to be real {\em further} reduces the errors on neutralino masses,
since the red domain in Fig.~\ref{fign1n2} is to be excluded. 
In the most general MSSM case, $\mu$ may have a
non-zero phase, but if we restrict our analysis to a real parameter space,  
this is an 
interesting additional constraint, resulting solely from the
consistency of Eqs.~(\ref{b4}), (\ref{b5}). The two other domains correspond to
$\mu>0$ and $\mu<0$ respectively, so that the latter cannot be excluded
given these SPS1a accuracies on the two neutralino masses. We notice that if
neutralino mass accuracies could be reduced by a factor of about 2, 
the $\mu <0$ solution would disappear altogether (as well as the complex 
$\mu$ possibility).

Next the corresponding constraint in the $\tb, \mu$ plane are shown
in Figs.~\ref{figtbmu+} and \ref{figtbmu-} respectively for $\mu>0$ and
$\mu<0$. We observe that $\tb$ is practically unconstrained, especially
for large $\tb$, and $\mu>0$
poorly constrained, for these accuracies on neutralino and gluino masses.
However for $\mu>0$ the two parameters appear strongly 
correlated, as shown by the contour shape:
e.g. for $\mu>0$, large $\mu\sim 600-700$ GeV
is only possible for small $\tb\sim 2$. This correlation is not an 
artifact of our simple random scan, but a simple
consequence of the $\mu$ and $\sin 2\beta$ dependence within 
Eqs.~(\ref{b4}), (\ref{b5}) (see explicitly Eq.~(\ref{tb2N}) in Appendix A). 
Note also that the sign of $\mu$ and of $\sin 2\beta$ are partly correlated 
(see Eq.~(\ref{tb2N}), so that our convention $\tb >0$ impose constraints
that are quite different for the $\mu >0$ and $\mu <0$ cases.  
More precisely we find:
\be
\tb \gsim 1 \;,\;\;\; 280\: \mbox{GeV} \lsim \mu \lsim 750 \:\mbox{GeV}
\ee
for $\mu>0$ and 
\be
\tb \gsim 2.7 \;,\;\;\; 210 \:\mbox{GeV} \lsim |\mu| \lsim 370 \:\mbox{GeV}
\ee
for $\mu <0$, and in both cases
no upper limits on $\tb$, which is simply understandable because 
only $\sin 2\beta$ appears in any of the relations above in 
Eqs.~(\ref{4inv}--\ref{b5}), so that for large $\tb$ 
any sensitivity on $\tb$ disappears. 

These results are thus consistent with what was anticipated from the previous
analysis illustrated in Fig.~\ref{figm1m2}, where the gaugino 
``mSUGRA" universality band 
is crossing all domains of the chosen $\tb, \mu$ range of variation:
in particular the green region where $\tb$ was essentially arbitrary. 
The fact that $\mu$ is better
constrained for $\mu<0$ is understandable from Fig.~\ref{fign1n2} where
the domain corresponding to $\mu<0$ is smaller than the $\mu>0$ one, 
moreover the central value
$|\mu| \sim 357$ GeV is excluded on the plot Fig.~\ref{figtbmu-}
This is due to the partly correlated sign of $\mu$ and $\sin 2 \beta$
in this inverted determination (see Eq.~(\ref{tb2N})), but without
knowing the true SPS1a value of $\mu$ we could not exclude $\mu<0$ solutions
solely from these cascade decay mass accuracies.  
\subsection{Scenario S3: $\mu$, $\tb$ from three neutralino with
or without universal gaugino masses}\label{s36}
Let us finally consider another (more optimistic) scenario S3 
where three neutralino masses could be determined  
according to the input $S_3$ in Table \ref{tabinput} above.
As explained above in this case one gets from Eqs.~(\ref{4inv}) an extra
relation, Eq.~(\ref{mu3N}), resulting in a determination 
of $\mu$ independent of $\tb$, which is valid both for the general MSSM
case, or assuming additional gaugino mass relations (like universal ones typically). For the
general MSSM, the much improved determination of $M_1, M_2$ was
illustrated by the red domains in Fig.~\ref{figm1m2}. Here for completeness
we illustrate in Fig.~\ref{figm1mu3} the corresponding domains in the $(M_1,\mu)$ planes, for the
two possible case $M_1 < M_2$ (in orange) corresponding to SPS1a, and also for
the alternative solution with $M_2 < M_1$ (in red). $|\mu|$ is determined with
an accuracy of about $\Delta\mu\sim 15$ GeV, but as already mentioned
the sign of $\mu$ remains undetermined.  Moreover, in a most general MSSM, 
without any prior knowledge on gaugino and Higgsino mass relative values,
the sole knowledge of three neutralino masses does not
determine the relative hierarchy among $M_1$, $M_2$, and $|\mu|$. So strictly
speaking there is a six-fold ambiguity in this case,
considering all possible ordering of these three parameters (see Appendix A).  
        \begin{figure}[h!]
  \begin{center}
  \includegraphics[width=8cm,angle=-90]{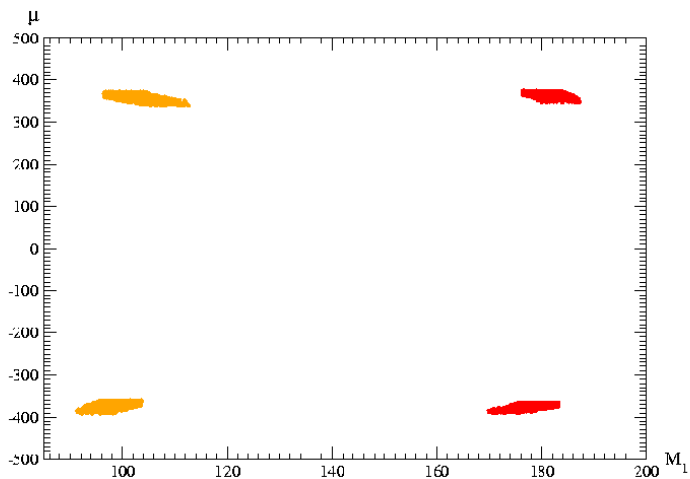}
\caption{Constraints on $M_1$, $\mu$ (at $Q_{EWSB}$ scale) in general MSSM 
for three neutralino mass measurements in gluino + squark cascade decay. } 
\label{figm1mu3}
  \end{center}
 \end{figure}

We now consider the gaugino mass universality case with three neutralino
mass input. The ambiguity on the relative magnitude
of $M_1$, $M_2$ and $\mu$ obtained in a general MSSM is of course resolved 
in this case since
the hierarchy at the low scale is entirely determined from universal 
initial values of $M_i$.
The resulting constraints on $\mu, \tb$ are illustrated in 
Fig.~\ref{figtbmu3}, where 
one observes that the $\mu<0$ solution has disappeared (and the equivalent of
Fig.~\ref{fign1n2} would show that only a smaller part of the red 
contour $\mu>0$ is surviving).  However, only $\mu$ is much more
constrained, while apart from a slightly more interesting lower bound, $\tb$ 
remains essentially unconstrained for large $\tb$. This is simply due  
to the $\sin 2\beta$ only dependence in all these
relations, whatever the number of input neutralino masses. More precisely:
\be
\tb \gsim 2.7 \;,\;\;\; 350\: \mbox{GeV} \lsim \mu \lsim 372\: \mbox{GeV}\;.
\ee
 These results are
consistent with general expectations, namely that the gaugino sector alone
can hardly constrain $\tb$ at the LHC, even in the mSUGRA (gaugino
universality) case, but these features are perhaps very simply illustrated
here analytically. 
%
        \begin{figure}[h!]
  \begin{center}
  \includegraphics[width=8cm,angle=-90]{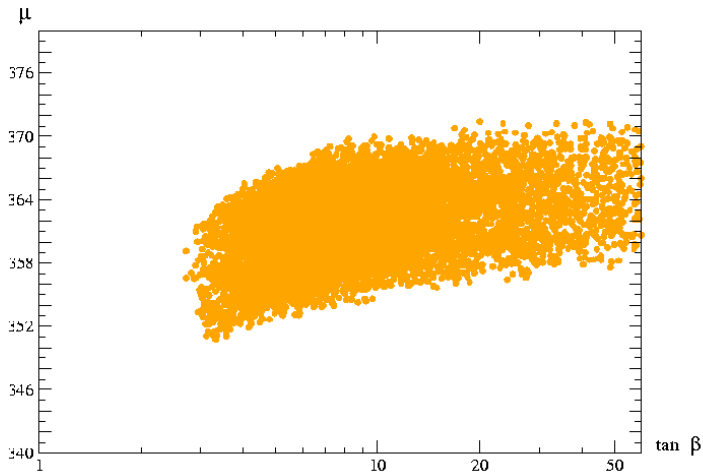}
\caption{Constraints on $\mu$, $\tb$ from gaugino universality and
gluino plus three neutralino mass measurements in gluino + squark cascade decay. } 
\label{figtbmu3}
  \end{center}
 \end{figure}

Concerning the inclusion of radiative corrections 
into this essentially tree-level
determination, we have already mentioned the use of the 
$\overline{DR}$ parameters $\bar s^2_W$, $\bar m_Z$ in all these
relations, 
which already incorporates a part of the corrections. The bulk of 
these $\overline{DR}$ 
corrections with respect to tree-level values come from standard model
contributions, and supersymmetric contributions, though not negligible,
are strictly speaking inducing some theoretical uncertainties if
loop contributions from unknown sector are taken into account.
Next, the corrections from the running to the pole
neutralino masses remain moderate in the MSSM, so that one may neglect
them in a first approximation. Actually, 
most of these radiative corrections
can be incorporated in the form of corrections to the ``bare" 
parameters $\mu$, $M_1$, $M_2$ to very good approximation\cite{bpmz}, 
while additional corrections in the neutralino mass sector
are smaller\cite{bpmz}. Our procedure to determine e.g. $M_1$, $M_2$ or $\mu$ 
above remains thus correct, up to an implicit shift on these parameters, provided that one has some
knowledge on these radiative corrections. 
In fact gluino mass corrections
are dominantly due to squarks/quark, and quite similarly for
neutralinos, so they
can be evaluated within a reasonable approximation, 
without assuming more knowledge than the available cascade 
decay input\footnote{NB the radiative corrections to $M_1$, $M_2$
and $\mu$ could be used in principle to try to reduce  
the $M_1 <M_2$ versus $M_2 < M_1$ etc reconstruction ambiguities in the unconstrained MSSM. However these corrections are far too small to disentangle
these ambiguities, at least for the prospected LHC neutralino mass accuracies.}. Though it is
not completely straightforward to incorporate consistently all these corrections
within these parameters, it is very similar to the kind of
iterative procedure required for a standard top-down calculation
of MSSM spectrum\cite{suspect}, as discussed in sub-section 2.3. 
This is taken into account in the above illustrations.    

%
\begin{table}[h!]
\begin{center}
\caption{\label{minuit_m3mutb} Constraints on gaugino-Higgsino sector parameters 
(at $Q_{EWSB}$ scale) from a standard top-down $\chi^2$ MINUIT fit of neutralino and gluino masses under different universal or 
non-universal gaugino mass assumptions.} 
\begin{tabular}{|c||c|c||c|}
\hline\hline
Data $\&$ fitted parameter   & MIGRAD ($68\%$C.L.)   & MINOS ($68\%$C.L.)       & nominal \\
(+ model assumptions)  &  &  &  SPS1a value \\  
\hline
$m_{\t N_1}, m_{\t N_2}, m_{\t g}$ &  (convergent)  & (problems)   &  \\ 
(Non-universal $M_i$;  &  &      &  \\
1-loop RGE; no $\t N_i$ R.C.) &   &             & \\
  $M_1$  & $\sim$97--126 (from fit)   &   &  108.8  \\
 $M_2$  &  $\sim$181--381 (from fit)  &   &  208.9  \\
 $\mu$  & 200--1500 (scanned)   &           &  340.6  \\
 $\tb(Q_{EWSB})$  & 1--50 (scanned)  &  & 9.74 \\
 \hline
 $m_{\t N_1}, m_{\t N_2}, m_{\t N_4}, m_{\t g}$ & (convergent) & (convergent) & \\
 $M_1$  &  108.8 $\pm 7.2 $   & 108.8 $\pm 7.2 $  &  108.8  \\
 $M_2$  &  208.9 $\pm 8.2 $   &  208.9 $\pm 8.2 $ &  208.9  \\
 $\mu$  &  340.6 $\pm 11.6$    &  340.6 $\pm 11.6$         &  340.6  \\
 $\tb(Q_{EWSB})$      & 9.74 (fixed)  &   9.74 (fixed) & 9.74 \\
\hline\hline
$m_{\t N_1}, m_{\t N_2}, m_{\t g}$ & (convergent) & (problems with $\tb$)& \\
 (Universal $M_i(Q_{GUT})$;   &        & &  \\
 1-loop RGE; no $\t N_i$ R.C.)   &    &    &\\
 $M_3$  &  603.8 $\pm 13.5 $   &  603.8 $\pm 13.5 $ &  603.8  \\
 $\mu$  &  341.6 $\pm 293$    &  341.6 $\pm 293$         &  340.6  \\
 $\tb(Q_{EWSB})$      &   9.73 $\pm 52.6$  & (not calculated) 
 & 9.74 \\
            &                   &                        &   \\
\hline
$m_{\t N_1}, m_{\t N_2}, m_{\t N_4}, m_{\t g}$ & (convergent) &
 (problems with $\tb$)   &  \\
 $M_3$  &  603.8 $\pm 13.4 $   &  603.8 $\pm 13.4 $ &  603.8  \\
 $\mu$  &  340.6 $\pm 13$    &  340.6 $\pm 13$         &  340.6  \\
 $\tb(Q_{EWSB})$      &   10.0 $\pm 15.3$  & (not calculated) 
 & 9.74 \\
            &                   &                        &   \\
\hline\hline
\end{tabular}
\end{center}
\end{table}
In order to cross-check the inverse analytical determination above, 
we perform alternatively 
standard top-down $\chi^2$ minimization fits, using
MINUIT. The best fit
results, with or without gaugino mass universality assumptions and for  
different assumptions on $\mu$ and $\tb$, are shown in Table
\ref{minuit_m3mutb} both for two and three neutralino mass input.   
Actually, in the non-universal 
gaugino mass case, two neutralino masses are clearly not enough input 
to constrain  
the four parameters $M_1$, $M_2$, $\mu$ and $\tb$. Thus to compare 
as much as possible with the previous  reconstruction results, 
in this case we also 
perform scans over $\mu$ and $\tb$ (taking $\mu \gsim 200$ GeV to avoid
the pole region discussed above). The corresponding results obtained for the 
``envelope'' of best fit values, with confidence level domains for $M_1$ and $M_2$, are shown 
in the second column of  Table \ref{minuit_m3mutb} (in the first two entries).     
The rather bad
constraints obtained in this case are roughly consistent
with the above analysis of the $\mu$ dependence of $M_2$ and $M_1$
illustrated in Fig.~\ref{figm1m2}, and already discussed above.  
(Note that our simple top-down fit is not able to exhibit the $M_1, M_2$
reconstruction ambiguities in the non-universal gaugino mass case,
discussed in sub-sec. 3.1, 3.4 and Appendix A, though these ambiguities may be
implicitly responsible for the large errors found for $M_1$, $M_2$, when $\mu$
is varied in a wide range). Similar ambiguities 
are however more clearly exhibited when performing more sophisticated 
minimization, where typically extra ``best fit solutions'' appear\cite{dzerwas}).  
In contrast, the determination of $M_1$, $M_2$ 
is much improved if $\mu$ is better constrained,  
and accordingly there is a substantial improvement on the $\mu$ determination 
if the third neutralino mass input $m_{\t N_4}$ is available, as shown in the
Table.   
Those results from Table \ref{minuit_m3mutb} are thus 
roughly consistent with the analytic behaviour
illustrated in Figs.~\ref{figm1m2mu}, \ref{figm1mu3}, and \ref{figtbmu3}. 
The symmetrical (MIGRAD) minimization, however, does not
reflect very well the true sensitivity to some of the parameters, most notably
for $\tb$. Indeed the unsymmetrical non-linear MINOS minimization 
does not give any useful constraint for $\tb$, even for universal gaugino
mass assumptions. 
Note that in the latter case $m_{\t N_1}, m_{\t N_2}$ 
are essentially Bino-like 
and Wino-like respectively, while $m_{\t N_4}$ is essentially Higgsino-like, which
also explain the drastic improvement on $\mu$ accuracy from a third neutralino
mass input. 
\subsection{Reconstructing gaugino masses at GUT scale}
        \begin{figure}[h!]
  \begin{center}
  \includegraphics*[width=8cm,angle=-90]{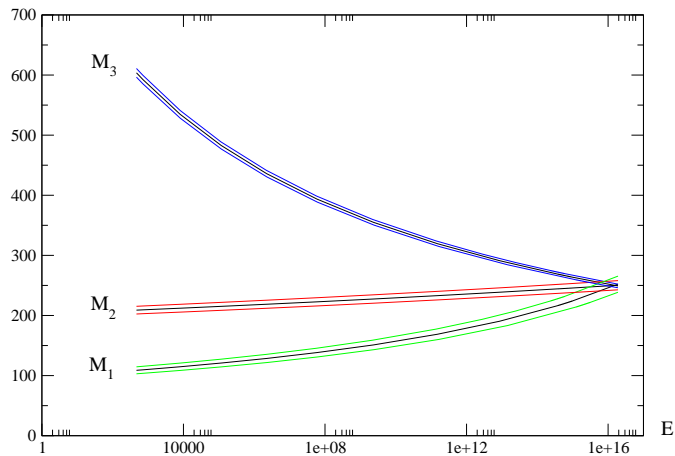}
\caption{Bottom-up RGE of gaugino masses from EWSB to GUT scale, including
error propagations ($\mu$ and $\tb$ are fixed to their SPS1a values). } 
\label{rgem123}
  \end{center}
 \end{figure}
Finally we perform a bottom-up RGE of the gaugino mass parameters in the non-universal case,
in order to check eventually
for their unification at a high GUT scale, following ref. \cite{bpzm123}.
(NB the bottom-up RGE procedure for gaugino masses and other basic MSSM parameters
from given values at EWSB scale is explained in Appendix B). 
This is first shown in Fig.~\ref{rgem123}, taking  
the best $M_i(Q_{EWSB})$ determination above (i.e. corresponding to three
neutralino mass input), and in Fig.~\ref{rgem123worst} 
taking the worst $M_i(Q_{EWSB})$ determinations (i.e. a scenario with
two neutralino mass input and no known constraints on $\mu$).  
One can see that a very good check of GUT scale universality is possible
as long as the initial $M_i(Q_{EWSB})$ accuracies are reasonable: 
in other words, the error
dispersion from the gaugino mass RGE remains small (which is clearly 
explained from the fact that their RGE content only depend
on gauge couplings at the one-loop level). As expected this is qualitatively consistent with the former results of ref. \cite{bpzm123} (though it appears 
that the $M_1$, $M_2$
LHC accuracies considered at that time were slightly more 
optimistic than the ones we obtain here from our analysis). 
Note that the dispersion
due to the bottom-up RGE can be much more important for other parameter 
sectors, in particular for the $m_{H_u}$ parameter in the Higgs
sector (see the discussion in Appendix B).  
Nevertheless, the possibility of checking
universality at GUT scale may become elusive even for gaugino masses  
in the extreme case where almost nothing is known on $\mu$,
such that the $M_1$, $M_2$ low scale values errors are large. In this case only
the $M_3$ error remains under control, as illustrated in Fig.~\ref{rgem123worst}.
We also remark once more that in addition to performing a bottom-up RGE, mSUGRA 
gaugino universality could
be checked efficiently from plots as illustrated in Fig.~\ref{figm1m2} 
(where it can also be eventually distinguished from other high scale 
SUSY-breaking models like AMSB). 
        \begin{figure}[h!]
  \begin{center}
  \includegraphics*[width=8cm,angle=-90]{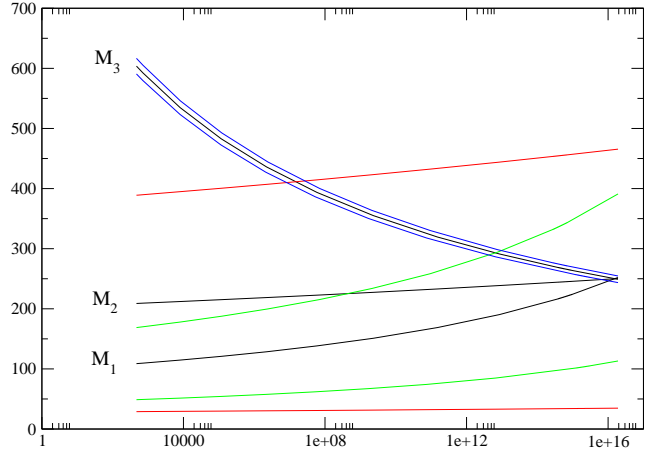}
\caption{Bottom-up RGE of gaugino masses from EWSB to GUT scale, including
error propagations for the worst case ($\mu$ and $\tb$ undetermined). } 
\label{rgem123worst}
  \end{center}
 \end{figure}
%
\section{Squark/slepton parameter determination (first two generations)}  
We now consider a bottom-up reconstruction approach for the (first
and second generation) squark and/or slepton sector parameter of the MSSM. 
Because of the negligible mixing for the first and second generations,
as well as the simplified RGE structure of this sector, we can elaborate 
a rather general strategy to reconstruct the relevant parameters up to an 
(eventual) GUT scale. 
Similarly to the gaugino/Higgsino sector, we shall consider
two different model assumptions: a general MSSM sfermion sector, or
a more constrained scenario assuming universality of slepton and squark masses. 
\subsection{General MSSM and a simple squark/slepton mass sum rule}
The masses of the sfermions  
   $\tilde q (\ne \t b,\t t)$ and $\tilde l$ which participate to the
   cascade decay in Eq.~(\ref{casc}) 
    obey (at the EWSB scale) well-known relations  (at tree-level)
    e.g for the up squark and selectron:
  \beq
\nn
 m^2_{\t u_1} &=&  m^2_{\t u_L}+(\frac{1}{2}
-\frac{2}{3} s^2_W) m^2_Z  \cos 2\beta \\ 
 m^2_{\t e_2} &=&  m^2_{\t e_R}-s^2_W m^2_Z 
 \cos 2\beta 
\label{mu1me2}
\eeq
which are valid in the general MSSM~\footnote{$s^2_W$ in Eq.~(\ref{mu1me2}) 
and further equations below
should be understood as the $\overline{DR}$-scheme parameter $\bar s^2_W$.}. 
Similar relations hold for the other squark and slepton flavors. 
Eqs.~(\ref{mu1me2}) simply relate the 
physical masses to the soft breaking scalar masses via the additional D-terms.  
These relations are particularly simple for the first two 
squark and slepton generations due to negligible mixing.  
The main idea is to consider  
specific linear combination (``sum rules") in order 
to eliminate the $\tb$ dependence:
\begin{equation}
s_W^2  m_{\t u_1}^2+(\frac{1}{2}-\frac{2}{3}s_W^2)  
 m_{\t e_2}^2 =s_W^2 { m_{\t u_L}^2}+(\frac{1}{2}
-\frac{2}{3} s_W^2) { m_{\t e_R}^2}\;.
\label{sr1}
\end{equation}
Then taking into account the available accuracy on the physical masses
$m_{\t u_1}$ and $m_{\t e_2}$ provides constraints on the MSSM soft-breaking
scalar parameters
$m_{\t u_L}, m_{\t e_R}$ {\em independently} of $\tb$ values. 
So, even if   
 $\tb$ is largely undetermined (as is the case from the
neutralino sector alone at LHC, illustrated in previous section), Eq.~(\ref{sr1}) gives a (model-independent)  
determination of the linear combination of basic parameters which will be 
roughly of the order of magnitude of the physical mass accuracies,
i.e. a few percent for typical LHC prospects. 
More precisely, a straightforward calculation from the
experimental accuracy in Table \ref{tabexp} gives
a $\sim 2.3\:\%$ ($\sim 2\: \%$) relative accuracy for 
the linear combination in Eq.~(\ref{sr1}), if we combine 
the $m_{\t u_1}$
and $m_{\t e_2}$ errors linearly (quadratically), respectively. 
(NB we have neglected at the moment for simplicity  
the error on $\bar s^2_W$: the effect 
of $\bar s^2_W$ uncertainties will be studied below).  
Another advantage of this linear combination is 
that, even in a general non-universal MSSM case, the 
RG evolution of the relevant parameters ($m_{\t u_L}(Q)$, $m_{\t e_R}(Q)$) 
depends only on the gauge couplings $g_i$ and the gaugino 
masses $M_i$~\footnote{This is only true at the one-loop level RGE,
since at two-loop level practically all MSSM parameters enter the RG
evolution of squark and slepton mass terms\cite{rge,RGE2}. 
We shall discuss below how the
inclusion of two-loop RGE effects affect our results, but we anticipate
that neglecting these higher loop effects do not change drastically the
obtained constraints.}. 
The linear combination Eq.~(\ref{sr1}) can thus be RG-evolved within a 
restricted set of input parameters solely determined
from the gluino cascade, in order to  
obtain the soft scalar term values at the GUT scale, where  
one may check for eventual universality relations.
Before doing this RG evolution, it is necessary  
to subtract out radiative corrections linking the running to the pole
masses, which are not negligible for squarks, in the
way discussed in section 2.3. More precisely we have
\be
M^{pole}_{\t u_1} = m_{\t u_1}(Q) +\Delta m_{\t u_1}(Q,m_{\t q},
m_{\t g},...)
\ee
where these corrections are largely dominated by
squark/gluino contributions at one-loop. For a typical mSUGRA scenario like 
SPS1a we have $ \Delta m_{\t u_1}(Q_{EWSB},m_{\t q},
m_{\t g}) \sim 19$ GeV, which can be consistently subtracted out 
to define the running mass $m_{\t u_1}$. 
Concerning radiative corrections linking the running 
to the pole slepton mass, they are generally much smaller and we shall
neglect them in our analysis. \\
Next, it is a straightforward exercise
to work out the RG evolution of the linear combination
(\ref{sr1}): 
\beq \nn 
  \frac{d}{dt} \left[s_W^2 { m_{\t u_L}^2}+(\frac{1}{2}
-\frac{2}{3} s_W^2) { m_{\t e_R}^2}\right] = &&  
  s^2_W \frac{d m^2_{\t u_L}}{dt} +(\frac{1}{2}-\frac{2}{3} s_W^2) 
 \frac{d m^2_{\t e_R}}{dt}  \\  && + \frac{d s^2_W}{dt} 
(m^2_{\t u_L} -\frac{2}{3}  m^2_{\t e_R})
\label{rgsr}
\eeq
where  $t \equiv \ln Q$ and the standard RG evolution of the relevant
parameters $\t m^2_{u_L}$, $\t m^2_{e_R}$ is used (which as mentioned
only depend on $g_i$, $M_i$). We also have:  
\be
\frac{d s^2_W}{dt} = (\frac{3}{5}g^2_1 +g^2_2)^{-1}\:
\left(\frac{3}{5} c^2_W \frac{d g^2_1}{dt} -s^2_W 
\frac{d g^2_2}{dt}\right)
\label{rgsw}
\ee
with $s^2_W(t) \equiv \frac{3}{5}g^2_1(t)/(\frac{3}{5}g^2_1(t)+g^2_2(t))$
and the 
factor $3/5$ is due to the standard normalization of the $U(1)_Y$
coupling $g_1$ in the MSSM RGE. Eqs.~(\ref{rgsr}) and (\ref{rgsw}) 
take into account the RG evolution
of $\bar s^2_W$. (Note that the latter is not at all negligible since it is related
to the running of gauge couplings, which change substantially
from $m_Z$ input values to GUT scale values).  

As already mentioned one important feature
of this bottom-up RG evolution is that some of the low scale
parameter uncertainties are amplified once evolved to a large scale, 
depending on the structure
of RG beta functions for some of the relevant parameters: 
this is the case to some extent with
the evolution of Eq.~(\ref{sr1}, as we will see later. 

\subsection{Constrained MSSM with squark, slepton universality}
If we now assume squark and slepton mass universality 
at the GUT scale,
Eq.~(\ref{rgsr}) immediately determines $m^{q,l}_0(Q_{GUT})$:
\be
s_W^2 { m_{\t u_L}^2}+(\frac{1}{2}
-\frac{2}{3} s_W^2) { m_{\t e_R}^2} (Q_{GUT}) \equiv 
\frac{5}{8}\:(m^{q,l}_0)^2
\label{mql0univ}
\ee
where the gauge coupling universality relation at the GUT scale:
$s^2_W(t=\ln Q_{GUT})=3/8$ has been used. $m^{q,l}_0$ indicates that we 
only assume universality
for the (first two generation) squark and slepton sector at this stage, 
i.e. not necessarily for the third generation sfermions, nor for Higgs scalar
terms like in mSUGRA models. 
\subsection{Explicit reconstruction test for the SPS1a input}
We now determine explicit constraints on the squark 
and/or slepton sector parameters
from the specific SPS1a blind input with the expected accuracy on the masses
of $\t q_L$ and $\t l_R$ from Table \ref{tabexp}. As already
mentioned in section 2, there is at present no way to tag 
the charge and flavor of the relevant (first two generations) 
squark at LHC. Accordingly the resulting mass accuracies
of say, an up or down squark, are assumed to be identical\cite{cascade2}, so that 
there is no need to combine their errors in a statistically elaborated
manner, and we thus assume that it is sufficient for our purpose to take the 
average of two (identical) errors in our analysis.   
By combining thus the accuracies on the measured 
$\t u_1$ and $\t e_2$ masses in Table \ref{tabexp}, we obtain for SPS1a from
Eqs~(\ref{sr1}), (\ref{mql0univ}):
\be
84\:(86)\: \mbox{GeV} \lsim m^{q,l}_0 \lsim 116 \:(112)\:\mbox{GeV}\;.
\label{m0res}
\ee
Note that the first limits are for linearly combined 
mass uncertainties (while those in parenthesis
are for quadratically combined mass uncertainties). 
We emphasize that the bounds in Eq.~(\ref{m0res}) are  
independent of $\tb$ values. 
However, there is a rather important
amplification of the initial low scale uncertainty $\sim 2\%$ 
due to error propagation in the 
bottom-up RG evolution, and because of the additional terms proportional
to $d s^2_W/dt$ in Eq.~(\ref{rgsr}). 
To better trace the origin of the resulting $m_0$
uncertainties, it is illustrative to consider independently the squark and slepton 
mass uncertainties: this gives 
\be
86.6\: \mbox{GeV} \lsim m^{q,l}_0 (\Delta m_{\t u_1})\lsim 112 \:\mbox{GeV}
\label{m0_u1}
\ee
and
\be
97.1\: \mbox{GeV} \lsim m^{q,l}_0 (\Delta m_{\t e_2})\lsim 103 \:\mbox{GeV}
\label{m0_e2}
\ee
respectively. Thus the final uncertainty on $m_0$ is largely dominated
by the $m_{\t u_1}$ initial accuracy.
Actually, the latter bounds on $m^{q,l}_0$ were calculated while 
fixing the gaugino mass terms $M_i$. But since  
the RG evolution of $m_{\t u_L},m_{\t e_R}$ 
depends on gaugino masses, 
the $m^{q,l}_0$ accuracy should be sensitive to $M_i$ uncertainties, 
mainly those of $M_3$ which are enhanced in the $m_{\t u_L}$ RGE 
by the strong coupling:
$d m^2_{\ u_L}/dt \sim -\frac{8}{3\pi} \alpha_S M^2_3$\cite{rge,RGE2}.
This leads to an important
amplification of $m^{q,l}_0$ final uncertainty due to the 
$\Delta M_3\sim 7$ GeV uncertainty (although
the latter effect is damped somehow by $\bar s^2_W \sim .238$ in
the first term of the RHS of Eq.~(\ref{rgsr}), while other 
terms in Eq.~(\ref{rgsr}) are not much
sensitive to $M_i$ uncertainties). 
One thus obtains, in the conservative case of
uncorrelated and linearly combined errors,   
a maximal uncertainty on $m^{q,l}_0$ of about $\pm \sim 32$ GeV 
(respectively $\pm \sim 22$ GeV when combining errors quadratically) instead of 
the bounds shown in Eqs.~(\ref{m0res})--(\ref{m0_e2})~\footnote{Note also
that the dominant uncertainties from $m_{\t u_1}$ and $m_{\t g}$ (i.e. $M_3$)
have the opposite (anti-correlated) effect: from the RG structure, increasing 
(resp. decreasing) $m_{\t u_1}$ makes $m^{q,l}_0$ to increase (resp.
decrease), while the opposite behaviour is obtained 
from $M_3 \pm \Delta M_3$.}.      
However, the linear combination Eq.~(\ref{sr1}) does not 
use the full information from the two independent mass relations
in Eq.~(\ref{mu1me2}): we shall illustrate below how this additional 
information improves rather substantially the limits on $m^{q,l}_0$.  

Another potentially interesting question is whether one 
can derive at the same time any useful limits on $\tb$, once using   
the complete information from both squark and slepton masses. 
At first sight one may naively expect to obtain some upper bounds on $\tb$, since
the relations (\ref{mu1me2}) are sensitive to $\cos 2\beta$.  
However a simple estimate immediately indicates that 
interesting $\tb$ upper bounds from 
this squark, slepton sector 
are hardly expected for the given LHC $\t e_2, \t u_1$ mass accuracies: in fact
for the SPS1a point with $\tb(Q_{EWSB})\sim 9.74$, $\cos 2\beta \sim -0.979$ 
i.e. very close to $-1$, so that one would need at least an accuracy
$\lsim 0.02$ on $|\cos 2\beta|$ to put useful upper limits on $\tb$. 
In contrast, a simple calculation of uncertainties from both Eqs.
(\ref{mu1me2}) gives:
\be
\Delta |\cos 2\beta| \sim 0.6
\label{crudetb}
\ee
even in the optimistic case where we combine the $m_{\t u_1}, m_{\t e_2}$
uncertainties quadratically, and neglect the errors on $s^2_W$. 
The above estimate, however, does not take into account other possible
constraints on $\tb$, that may come from other sectors, or from theoretical
consistency. For instance the obvious constraint: $|\cos 2\beta| \le 1$ 
puts additional limits 
on $m^2_{e_R} -m^2_{\t e_2}$ via  Eqs.~(\ref{mu1me2}), so indirectly
on $m^{q,l}_0$. Furthermore, for $m^{q,l}_0
\sim 100$ GeV it is easily checked that 
$\tb$ cannot be larger than $\tb \lsim 35-36$, since beyond this value
the lightest stau $m_{\t \tau_1}$ mass becomes tachyonic 
due to the large stau mixing $\bar m_\tau \,. \mu \tb$ term. 
Moreover, the lightest
Higgs mass becomes inconsistent (or too low) 
for small $\tb \lsim 2.2$ approximately, and for low $\tb$ values 
the LEP lower bounds on $m_h$\cite{LEPmh} 
actually put a tighter constraint $\tb \gsim 8-9$.
Yet the latter limits are theoretical and model-dependent
(or experimental and model-dependent in the case of $m_h$ bounds), 
and specific of the SPS1a benchmark\cite{benchmark}, 
which was chosen on purpose to satisfy the present experimental constraints. 
If we push $m^{q,l}_0$ sufficiently above 
the SPS1a central values, the upper bound $\tb \lsim 35$ from tachyonic 
$\t \tau_1$ is easily evaded, though in this case one should also
take into account the lower bound $m_{\t \tau} \gsim 104$ GeV 
from LEP limits\cite{LEPlim}. (For example, for $m_0 \sim 200$ GeV
$\tb \sim 50$ is not excluded by tachyonic stau, while the gluino, squarks
cascade decays would not be drastically different from the SPS1a one). 
Similarly, the $\tb$ lower bound due to LEP $m_h$ lower limits
could easily be evaded in an unconstrained MSSM\cite{adkps}.  
We will thus not apply such direct (or indirect) experimental limits 
which are much dependent on the
specific SPS1a choice, since our main aim is to present 
a reconstruction strategy expected to be valid beyond this particular benchmark choice. 

Accordingly a question that we examine in some detail next  
is whether the sole $m_{\t u_1}, m_{\t e_2}$ mass measurements could 
put some extra {\em model-independent} limits on $\tb$.    
From the previous estimate it appears 
that to obtain stringent such experimental 
constraints on $\tb$, one  would require an accuracy about an 
order of magnitude
better on the squark and slepton masses than the one
prospected at the LHC. Incidentally this is roughly
the accuracy expected at the ILC 
(though only for the sleptons), where both $l_R$ and $l_L$ 
masses could be measured at the per mille 
level\cite{LHCstudy,SPA}. 
However, a detailed ILC analysis is beyond the scope of the present paper
and left for future work. \\
We anticipate that (model-independent) limits on $\tb$ 
will be indeed absent, or very marginal if using solely the first two generation squark 
and slepton mass accuracies. This is consistent with general 
prospects\cite{LHCstudy,kane}.
We found however useful to examine this issue in
some detail, since our construction is not limited to the LHC
mass accuracies here considered: thus tracing analytically
the sensitivity on parameters can help to understand
better what determines the constraints in a more elaborated analysis.

To begin, there is a subtlety that is not taken into account in the   
above crude estimate of error combination in Eq.~(\ref{crudetb}), 
such that it may underestimate the $\tb$ sensitivity: 
as emphasized previously the RG equations for $m_{\t u_L}, m_{\ e_R}$ only
depend on gauge couplings and gaugino masses, so in particular 
do not depend on $\tb$ (at one-loop level). But
there is in fact an indirect dependence on $\tb$ even at one-loop 
(though very moderate): it  
originates from the boundary conditions on the RGE, namely
the initial values of gauge couplings $g_i$ in the $\overline{DR}$ scheme,  
as well as their values at the GUT scale (if gauge unification is imposed),  
depend slightly on $\tb$ through radiative corrections (see e.g. ref.
\cite{bpmz}). Although these effects are  strictly speaking small higher
order corrections on $g_i(m_Z), g_i(m_{GUT})$ values, they are enhanced
when running parameters from low to high GUT scale over more than 13 
orders of magnitude. Indeed most realistic calculations of MSSM spectra 
do take into account this dependence 
consistently\cite{isasugra,softsusy,spheno,suspect}. 
Also, these values are notoriously different if using the RGE at
the full two-loop level, or in a one-loop approximation, as will be illustrated
below.  (This is somewhat similar to the impact of precise initial 
gauge couplings and RGE approximations on the GUT scale value $\sim 2 \times 
10^{16}$ GeV). 
       \begin{figure}[h!]
  \begin{center}
  \includegraphics[width=8cm,angle=-90]{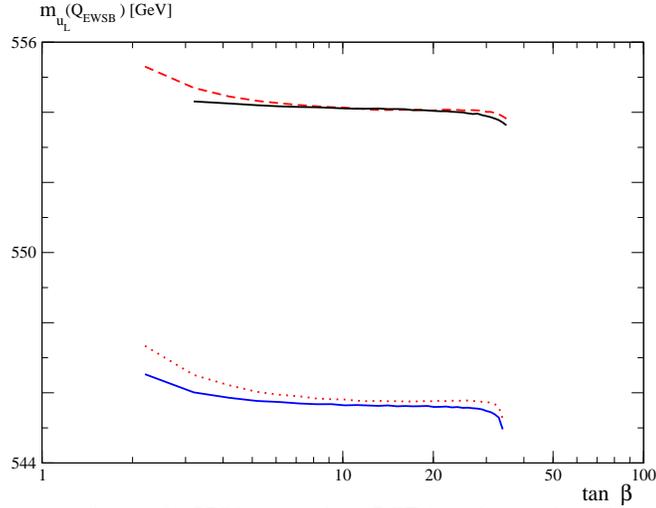}
\vspace{-1.cm}
\caption[long]{\label{muLtb} $m_{\t u_L}(Q_{EWSB})$ as function of $\tb$ 
for SPS1a in one-loop RGE (top plots) and two-loop RGE (bottom plots).
Red dashed line: one-loop RGE, default $Q_{EWSB}= (m_{\t t_1} m_{\t t_2})^{1/2}$.
Black line: fixed $Q_{EWSB}=468$ GeV. 
Orange dotted line: two-loop RGE, defeat $Q_{EWSB}= (m_{\t t_1} m_{\t t_2})^{1/2}$.
Blue line: fixed $Q_{EWSB}=468$ GeV. }
  \end{center}
  \end{figure}
Moreover, the low energy EWSB scale at which all soft parameters are evaluated
also depend in principle on $\tb$ values, and more generally it is not 
fixed by first principles: in most analysis the default EWSB scale
is often fixed to $Q_{EWSB} \sim (m_{\t t_1} m_{\t t_2})^{1/2}$ (which
is know to minimize the scale dependence\cite{Vscale} of the MSSM one-loop 
effective scalar potential). For the SPS1a point this is a well-defined value, but 
when relying only on the gluino cascade decay data, the stop masses are not 
assumed to be know in a general MSSM case. 
Thus we can in principle vary this EWSB scale, which can affect the final
value of soft scalar masses, since this scale determines the endpoint 
of the RG evolution. In our calculation we use either a fixed value (close
to the true SPS1a for definiteness), or adopt the above default value 
in universality cases.  In fact $m_{\t u_L}$ is expected to be specially
sensitive to such variations, due to the large coefficient $\sim \alpha_S M^2_3$
within its RGE, which makes it to run much faster than $m_{\t e_R}$.
(E.g. for SPS1a $m_{\t u_L}(Q_{GUT})=100$ GeV evolves to $m_{\t u_L}(Q_{EWSB})
\sim 560$ GeV.)

For illustration in Fig.~\ref{muLtb} we vary 
$m_{\t u_L}(Q_{EWSB})$ as function of $\tb$,
at one- and two-loop RGE order, both 
with fixed and default EWSB scale, using SuSpect 2.41.     
The variation with $\tb$ is less than 
2 GeV for $m_{u_L}$ (and we checked that it is completely negligible
for $m_{e_R}$) for the whole $\tb$
theoretically allowed range. This is accordingly 
below the prospected LHC accuracy on $m_{\t u_L}$, and thus rather negligible 
for our analysis. In contrast, 
there is a large difference between one and two-loop RGE, but this
is usually the case for the
whole MSSM spectrum, as it is well known. This illustrates that  
for such reconstruction (or in fact any other reconstruction methods)  
one should be careful to be consistent with the RGE approximation used.  
Concerning now the small $\tb$ dependence it should be kept in mind that any more elaborated fit (as could be performed from a $\chi^2$ minimization using
MINUIT) will be eventually sensitive to such effects, since these
cannot be easily ``switched off'' from the fitting procedure. Moreover, such
indirect dependence on $\tb$ will be relevant anyway once reaching a better
accuracy on slepton masses as is prospected at the ILC.   \\

All these features are examined more quantitatively from a 
systematic scan over parameters, where we took into account 
the propagation of $m_{\t u_1}, m_{\t e_2}$ uncertainties in both relations
(\ref{mu1me2}) after evolving from low to high scale, and back. More precisely
we scanned 
over allowed mass bounds, considering the two mass uncertainties as 
independent (and uncorrelated) for simplicity. We also took into account
any additional small dependence e.g. on $\tb$ such as the one illustrated
in Fig.~\ref{muLtb}. The ($m^{q,l}_0, \tan\beta$) distributions obtained
are shown first in Fig.~\ref{m0tb_fig1}
(the scan was performed with $3000$ uniformly distributed random numbers).
       \begin{figure}[h!]
  \begin{center}
  \includegraphics[width=8cm,angle=-90]{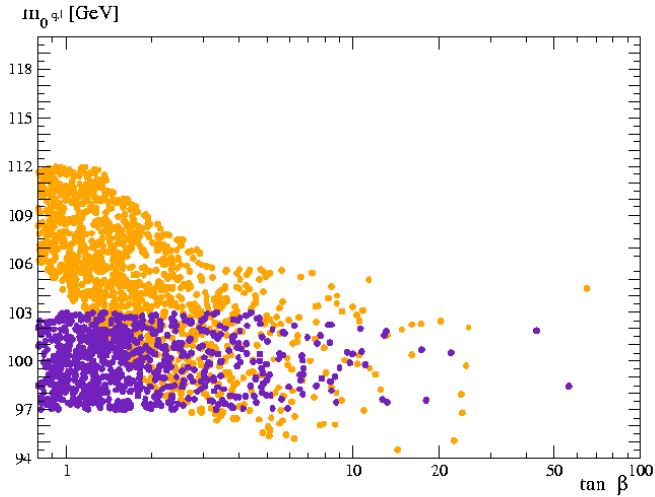}
\vspace{-1cm}
\caption[long]{\label{m0tb_fig1} Constraints on $m_0\equiv m^{q,l}_0(Q_{GUT})$ 
and $\tb$ from
first two generation squark and slepton in gluino cascade decays after
bottom-up RG evolution. The orange (respectively magenta) points 
correspond to constraints
  obtained from the squark $ m_{\t u_1}$ 
  (respectively $m_{\t e_2}$) relations in Eqs.~(\ref{mu1me2}).}
  \end{center}
  \end{figure}
 Comparing with Eq.~(\ref{m0res}) one notes that the lowest values of 
 $84 \lsim m_0 \lsim 94 $ GeV have been excluded simply  
from the constraint $|\cos 2\beta| <1$.  
However, arbitrary large values of $\tb$ are possible (provided
that a sufficiently large number of scan points are taken), confirming
the simple error estimate above in Eq.~(\ref{crudetb}).  
The fact that large values
of $\tb$ are very few in Fig.~\ref{m0tb_fig1}, while low 
$\tb$ values appear very much
favored, is actually an artifact of the (uniform) 
scanning procedure
where we basically scanned over the $ m_{\t u_1}$, $m_{\t e_2}$ masses, 
which determine $\cos 2 \beta$ via Eq.~(\ref{mu1me2}) 
rather than $\tb$. So the distribution of points 
in Fig.~\ref{m0tb_fig1} is simply resulting from the transformation
from $\cos 2\beta$ to $\tb$ and has not much   
statistical meaning as ``most likely" values of $\tb$.
As already mentioned we shall refrain to enter into a fully   
realistic statistical treatment of
uncertainties, which would require to take into account\cite{dzerwas}, 
among other things, the non-trivial
correlations implied by the cascade decay mass measurements.   
It is yet tempting in the present case to proceed one step
further with a little more elaborated analysis.
We thus perform a different scan, where instead of uniformly distributed
``flat prior'' random numbers, we exploit
simple trigonometric relations to match
more faithfully the true $\tb$ distribution. 
In addition we start with Gaussian-distributed 
random numbers (assuming thus that the two independent
errors are purely statistical, which is actually not 
really correct\cite{cascade1,cascade2}).   
We then calculate a ``theoretical" $\chi^2$ (where
$\chi^2_{min}= 0$ trivially since we use the correct central
values of the masses) just to obtain well-defined confidence levels (C.L.)
for the (joint) estimation of the two parameters $m^{q,l}_0$ and $\tb$, 
assumed to be independent and uncorrelated.  
        \begin{figure}[h!]
  \begin{center}
  \includegraphics[width=9cm,angle=-90]{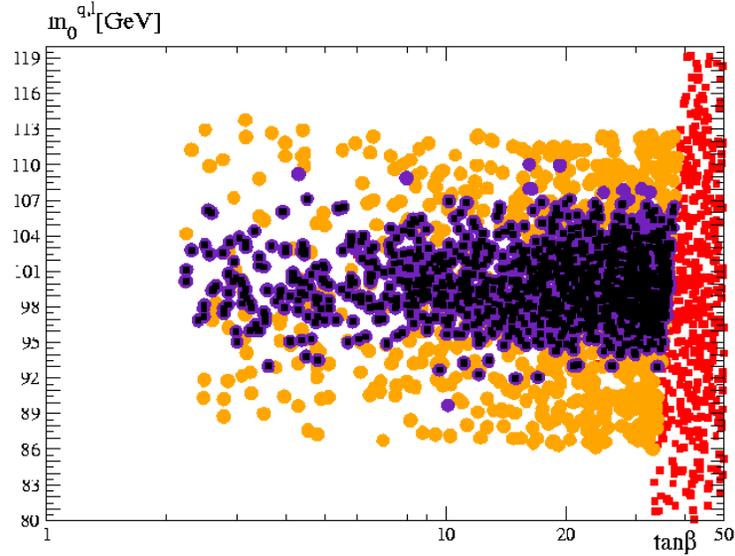}
\vspace{-1cm}
\caption[long]{\label{m0tb_fig3} Constraints on $m_0\equiv m^{q,l}_0(Q_{GUT})$ and $\tb$
with a scan over Gaussian random numbers:
orange disks: 2-$\sigma$ allowed points
($\chi^2 = \chi^2_{min}+6$) from $m_{\t e_2}$ relation; indigo disks:
2-$\sigma$ allowed points from $m_{\t u_1}$ relation; black squared: 
1-$\sigma$ allowed points ($\chi^2 = \chi^2_{min}+2.3$) 
from $m_{\t u_1}$ relation. Red squared: excluded by tachyon $\t \tau_1$.
(NB the square or disk sizes have no physical meaning).}
\end{center}
       \end{figure}
%
The result of this Gaussian scan is shown in Fig.~\ref{m0tb_fig3}. 
The domains obtained   
from the $m_{\t e_2}$ and $m_{\t u_1}$ relations in Eq.~(\ref{mu1me2}), 
at one-$\sigma$ level (more precisely 
$68\%$ C.L.) are shown in black, and the {\em additional} points allowed at
two-$\sigma$ ($95\%$) C.L. are shown in indigo and orange respectively.
The distribution of those allowed points is now essentially  
uniform in $\tb$ (though now the apparent concentration for relatively large 
$\tb$ is purely an artifact of the logarithmic scale), in contrast with the previous plot in Fig.~\ref{m0tb_fig1}. More importantly, 
$\tb$ is constrained by lower and upper bounds, but
these essentially originate from the theoretical (model-dependent) 
constraints, indicated in red for large $\tb$ 
which correspond to a tachyon $\t \tau_1$. (NB the white zone for $\tb \lsim 2.2$
corresponds to a very light $m_h$). 
More precisely we observe that even at the $68\%$ C.L., values of $\tb$ up to
the theoretical SPS1a constraint $\tb\lsim 35$ are not excluded. 
Moreover one sees that the $95\%$ C.L. additionnal points (indigo and orange) 
have only the effect of slightly
enlarging the $m^{q,l}_0$ determination but not much 
influence on the $\tb$ determination. 
Thus a  more realistic 
treatment of errors simply confirms our above 
crude estimate, indicating that no interesting model-independent upper limits
on $\tb$ can be derived from the prospected 
LHC squark and slepton masses accuracies. \\

It is instructive to compare at this stage those results
with a more standard top-down fit. We thus 
used MINUIT to perform a standard $\chi^2$ minimization 
starting from a MSSM model with universal
GUT $m^{q,l}_0$ and  
$\tb$ free parameters. First we fix the 
gaugino masses $M_1, M_2, M_3$ to their SPS1a values
and then fit this model to the ``data" consisting solely of  
$m_{\t u_1}, m_{e_2}$. This
two-parameter fit result is shown in Table
\ref{tabminuit_ql} for different MINUIT minimization options,
at the one-$\sigma$ level.  
\begin{table}[h!]
\begin{center}
\caption{\label{tabminuit_ql} Constraints on $m_0, \tb$ obtained
from a standard top-down $\chi^2$ fit with different model and input
assumptions and different 
level of MINUIT minimizations. } 
\begin{tabular}{|c||c|c||c|}
\hline\hline
Data $\&$ fitted parameter         & MIGRAD       & MINOS        & nominal \\
(+ model assumptions)  & minimization ($68\%$C.L.)& minimization ($68\%$C.L.)&  SPS1a value \\  
\hline
$m_{\t e_2}, m_{\t u_1}$ &         &                       & \\
 (1-loop RGE + no $\t q$ R.C.)   & (convergent)         & $\tb$ exceed
 limits          &  \\
 $m^{q,l}_0$  &  99.98 $\pm 8.8  $   &  99.98 $\pm$ 8.7  &  100  \\
 $\tb$      &   9.41 $\pm$ 29.8  & 9.41
 $^{+\mbox{\small no limits}}_{-\mbox{\small no limits}}$&
 9.74 \\
            &                   &                        &   \\
\hline
$m_{\t e_2}, m_{\t u_1}$ + &         &                       & \\
$m_{\t N_1}, m_{\t N_2}, m_{\t g}$ &    &                  &  \\
 (1-loop RGE + no $\t q$ R.C)   & (convergent)         & $\tb$ exceed
 limits          &  \\
 $m^{q,l}_0$  &  99.96 $\pm 7.9 $   &  99.9 $^{+10}_{-10.5}$ &  100  \\
 $m_{1/2}$  &   250 $\pm 4.3$    &  250 $\pm 4.3$         &    250 \\
 $\tb$      &   10. $\pm$ 29.8  & 10 $^{+29.8}_{-\mbox{\small no limits}}$
 & 9.74 \\
        &        &            &        \\
\hline\hline
\end{tabular}
\end{center}
\end{table}
%
In fact the simpler MIGRAD\cite{minuit} (symmetric error) minimization did 
converge, giving apparently a (marginal) upper $\tb$ limit 
at the one-$\sigma$ level. But it is not a very 
useful $\tb$ bound, 
being above the theoretical $\tb$ limits in this SPS1a case.
Moreover, the more 
elaborated MINOS minimization, taking into account properly
unsymmetrical errors and non-linearities\cite{minuit}, 
did not find any $\tb$ upper (nor lower)
limits. These results are thus qualitatively consistent with our more naive 
analysis above. 
Note that the $m_0$ bounds are also roughly   
consistent with our previous results.  
Now there are several reasons not to trust even the marginal upper bounds
found for $\tb$:
though it is difficult to trace the very details of the minimization steps,
the fit is much probably indirectly influenced by the 
 $\t \tau_1$ becoming very small (and ultimately 
 tachyonic) for $\tb \gsim 35-36$.
 More precisely, above those values, 
SuSpect still gives $m_{\t u_1}$, $m_{\t e_2}$ output (unless
explicit warning flags are switched on), but these
 are no longer very reliable (because the iterations needed to calculate
 a convergent spectrum are stopped in this case\cite{suspect}).   
 In particular,
there is an abrupt change of $m_{\t u_L}$ once $\tb\gsim 36$, see
Fig.~\ref{muLtb}  
(though the overall variation remains reasonable). 
Thus, comparing these minimization results and 
Fig.~\ref{m0tb_fig3} with the simple estimate Eq.~(\ref{crudetb}), we
can infer that the marginal upper bound on $\tb$
is principally determined by the indirect small higher order $\tb$ dependences  
(like typically the dependence $m_{\t u_L}(\tb)$ as illustrated in Fig.~\ref{muLtb}
originating mainly from the slightly varying EWSB scale, rather than directly from 
the $m_{\t u_1}$, $m_{\t e_2}$ accuracies). 
We have crosschecked this by redoing similar fits with a constant EWSB scale, or
with two-loop RGE, which both have the effect of smoothing somehow the
variation of $m_{\t u_L}(\tb)$ near the transition to tachyonic $\t \tau_1$
for $\tb \sim 36$, and the $\tb$ upper bound in Table  \ref{tabminuit_ql} tends
to increase (or even to disappear with non convergent minimizations).  
We have also further checked this by
increasing progressively the mass accuracies: 
while the corresponding $m_0$ bounds decrease, following the expected statistical behaviour, for $\tb$ one obtains either non convergent
minimizations, or extra odd solutions far from the SPS1a
values. There are anyway no improvements on $\tb$ bounds until a substantial
decrease of these experimental errors is set (about an order
of magnitude smaller than the LHC accuracies of Table \ref{tabexp}). 
 
For completeness we performed another minimization,  
taking into account in addition the neutralino and gluino masses with 
a three-parameter fit of $m_0, m_{1/2}, \tb$. The situation does not improve 
much as concerns $\tb$ limits, as illustrated by the corresponding
results in Table \ref{tabminuit_ql}. In this case the (unsymmetrical) MINOS
$m_0$ bounds
are slightly worse than for the two-parameter fit, which is expected 
since now $M_3$ is not fixed.
Accordingly the $M_3$ accuracy propagates to the $m_0$ determination
via the RGE, as explained above. On the other hand the accuracy on $m_{1/2}$
is very good.  
Finally  we also performed a similar fit for a benchmark
point like SPS1a except $m_0 =200$ GeV, in which case the transition 
to tachyonic $m_{\t \tau_1}$ happens only for very large $\tb > 50$, 
having thus potentially less
influence on the fit. Assuming the same mass 
accuracies than for the true SPS1a point we obtain, for a
two-parameter  (respectively three-parameter) fit: 
$\tb = 9.93 \pm 52.2$ (respectively $\tb = 10.1 \pm 50.7$), at 
one-$\sigma$ level, while errors for $m_0$ and $m_{1/2}$ are very comparable 
to the SPS1a ones. So this confirm the above analysis and indicates that
the  $\tb$ upper bounds are essentially inexistent. 

This also illustrates that a ``global'' top-down fit, whatever
elaborated with MINUIT algorithms, may be fooled and lead to misleading 
conclusions due to extra parameter dependences which originate from theoretical 
approximation artifacts (e.g. here the choice of EWSB scale, RGE approximation, 
etc). In the present case, one could of course easily
avoid such problems by simply adding protections within the 
minimization procedure, but that would amount to put 
explicitly the (model-dependent) upper bound on $\tb$ due to 
tachyonic (or more generally too light) $\t \tau_1$. \\
    
 We finally examine two questions related to theoretical uncertainties
that are relevant to the above analysis. First, as already mentioned in deriving
the $m^{q,l}_0$ constraints in Eqs.~(\ref{m0res}-\ref{m0_u1}) 
we had neglected for simplicity
the errors on $s^2_W$: the latter are actually
not quite negligible, since even a small uncertainty can in principle
affect our determination from Eq.~(\ref{sr1}). However, the bulk
of radiative correction contributions to the $\bar s^2_W$ parameter, 
in the $\overline{DR}$ scheme, originates from standard model and are thus
predictable in our framework. Additional supersymmetric
contributions are not negligible either\cite{bpmz},
but we checked that varying all MSSM parameters 
and sparticle masses form the SPS1a to models with arbitrary MSSM 
values produces a variation of $\bar s^2_W$
of about $0.6\%$ only. One may probably push parameters to extreme values
to find a slightly larger variation, so we conservatively consider a $1 \%$
uncertainty on $\bar s^2_W$. The impact on $m_0$ determination
is an (upper) shift by about $\sim 3$ GeV with respect to the numbers quoted
in Eqs.~(\ref{m0res}-\ref{m0_u1}). Note that the correct dependence
of $m_0$ and $\tb$ upon $\bar s^2_W$ is automatically 
taken into account in the MINUIT 
fit results in Table \ref{tabminuit_ql}.
 
Another potential question is
that the determination of $m^{q,l}_0$ via Eq.~(\ref{sr1}) together with RGE,  
depends only on the cascade masses if restricting the RGE to 
one-loop order. At the two-loop order, practically all 
other MSSM parameters are entering the RGE.
Nevertheless, it is possible to study the impact of this uncertainty
by assuming, within the two-loop RGE level, simple (e.g. universal) relations
for the unknown parameters, and to redo our analysis: while central values
are evidently shifted, the impact on error propagation is rather negligible, 
with minor quantitative changes on e.g. the obtained $m^{q,l}_0$ constraints.  
This is also consistent with some comparisons we made of two-loop versus one-loop RGE fit results using MINUIT.

We thus conclude  
that $\tb$ is essentially unconstrained from the data we used at this 
stage, which is not much a surprise and consistent with general expectations on 
LHC prospects\cite{LHCstudy,kane}. 
Model-independent constraints on $\tb$, though moderate, may be obtained however 
from other sectors as we shall see in next sections, either considering the 
information from bottom squarks in the cascade, or adding  
the lightest Higgs mass measurements.
%
%
\section{Third generation squark parameter determination}
We will now consider the possible determination of some of the
third generation squark parameters. As discussed in section 2,
the sbottoms enter the gluino cascade and both mass 
eigenstates may be measured
to some extent (though with less accuracy for the heaviest $\t
b_2$)\cite{LHCstudy,cascade2}. 
We will examine what additional information they provide,
both in unconstrained MSSM or assuming universality relations, as in
previous analyses. 
\subsection{Scenario S5: constraints from the sbottom masses in 
non-universal MSSM}
So far, we have obtained from the gluino cascade useful 
constraints on the gaugino/Higgsino parameters and $m^{q,l}_0$,
while $\tb$ is very poorly determined. 
The 
sbottom sector may provide information on the missing
third generation soft scalar terms, or eventually on $\tb$.
To set up signs and other conventions, we recall the sbottom 
mass squared matrix:
  \begin{equation}
    M_{\tilde b}^2=
    \left(
      \begin{array}{cc} 
         { m^2_{Q3L} }+ m_b^2+(-\frac{1}{2}+\frac{1}{3}s^2_W)\, m^2_Z\cos 2\beta  & 
m_b\, ({ A_b} -\mu\tan\beta) \\
         m_b  ({ A_b} -\mu\tan\beta) &
 {m_{b_R}^2}+m_b^2 -\frac{1}{3}\, m^2_Z\cos2\beta
     \end{array} \right) 
     \label{sbmass}
  \end{equation}
  where again all parameters are implicitly understood to be 
  in the $\overline{DR}$ scheme. 
It is immediate that the trilinear term $A_b$ will be 
very badly determined, being largely suppressed by the 
bottom mass $m_b$, with $\bar m_b(Q_{EWSB})\sim 3$ GeV e.g. in SPS1a, and
similar values in more general SUSY scenarios.  
 However, this also means that we can practically neglect $A_b$. 
In this case the sbottom masses determine, to a good approximation,
  the third generation
 parameters $m_{Q3_L}$, $m_{b_R}$ in a simple way directly from 
 (the trace of) (\ref{sbmass}):
\be
m^2_{\t b_1} +m^2_{\t b_2} +\frac{m^2_Z}{2}\cos 2\beta -2 m^2_b \equiv S = m^2_{Q3_L}+m^2_{b_R}\;. 
\ee  
Now, taking one of the mass eigenvalue (say $m_{\t b_1}$) as a given function
 of the other parameters $\mu, \tb$ etc, we can solve a simple
 relation for $m_{Q3_L}$, $m_{b_R}$: 
\beq
      m_{Q3_L(b_R)} & = & \left[\frac{S+(-)D}{2}\right]^{1/2} \nn \\
 D  &= & -Y+\left[(m^2_{\t b_2}-m^2_{\t b_1} -2 m_b X_b)
      (m^2_{\t b_2}-m^2_{\t b_1} +2 m_b X_b)\right]^{1/2}\;,
\label{sbeq}
\eeq 
where $Y= (-\frac{1}{2} +\frac{2}{3} s_W^2)m^2_Z \cos 2 \beta$
and $X_b = A_b -\mu \tb$. 
There is in fact a twofold ambiguity, namely
a second solution where $m_{Q3_L} \leftrightarrow m_{b_R}$ in Eq.~(\ref{sbeq}), since it is a second order equation. For SPS1a we know that
 $ m_{Q3_L}(Q_{EWSB}) > m_{b_R}(Q_{EWSB})$ resolves this ambiguity,
 but for unconstrained MSSM one has to deal in principle 
 with the two possibilities in the absence of further information other
 than the sbottom masses.  
 In addition one also expects radiative corrections to this 
 tree-level determination
 which might spoil its simplicity. Radiative corrections
 to sbottom masses are  not negligible in general, 
 but to a very good approximation they are largely dominated
 by gluino and squark contributions, as already discussed in sub-section 2.3. 
 So we have a reasonable knowledge  
 of these corrections, such that the running masses in the
 $\overline{DR}$ scheme involved in Eqs (\ref{sbeq}) may be obtained
 from the pole masses after appropriate subtractions. 
 \subsection{Explicit reconstruction: SPS1a test case}
 The resulting 
 $m_{Q3_L}$, $m_{b_R}$ determination for the SPS1a test case
 with corresponding input mass accuracies is illustrated in Fig.
 \ref{figsb1}. We plot the allowed range for  $m_{Q3_L}$, $m_{b_R}$
 obtained from a (uniform) scan over mass accuracies from Table \ref{tabexp} 
 and the other relevant parameters, as indicated. Different cases are illustrated
 when varying $\tb$ and $\mu$ within the previously
 obtained limits, i.e. $\mu\sim 360\pm 200$,
 or more constrained situations, $\Delta\mu \pm 10$ 
 (the latter corresponding roughly to the case 
 where three neutralino masses can be measured), and $3\lsim \tb\lsim 35$
 (using the theoretical upper bound for SPS1a).  
 Since $A_b$ is essentially unknown at this stage, we vary it widely in the
 range: $-1$ TeV to $1$ TeV. 
          \begin{figure}[h!]
    \begin{center}
  \includegraphics[width=7cm,angle=-90]{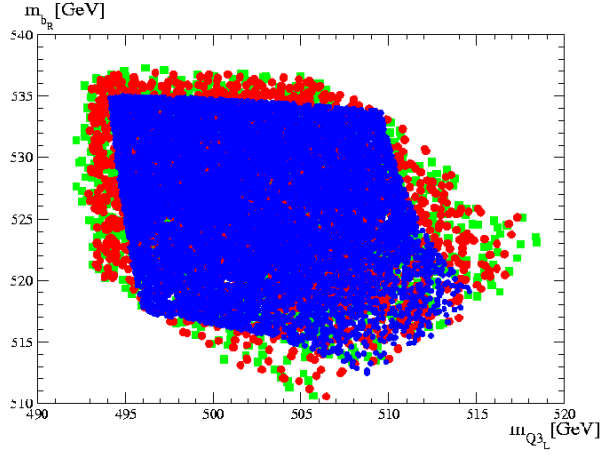}
\caption[long]{\label{figsb1} 
Constraints in general MSSM on $m_{Q3_L}$, $m_{b_R}$ (at $Q_{EWSB}$ scale) from
$m_{\t b_1}, m_{\t b_2}$ measurements in the gluino cascade decays.
Radiative corrections linking the pole to running masses have been subtracted out. 
Experimental errors on sbottom masses are taken from Table \ref{tabexp}.
Blue region: fixed nominal values of $\tb \sim 9.74$, $\mu\sim 357$ GeV (and $A_b=0$). Red region: 
$3 \lsim \tb \lsim 35$, $\Delta\mu\sim 10$ GeV, $ -100 GeV < A_b < 100 GeV$;
green region: $3 \lsim \tb \lsim 50$, $\Delta\mu\sim 200$ GeV,  
$-1 \mbox{TeV} < A_b< 1 \mbox{TeV}$. Nominal SPS1a values are $m_{Q3_L} = 503$ GeV, 
$m_{b_R}= 525$ GeV (at one-loop RGE level).}
  \end{center}
 \end{figure}
          \begin{figure}[h!]
    \begin{center}
 \includegraphics[width=7cm,angle=-90]{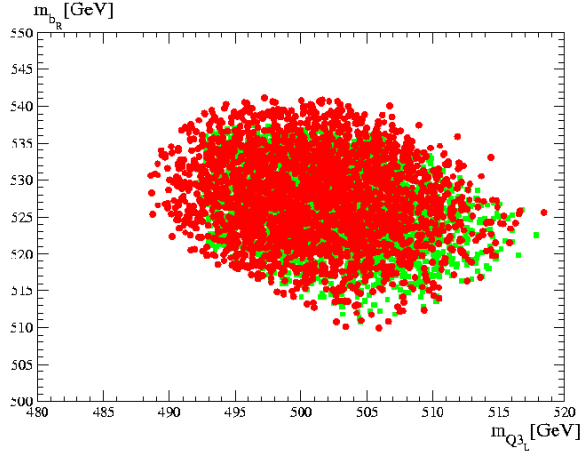}
\caption[long]{\label{figsb2} Comparison of uniform (in green) versus Gaussian
(in red) scan with one-$\sigma$ ($68\%$ C.L.) level contours. The variation 
of extra parameters
correspond to the less determined range in Fig.~\ref{figsb1}, i.e.:
$3 \lsim \tb \lsim 50$, $\Delta\mu\sim 200$ GeV,  
$-1 \mbox{TeV} < A_b< 1 \mbox{TeV}$.}  
  \end{center}
 \end{figure}
As one can see, the determination of $m_{Q3_L}$, $m_{b_R}$ is quite accurate 
 even in a non-universal MSSM case, and for largely unknown $A_b$. 
 This is clearly  
 explained by the strong suppression of $A_b$, which has thus practically 
 no influence on the scalar mass parameter determination. Moreover, 
 the variation of
 $\mu$, and even the large variation of $\tb$, have no strong influence either, 
 which is explained by the fact that they are both very suppressed 
 also by $m_b$ as compared to other parameters. To examine the impact of
  radiative corrections, we did a similar scan but neglecting on purpose
 the radiative corrections, i.e. using as central values the pole
 sbottom masses. As a result  
 the $m_{Q3_L}$, $m_{b_R}$ central values are substantially 
 shifted with respect
 to nominal SPS1a values, but the accuracy obtained on $m_{Q3_L}$, $m_{b_R}$
is very similar.  

Next, in Fig.~\ref{figsb2} we compare the previous results obtained 
from a simple scan
with uniformly distributed random number with those from a scan made with
Gaussian-distributed random numbers, defining a $\chi^2$
domain similarly to the analysis in section 4.
One can see that the corresponding one-$\sigma$ limits show the expected
statistical behaviour, having a more regularly-shaped contour than the one 
in the uniform case, but with no drastic changes in the limits obtained on 
$m_{Q3_L}$, $m_{b_R}$.
\subsection{Scenario $S^{\prime}_5$: constraints from sbottoms in MSSM with sfermion mass universality}
 Prospects for the third generation scalar terms are
 better in a more constrained MSSM i.e. with universality relations assumed 
 in the scalar sector. In this case, one has:
 \be
  m^{q,l}_0
 \equiv m_{Q3_L}(Q_{GUT})= m_{b_R}(Q_{GUT}) 
 \label{sqsluniv}
 \ee
 Then, using the above results from section 4 on $m^{q,l}_0$ limits, 
 one can determine from a top-down RGE the corresponding accuracy
 obtained on $m_{Q3_L}(Q_{EWSB})$ and $m_{b_R}(Q_{EWSB})$:
 \be
 m_{Q3_L}(Q_{EWSB}) \sim 498 \pm 1.2 \pm 7 \mbox{GeV},\;\;\;
  m_{b_R}(Q_{EWSB}) \sim 521 \pm 1.8 \pm 6 \mbox{GeV}\;,
 \label{mqlbruniv}
 \ee 
 where the first errors correspond to a variation of $m_0$ for fixed
 $M_3$, while the second additional errors take into account the 
 $m_{\t g} \sim M_3$ uncertainty (which 
 dominates the final uncertainties). Alternatively, one may use here
 the limits on $m_0$ obtained from the $\chi^2$ minimization
 in Table \ref{tabminuit_ql}, which would gives results roughly 
 comparable to the bounds in Eq.~(\ref{mqlbruniv}) (except that it is
 somewhat more difficult to disentangle the effect of the $M_3$ uncertainties
 from the fit).
 Comparing with the unconstrained MSSM
determination from the sbottom masses in Figs.~\ref{figsb1}, \ref{figsb2} one
can see a definite improvement on $m_{Q3_L}(Q_{EWSB})$ and $m_{b_R}(Q_{EWSB})$ accuracies by about a factor two. 
This may be sufficient to resolve the
twofold ambiguity between $m_{Q3_L}$ and $ m_{b_R}$ discussed
above in the non-universal sfermion case. 
    
 Perhaps more interestingly, one may expect to derive an independent
 determination of $\tb$ in this sfermion universality scenario, by combining all information
 from $m_0$ and the sbottom masses. More precisely, turning the other
 way round Eqs.~(\ref{sbeq}) one can determine very simply
 $X_b =Ab-\mu \tb$ in terms of the know parameters, as follows:
 \be
 2\,m_b\,X_b = -\left[ (m^2_{\t b_2}-m^2_{\t b_1})^2 -
 (m^2_{\t Q3_L}-m^2_{\t b_R} +Y)^2 \right]^{1/2}.
 \label{Xbsol}
 \ee
 (Actually Eq.~(\ref{Xbsol}) comes from a second order equation, 
 so in principle there is again a twofold 
 ambiguity: $2m_b X_b = \pm [\cdots]^{1/2}$.
 But within our sign convention, if $\mu>0$, $X_b<0$ necessarily,
 even for general MSSM, unless $|\mu| \tb < A_b$, which is not a very common 
 situation except perhaps for very small $\tb$ values.)~\footnote{Note also
 that $\tb$ enters both sides of Eq.~(\ref{Xbsol}) via 
 the D-term $Y= (-\frac{1}{2} +\frac{2}{3} s_W^2)m^2_Z \cos 2 \beta$.
 Thus we have to iterate on $\tb$, and this
 is converging fastly.}   
 
        \begin{figure}[h!]
  \begin{center}
  \includegraphics[width=7cm,angle=-90]{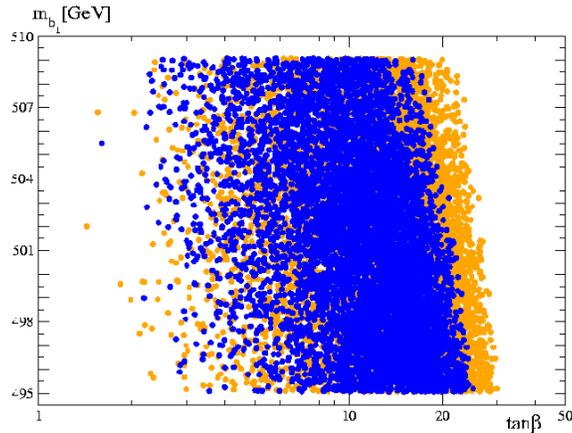}
\vspace{-1cm}
\caption[long]{\label{Xbtbfig1} Constraints on $\tb$ 
with uniform scan from $m_{\t b_1}, m_{\t b_2}$ accuracies.  
$A_b$ is assumed to be almost undetermined ($A_b = A_b(\mbox{SPS1a})\pm 1$ TeV). 
In blue : $\Delta m_{\t b_2} =7.9$ GeV. In orange: $\Delta m_{\t b_2} =16$ GeV.}
\end{center}
       \end{figure}
The resulting determination of $X_b$ and $\tb$ from Eq.~(\ref{Xbsol}) 
 is first illustrated in Fig.~\ref{Xbtbfig1}: while $X_b$ is directly 
 determined from Eq.  (\ref{Xbsol}), 
 for $\tb$ one has to take into account the additional uncertainties
 in the determination of $\mu$ (and $A_b$, which is assumed to be essentially
 unknown at this stage). However it is clear that this 
 has a moderate impact on the 
 $\tb$ determination, 
 since even a large variation $-1 \:\mbox{TeV} < A_b < 1 \:\mbox{TeV}$ 
 has a moderate effect, 
 $\sim \Delta A_b/\mu$ (unless if $|\mu|$ would be very small). 
 So even at this stage   
 where $A_b$ is completely undetermined, reasonable  
 constraints: $\tb \lsim 27-28$ are obtained, as shown by the blue contour. 
 Moreover, even when increasing the $m_{\t b_2}$ uncertainty by a factor of two,
 i.e. $\Delta m_{\t b_2} \sim 16$ GeV, one still obtains some reasonable 
 constraints on $\tb \lsim 30$, see Fig.~\ref{Xbtbfig1}.  
 (But the upper bound on $\tb$ disappear if the $\t b_2$
 is not measured at all). 
 Like for the previous section analysis, we also show the difference 
 between uniform and Gaussian scans, with corresponding
 one- and two-$\sigma$ contours in Fig.~\ref{Xbtbfig2}. For completeness
 the joint (correlated) determination of $(\tb, X_b)$ is also shown 
 in Fig.~\ref{Xbtbfig3}.   
 
 We thus observe that, provided the two sbottom masses can be measured with this
 accuracy at the LHC, the bounds on $\tb$ may start to be rather interesting, 
 at least at the one-$\sigma$ level. 
 These results are also confirmed
 by a $\chi^2$ minimization
 fit using MINUIT, with more or less comparable bounds obtained, as 
  shown in Table \ref{minuit_sbtb}.
        \begin{figure}[h!]
  \begin{center}
  \includegraphics[width=7cm,angle=-90]{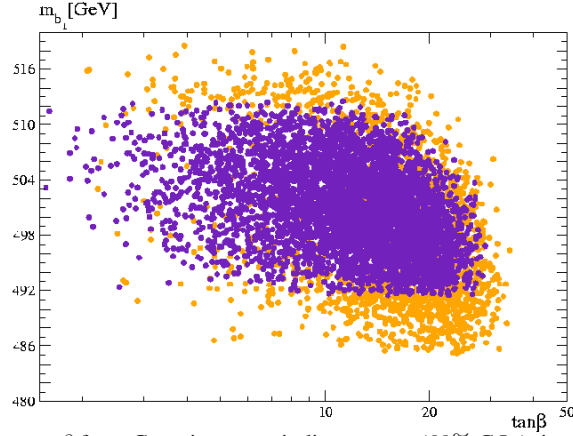}
\vspace{-1cm}
\caption{\label{Xbtbfig2} Constraints on $\tb$ 
from Gaussian scan: indigo: one-$\sigma$ ($68\%$ C.L.); in orange:
two-$\sigma$ ($95\%$ C.L.).}
\end{center}
       \end{figure}
        \begin{figure}[h!]
  \begin{center}
  \includegraphics[width=7cm,angle=-90]{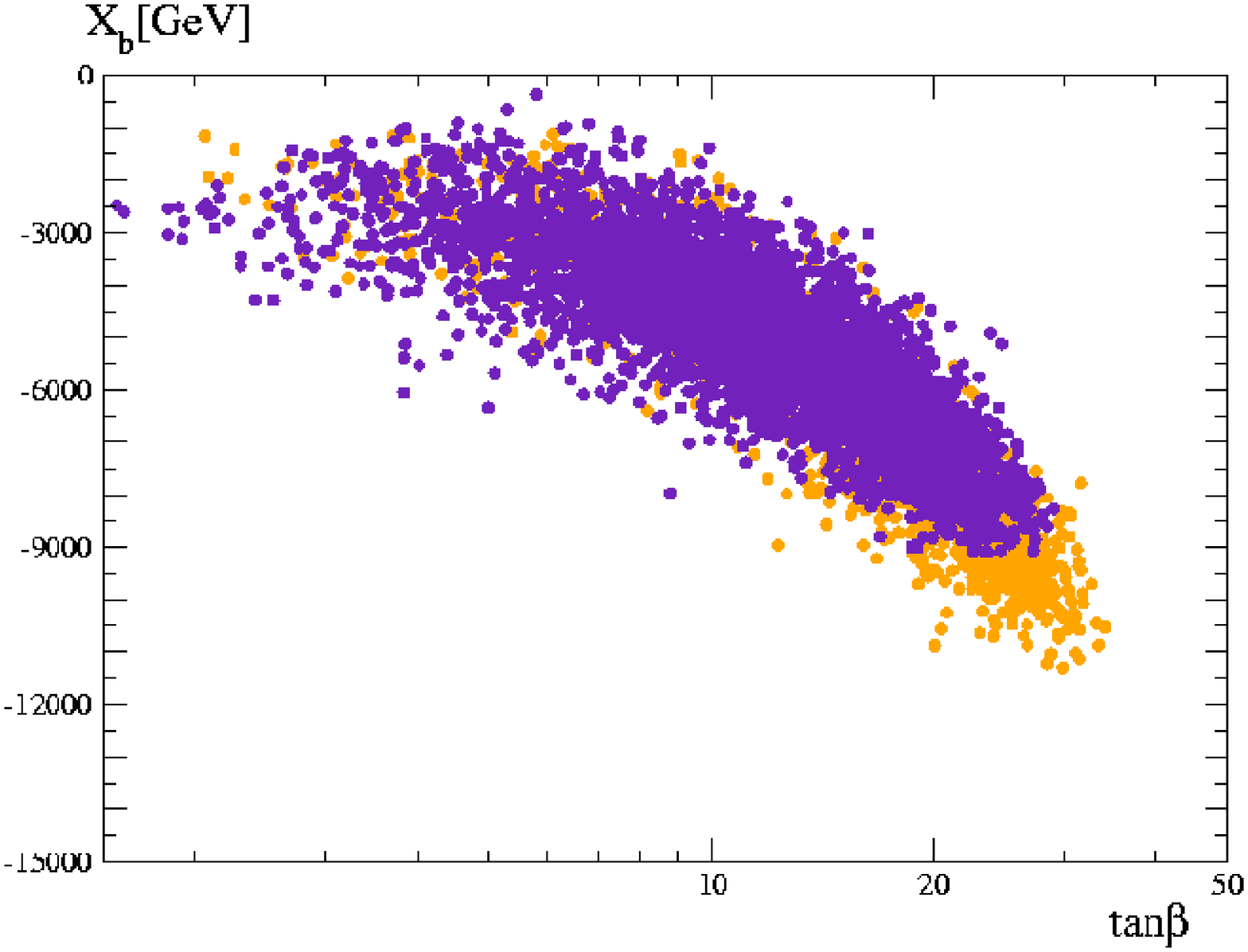}
\vspace{-1cm}
\caption{\label{Xbtbfig3} Constraints on $\tb, X_b =A_b-\mu \tb$ 
from Gaussian scan: indigo: one-$\sigma$ ($68\%$ C.L.); orange:
two-$\sigma$ ($95\%$ C.L.).}
\end{center}
       \end{figure}
\begin{table}[h!]
\begin{center}
\caption{\label{minuit_sbtb} Constraints on $m_0, \tb$ obtained
from a standard top-down $\chi^2$ fit of squark +sbottom masses 
and neutralino/gluino masses, for different 
options of MINUIT minimizations. } 
\begin{tabular}{|c||c|c||c|}
\hline\hline
Data $\&$ fitted parameter         & MIGRAD       & MINOS        & nominal \\
(+ model assumptions)  & minimization ($68\%$C.L.)& minimization ($68\%$C.L.)&  SPS1a value \\  
\hline
$m_{\t e_2}, m_{\t u_1}$, $m_{\t b_1,b_2}$ + &         &                    & \\
$m_{\t N_1}, m_{\t N_2}, m_{\t g}$ &    &                  &  \\
 (1-loop RGE + no $\t q$ R.C)   & (convergent)         & $\tb$ exceed
 limits          &  \\
 $m^{q,l}_0$  &  99.96 $\pm 9.9 $   &  99.9 $^{+10}_{-10.4}$ &  100  \\
 $m_{1/2}$  &   250 $\pm 3.6$    &  250 $\pm 3.6$         &    250 \\
 $\tb(m_Z)$      &   9.98 $\pm$ 13.6  & 9.98 $^{+13.6}_{-\mbox{\small no limits}}$
 & 10 \\
            &                   &                        &   \\
\hline\hline
\end{tabular}
\end{center}
\end{table}
 We finally mention that squark and slepton universality 
 relations as in Eq.~(\ref{sqsluniv}) is yet not enough to have any
 interesting determination of $A_b$, which is quite obvious from
 its very suppressed contribution to the mass matrix (\ref{sbmass}).  
 One thus needs a different strategy
 to determine, if possible, the remaining trilinear couplings. The  prospects
 for the latter would be certainly much better if one could measure the stop
 masses in similar cascades. This is not the case for the SPS1a, but 
 we will examine in the next section how the Higgs sector parameters
 can help to determine the stop sector trilinear couplings.

\section{Bottom-up reconstruction of Higgs sector parameters} 
In the Higgs sector, the most relevant parameters (at tree-level) 
are the scalar mass terms $m_{H_u}$, $m_{H_d}$ and $\tb$. 
These may be replaced equivalently, using the 
EWSB conditions, by $\mu$, the pseudoscalar mass $m_A$, and $\tb$. 
More precisely, the EWSB minimization equations
$\partial V_{\rm Higgs} / \partial H_d = 0$; $\partial V_{\rm Higgs} /
\partial H_u = 0$ (where $V_{\rm Higgs}$ designates the MSSM scalar
potential)  can be solved for $\mu^2$ and $B \mu$:
\begin{eqnarray} \label{eq:ewsb}
\mu^2 &=& \frac{1}{2} \left[ \tan 2\beta (m^2_{H_u} \tan \beta
- m^2_{H_d} \cot \beta) -m_Z^2 \right] \nn \\
B\mu &=& \frac{1}{2} \sin 2\beta \left[ m^2_{H_u} + m^2_{H_d} + 2
\mu^2 \right] 
\end{eqnarray}
which eventually includes radiative corrections to the scalar
potential in the form of tadpole corrections $t_u$, $t_d$ \cite{potential,bpmz}: 
\be
m^2_{H_u} \to m^2_{H_u} -t_u/v_u \ {\rm and} \  m^2_{H_d} \to m^2_{H_d}
-t_d/v_d \;.
\ee
Next, the running $m_A(Q)$ is defined as:
\begin{equation}
    \bar m^2_A(Q)= m^2_{H_d}(Q)+ m^2_{H_u}(Q) +2 \mu^2(Q) 
\label{mar}
  \end{equation}
(where as previously all parameters are implicitly understood to be in the
$\overline{DR}$ scheme). The $A$ pole mass is then related to $m_A(Q)$
via additional radiative corrections\cite{mh1-2loop,bpmz}.  
It is well-known, however, that in the MSSM the scalar top
sector contributes largely to radiative corrections to $m_h$ 
and $m_A$. Thus 
even if some parameters are
available from other sector analysis (though not precisely for $\tb$ as we have
seen), 
only an independent measurement of $m_A$ (or alternatively 
an information on the stop sector, on $X_t\equiv A_t-\mu/\tb$) could
give a more useful information on the remaining parameters. 
For the determination of these Higgs parameter sector in a general
(non-universal) MSSM case, the prospects are thus not very good if the available
data are similar to those of the SPS1a benchmark, even 
if the lightest Higgs mass $m_h$ could be determined with a good accuracy. 
In a constrained MSSM with complete high scale universality 
of scalar soft terms, prospects
are possibly better. We can then use our previous analysis with in addition:
\be
m^{q,l}_0 \equiv m_{H_u}(Q_{GUT})= m_{H_d}(Q_{GUT})
\label{sfhuniv}
\ee   
to determine $\mu$ via Eqs.~(\ref{eq:ewsb}) as well as
the running $m_A$ value in Eq.~(\ref{mar}).
From Eq.~(\ref{eq:ewsb}) one has very roughly 
for $\tb \gg 1$: $\mu^2 \sim -m^2_{H_u}$ 
which is largely insensitive to $\tb$. (NB in our analysis we use of course 
the complete expressions from Eqs.~(\ref{eq:ewsb})). Thus the rather good determination
of $m_0$ from the squark and slepton sector, as analyzed in previous section, 
together with universality assumptions (\ref{sfhuniv}), give a more precise 
determination of $\mu(Q_{EWSB})$. However, $m_{H_u}$ has a strong sensitivity
to the trilinear stop coupling $A_t$ through its RGE via terms 
$\propto y^2_t A^2_t$, which restricts 
a very good determination of $\mu$.     
 Typically  
for $4\lsim \tb \lsim 50$ and a very moderate variation of $A_t$, $\Delta A_t\sim
100$ GeV, we obtain:
\be
330 \:\mbox{GeV} \lsim \mu(Q_{EWSB}) \lsim 360 \:\mbox{GeV}
\label{muuniv}
\ee
which is comparable to the accuracy obtained in a general MSSM
from neutralino mass measurements when a third neutralino can be measured,
see section~\ref{s36}. 
Now for less limited $\Delta A_t\sim 200$ GeV (which corresponds
approximately to the range of variation we obtain once using all
input, as we shall see later) 
those bounds are somewhat increased:
\be
300 \:\mbox{GeV} \lsim \mu(Q_{EWSB}) \lsim 410 \:\mbox{GeV}
\label{muuniv2}
\ee 
Note that these bounds are very insensitive to the $m_0$ variation 
within its accuracy range  
determined by squarks and sleptons in section 4. The limits (\ref{muuniv}),
(\ref{muuniv2}) from EWSB constraints 
are anyway improved as compared to those 
obtained solely from two neutralino mass input in section 3.
\subsection{Naive tree-level counting of parameters}
Let us first start with a simple tree-level analysis in order to   
delineate the parameters entering in game.
Concerning the lightest scalar Higgs mass, at tree-level, $m^{2,tree}_h$ is given by:
\beq
    m^{2,tree}_h =\frac{1}{2}\left[m^2_A+m^2_Z-
    \left((m^2_A+m^2_Z)^2-4 m^2_A m^2_Z \cos^2 2\beta\right)^{1/2}\right]\;.
\eeq
Inverting this relation gives a (unique) solution for $m_A$:
\beq
             \bar m^2_A = \frac{ \bar m_h^2 ( m_Z^2 - 
     \bar m_h^2)}{m_Z^2 \cos^2 2 \beta -  \bar m_h^2}
           \label{masol} 
    \eeq 
      from which one can also derive:
            \beq
            \label{MHuHd}
             m^2_{H_u} =\frac{ \bar m^2_A-(\mu^2+m^2_Z/2)(\tan^2\beta-1)}
           {\tan^2\beta+1} \;,\;\;\; 
            m^2_{H_d} =  m_A^2-m^2_{H_u} -2 \mu^2 \;. 
           \eeq
 So very naively one may have thought to equate Eqs.~(\ref{mar}) with 
 Eq.~(\ref{masol}), deriving from this 
 precise $\tb$ constraints typically.  
  But, as already mentioned, this tree-level analysis 
   would be very unrealistic since the
   lightest Higgs mass $m_h$ and $m_A$ get large radiative
   corrections, already at one-loop level, mainly from the 
   stop/top loops, which are enhanced by a $m^4_t$ dependence. 
  Rather, as we did for the neutralino and squark sectors, 
  we will try to incorporate the bulk of these large corrections consistently.
\subsection{Reconstruction of Higgs sector parameters in constrained MSSM}     
 The MSSM neutral Higgs mass-squared matrix reads,  including radiative 
 corrections:
\begin{eqnarray}
      \left(\!\!
      \begin{array}{cc}m^2_Z\,\cos^2\beta + { m_A^2}\,\sin^2\beta 
      +S_{11} &\! -(m^2_Z+{ m_A^2})\,\sin\beta\cos\beta +S_{12}\\
      -(m^2_Z+{ m_A^2})\,\sin\beta\cos\beta +S_{12} 
      &\! m^2_Z\,\sin^2\beta + { m_A^2}\,\cos^2\beta +S_{22}
      \end{array}\!\!\right)\! 
\label{mhfull}
\end{eqnarray}
where $S_{ij}$ designate generically loop self-energy contributions. 
   To explain our procedure, let us first  
consider a very simple approximation for $m_h$ (see e.g. \cite{haber}):        
  \begin{equation}
     m_h^2={m^{2,tree}_h}+\frac{3 g^2_2 \,m^4_t}{8\pi^2 m^2_W} 
    \left[\ln \left(\frac{{m_{\tilde t_1} m_{\tilde t_2}}}{m_t^2}\right)
    +\frac{{X_t^2}}{M_S^2}-\frac{{X_t^4}}{12 M_S^4}\right]
\label{hsimp}
  \end{equation}
where 
  \beq
    &&{ X_t}=A_t-\mu \cot\beta\\ \nn
    &&M_S^2=[m_Q^2m^2_{t_R}+m_t^2(m_Q^2+m^2_{t_R})+m^4_t]^{\frac{1}{2}}\\\nn
    &&{m_{\tilde t_1} m_{\tilde t_2}}=[M_S^4-4m_t^2 {X_t^2}]^{\frac{1}{2}}
\eeq
(and $g_2$ is the $SU(2)$ gauge coupling).
There is a large literature on various approximations to the MSSM Higgs mass(es) 
radiative corrections.
We emphasize that in our actual analysis, we will not use 
the simple Eq.~(\ref{hsimp}) but a more elaborated expression, 
yet giving compact expressions for the needed $S_{ij}$ in Eq.~(\ref{mhfull})\cite{Svenetal}. (This compact approximation is included 
in SuSpect as one possible Higgs 
mass calculation option). The latter incorporates some of the leading two-loop 
effects, but depends only on the very same parameters as in Eq.~(\ref{hsimp}) 
(apart from SM-like parameters).\\  
At this stage, if a rather precise information on the stop sector
would be available, we could use an improved version 
of Eq.~(\ref{masol}), incorporating
radiative corrections, to derive $m_A$ independently from the
   relation (\ref{mar})~\footnote{Note that even  
   the full one- (two)-loop Higgs radiative corrections, 
  i.e. with full expressions of the $S_{ij}$ above, do
    preserve a {\em linear} $ m^2_A$ solution in terms of $m_h$
    similarly to Eq.~(\ref{masol}) but with appropriate corrections
   from $S_{ij}$.}.    
   However, within the SPS1a input assumptions, this would not give
   any useful constraints (given also the poorly known $\tb$ constraints). 
   We thus rather assume universality relations (\ref{sfhuniv}), such that  
   Eq.~(\ref{mar}), together with the 
 $m_h$ measurement, can determine both $X_t$ and $\tb$. 
   Since we also now assume complete universality
   of the squark sector, we can use the previous $m_0$ determination
   to calculate also the stop sector soft terms $m_{\t t_R}$ and 
   $m_{\t Q3_L}$ (the latter already obtained from the sbottom sector),
   as needed in Eq.~(\ref{hsimp}) (or its more complete generalization
   \cite{Svenetal} actually used in the numerics below). 
 The simplified expression of $m_h$ we used is remarkably close to the full 
one-loop (plus leading two-loop) values\cite{bpmz,bdsz,mh1-2loop} (in the
$\overline{DR}$ scheme), 
with at most a $2$ GeV discrepancy (and often much less) for a large 
range of MSSM parameters. (For the relevant SPS1a case, one finds  
$m_h^{simp} =111.28$ GeV, while $m_h^{2-loop} =110.90$ GeV.) 
\subsection{Application to SPS1a test case}
        \begin{figure}[h!]
\vspace*{-1cm}
  \begin{center}
  \includegraphics[width=8cm,angle=-90]{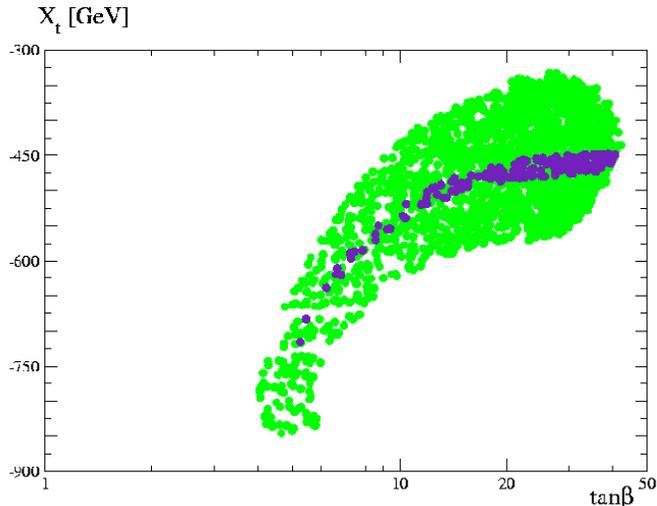}
\vspace{-1cm}
\caption{\label{figtbxt1} Constraints on $\tb$ and $X_t \equiv A_t -\mu/\tb$: 
indigo: one-$\sigma$ ($68\%$ C.L.) from a gaussian scan; 
green: same contour but for theoretical uncertainties $\Delta m_h =2$ GeV instead.}
\end{center}
       \end{figure}
        \begin{figure}[h!]
\vspace*{-1cm}
  \begin{center}
  \includegraphics[width=8cm,angle=-90]{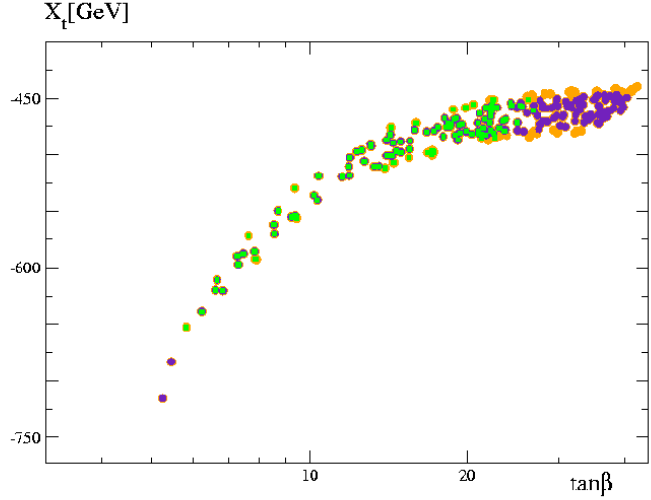}
\vspace{-1cm}
\caption[long]{\label{figtbxt2} Constraints on $\tb$ and $X_t \equiv A_t -\mu \tb$: 
indigo: one-$\sigma$ ($68\%$ C.L.); 
orange: additional points at two-$\sigma$ ($95\%$ C.L.); green squares: 
remaining points (one-$\sigma$ level) when adding sbottom mass measurements.}
\end{center}
       \end{figure}

We now apply the previous analysis to the reconstruction of some
of the parameters related to the Higgs sector for the SPS1a test case
with corresponding input mass accuracies. We perform again scan over parameters, taking into account
 the previous constraints on $m_0$ available from squark and slepton sector 
 (and letting $X_t$ arbitrary), deriving
joint constraints on $X_t, \tb$. The variation of other parameters
which it implies (like e.g. the variation of $\mu$ from EWSB relations (\ref{eq:ewsb})) 
is consistently taken into account. 
This is illustrated first in Fig.~\ref{figtbxt1}, where
the range of points obtained from the prospected experimental accuracy:   
 $\Delta m_h(\mbox{exp}) \sim 0.25$ from Table \ref{tabexp} 
are shown. Also shown on the figure is the similar range now corresponding 
rather to the present theoretical 
uncertainties in the Higgs mass calculations (we take the latter as
$\Delta m_h(\mbox{th})\sim 2$ GeV\cite{mh1-2loop,adkps}). Note again that
the density of points in the $\tb, X_t$ plane has no precise statistical
meaning: the larger ``density'' seen for rather large $\tb$ is an artifact of
the logarithmic scale. Also the one-$\sigma$ band which is thiner for low $\tb$
and small $X_t$, is easily explained from the $m_h$ dependence on $\tb$
and $X_t$: more precisely for fixed $\tb$, $m_h(X_t)$
reaches a maximum for a certain $X_t$ (see e.g. \cite{mh1-2loop}), and it is
more difficult to match the approximate SPS1a value of $m_h$ for both 
small $\tb$ and large (negative) $A_0$. 
  The domain on $\tb$ values obtained from this experimental 
 Higgs mass measurement, at the one-$\sigma$ level, is 
 thus approximately determined as:
\be
5 \lsim \tb \lsim 40  \;.
\label{tbfromh}
\ee
These bounds are strongly correlated with $X_t$ values, as the figure shows.
For $X_t$ one finds approximately:
\be
-730 \:\mbox{GeV} \lsim X_t \lsim -450 \:\mbox{GeV}
\ee   
Note also that all the points shown with relatively large $\tb \sim 35-40$ are 
satisfying the previously mentioned theoretical constraint, with no tachyon 
$\t\tau_1$, since when departing from the SPS1a benchmark, larger values 
of $\tb$ can be compensated by values of $A_0 > -100$ GeV. (This is of course
not taking into account any additional experimental constraints on $m_h$,
$m_{\t \tau}$, etc which, as already emphasized, are omitted here
since they are more model-dependent constraints.)  
The resulting limits on $X_t$ and
$\tb$ are of course sensitive to $m_h$ theoretical uncertainties as shown
on the figures (though very little as concerns the $\tb$ upper bound).  
In this case we obtain: 
\be
4 \lsim \tb \lsim 40;\;\;\;-850\: \mbox{GeV} \lsim X_t \lsim -330 \:\mbox{GeV} \;.
\ee	   
\begin{table}[h!]
\begin{center}
\caption{\label{minuit_tbh} Constraints on mSUGRA parameters $m_0, m_{1/2}, 
A_0, \tb$ obtained
from a standard top-down $\chi^2$ fit of (first two generation) squark, sleptons,  
neutralino/gluino masses, plus the lightest Higgs mass, for different 
choices of MINUIT minimizations. } 
\begin{tabular}{|c||c|c||c|}
\hline\hline
Data $\&$ fitted parameter         & MIGRAD       & MINOS        & nominal \\
(+ model assumptions)  & minimization ($68\%$C.L.)& minimization ($68\%$C.L.)&  SPS1a value \\  
\hline
$m_{\t e_2}, m_{\t u_1}$, + &         &                    & \\
$m_{\t N_1}, m_{\t N_2}, m_{\t g}$ +$m_h$ &    &                  &  \\
 (1-loop RGE + no $\t q$ R.C)   & (convergent)         &
 (problems)          &  \\
 $m^{q,l}_0$  &  99.96 $\pm 11.8 $   &  99.9 $^{+11.5}_{-10}$ &  100  \\
 $m_{1/2}$  &   250 $\pm 4.9$    &  250 $\pm 4.9$         &    250 \\
 $A_0$  &   -100.5 $\pm 150.0$    &                    &    -100 \\
 $\tb(m_Z)$      &   9.97 $\pm 4.3$  & 9.97 $^{+\mbox{no limits}}_{-4.3}$
 & 10 \\
            &                   &                        &   \\
\hline\hline
\end{tabular}
\end{center}
\end{table}
However the contours shown in Fig.~\ref{figtbxt1} 
did not take into account yet the additional constraints from  
the sbottom mass measurements which were obtained in previous section 5. 
This is now added in Fig.~\ref{figtbxt2} (in green), where the
one and two-$\sigma$ domains obtained from $m_h$ measurement are compared. 
One can see that 
at the two-$\sigma$ level the $X_t$ range increases slightly, for large
$\tb$, while 
$\tb$ bounds are not much affected. On the other hand, the sbottom mass 
constraints put
further upper limits on $\tb \lsim 27-28$, in consistency with the previous 
analysis in section 5.

These results are also compared with a top-down MINUIT minimization in Table
\ref{minuit_tbh},
using as input data $m_h$ in addition to the gluino cascade sparticle masses 
(but not yet using the sbottom masses at this stage). One observes that 
the MIGRAD symmetric
error on $\tb$ appears to be very good, but comparing with the contours
in Figs.~\ref{figtbxt1}, \ref{figtbxt2}  one can suspect
that it is essentially influenced by the {\em lowest} limit on $\tb$,
and that the actual errors are particularly unsymmetrical due to the flat
behaviour for large $\tb$.  
This seems 
confirmed by the fact that the unsymmetrical MINOS positive error is not 
calculated by MINUIT.    
The lower $\tb$ bound, on the other hand, is very consistent with our alternative
finding in Eq.~(\ref{tbfromh}). This behaviour is well explained by the
flattest dependence for large $\tb$, as clearly illustrated by the plots
in Figs.~\ref{figtbxt1}, \ref{figtbxt2}. 
For completeness we also show in Fig.~\ref{tbA0} 
the corresponding one- and two-$\sigma$ domains but in the $(\tb,A_0)$ plane,
as well as the domain obtained once adding sbottom mass measurements. 
One can see that the sbottom
mass bound have some impact on $A_0$ bounds, at least for large $-A_0$. 
These results may be compared with the $A_0$ limits obtained 
from the MINUIT fit in Table \ref{minuit_tbh}. 
        \begin{figure}[h!]
  \begin{center}
  \includegraphics[width=8cm,angle=-90]{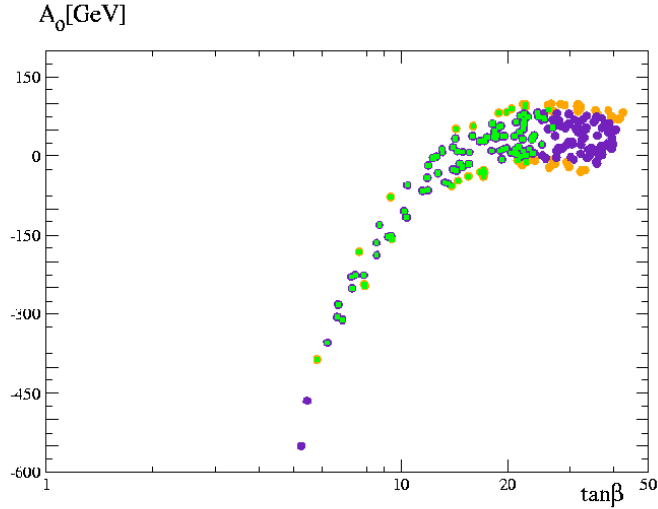}
\vspace{-1cm}
\caption{\label{tbA0} Contours from a Gaussian scan in the plane $(\tb, A_0)$: 
indigo: one-$\sigma$ ($68\%$ C.L.); 
orange: two-$\sigma$ ($95\%$ C.L.); green: 
remaining points (at one-$\sigma$) when adding sbottom mass measurements.}
\end{center}
       \end{figure}

Thus the $m_h$ measurement alone gives a mild upper bound on $\tb$ in
the case where only some knowledge on $m_A(Q_{EWSB})$ is assumed (through
sfermion-Higgs sector universality),  
but not on $X_t$. The improvement on $\tb$ limits 
is mainly coming from adding the sbottom mass measurements. 
However this mild sensitivity, as well as the strong $\tb-X_t$ correlation, is essentially determined
by the variation of $A_0$ which is assumed unknown at this stage. 
The uncertainties on $\mu$ (that are reasonable in this constrained MSSM case) 
have little influence (except perhaps for very small $\tb$),
since it is suppressed as $\mu/\tb$ within
$X_t$.  We shall see in the summarizing section \ref{summary} 
that overall, the bounds are a bit tighter
once combining all available informations, except those for $A_0$ which 
depend quite much on the level of approximation used
in RGE and radiative corrections.  
If an additional independent information on $A_0$ would be available 
(e.g. if the stop masses could be measured) the constraints in the $\tb, X_t$
plane could be evidently improved. Better constraints 
on $\tb$ are also 
prospected\cite{LHCstudy} clearly if some of the heavy Higgses $A, H, H^\pm$ 
could be measured,
that we do not assume in our `minimal' set of input mass measurements and
within the SPS1a benchmark case. (It would be rather straightforward, however, 
to extend our
above analysis in case such additional information would be available.)        

We finally mention that our results are not drastically changing once using
the more complete two-loop level radiative corrections for $m_h$
(assuming in such case universality/mSUGRA relations to calculate 
all higher order corrections), except 
for the bounds on $A_0$ that are substantially worsened   
in this case, as will be confirmed by further MINUIT fit results 
summarized in section \ref{summary}.
This comes from an observed flattest dependence of $m_h(A_0)$
for large $-A_0$ at the two-loop level, in the range considered for the other relevant parameters. 
As far as we can see, this may be explained by the fact that within the 
two-loop level radiative corrections to $m_h$\cite{mh1-2loop,bdsz}, there
is somewhat much room for possible cancellations of the $A_0$ dependence
entering in higher orders, most notably in all squark and stau
contributions both at the one- and two-loop level.   
This is another example of the caution needed to interpret results when performing 
minimization constraints using different level of theoretical approximations.   
\subsection{\label{summary} Combining all information and 
comparison with standard fits}
\begin{table}[h!]
\begin{center}
\caption[long]{\label{tabcomb} Combined best constraints from
all gluino cascade decay sparticles and lightest Higgs mass measurements
for the bottom-up approach in different MSSM scenarios. (NB a star in superscript
indicates discrete ambiguities in the reconstruction in the non-universal  
MSSM case, as discussed e.g. in sub-sec. 3.1 and Appendix A).} 
\begin{tabular}{|c||c|c||c|}
\hline\hline
Model or assumptions &Parameter & Constraint (GeV)& nominal SPS1a value \\
                     &          &                 & (from SuSpect 2.41)\\
\hline
general MSSM & $M_1(Q_{EWSB})^\star$ & $\sim$95--115   &  101.5 \\
" "          & $M_2(Q_{EWSB})^\star$ & $\sim$175--225  & 191.6 \\
" "          & $M_3(Q_{EWSB})$ &  $\sim$580--595 &  586.6\\
" "          & $(\frac{3}{8} m^2_{u_L}
+\frac{m^2_{e_R}}{4})^{1/2}(Q_{GUT})$ & $\sim$68--89 & 
$(\frac{5}{8})^{1/2} 100 \sim 79$  \\
" "          & $m_{Q3_L}(Q_{EWSB})^\star$ & $\sim$488--518     & 497     \\
" "          & $m_{b_R}(Q_{EWSB})^\star$  & $\sim$510--540     & 522          \\
" "          & $\mu(Q_{EWSB})^\star$ & $\sim$280--750    & 357  \\
" "          & $\tb(Q_{EWSB})$ & $\sim$1--36 (th. bounds)   &  9.74  \\
+ $m_{\t N_4}$  & $\mu(Q_{EWSB})^\star$ & $\sim$350--372    & 357  \\
" "          & $\tb(Q_{EWSB})$ & $\sim$2.7--36 (th. bounds)   &  9.74  \\
\hline
$\t q,\t l$-universality & $m^{q,l}_0(Q_{GUT})$ & $\sim$90--112  &  100  \\
" "          & $m_{Q3_L}(Q_{EWSB})$ & $\sim$490--506     & 497     \\
" "          & $m_{b_R}(Q_{EWSB})$  & $\sim$513--530     & 522          \\
$M_i, i=1,..3$-universality     & $M_i(Q_{GUT})$ & $\sim$ 245--255      & 250   \\  
  \hline
$\t b_1,\t b_2$ +universality & $\tb(Q_{EWSB})$ & $\sim$3--28  &  100  \\
  \hline
mSUGRA  & $m_0$ & $\sim$90-112  &  100 \\
        & $m_{1/2}$  & $\sim$245--255 & 250 \\
	&  $-X_t$ &  $\sim$450--730 & 530\\
	&  $-A_0$ & $\sim$ -100-350     &    100 \\
	&  $\tb(m_Z)$ & $\sim$ 5.5--28 &     10 \\
\hline\hline
\end{tabular}
\end{center}
\end{table}
We finally summarize and combine all previous constraints from the
different sectors, both in general MSSM or with universality
relations, in Table \ref{tabcomb}. By ``combined best constraints'' we
simply mean to evaluate the joint limits by crudely 
superposing the constraints obtained in the three different
sectors of gaugino/Higgsino, squarks/sleptons and Higgs, which eventually
results in slight improvements on some parameter limits. We did not attempt 
to perform a more elaborated statistical combination of the three
sector constraints within our present approach. 
However, these results are compared with those obtained
from a top-down $\chi^2$ fit of basic mSUGRA parameters to the same
mass measurements in Table \ref{tabfit}. 
As one can see, the results are overall qualitatively consistent, except perhaps for
the $A_0$ limits. The fact that $A_0$ bounds (and to some extent $\tb$
bounds as well) are worsen when
fitting parameters in mSUGRA at the full two-loop level instead of one-loop, 
was explained previously due to the possible cancellations of the $A_0$
dependence in higher order $m_h$ contributions. Note that for $\tb$, the MIGRAD
symmetric minimization error found here appears very optimistic as compared 
to the upper bound derived from the bottom-up reconstruction, $\tb\lsim 28$.
As already mentioned above it is possibly much influenced by 
the {\em lower } bound that results
from the combination with the Higgs sector measurements. 
The MINOS upper bound is however more consistent with the bottom-up result.   
\begin{table}[h!]
\begin{center}
\caption{\label{tabfit} Combined constraints on mSUGRA basic parameters
from a standard top-down $\chi^2$ fit with MINUIT of 
all gluino cascade decay sparticle masses plus $m_h$ measurements.} 
\begin{tabular}{|c||c|c||c|}
\hline\hline
Model and assumptions &Parameter & $68\%$ C.L. limits (GeV)& SPS1a value \\
\hline
mSUGRA  & $m_0$ & 99.96 $\pm$ 11.2  &  100 \\
2-loop RGE + full $\t q$ R.C. + 2-loop $m_h$ 
  &       & (99.95 $\pm$ 11.7) &        \\
(1-loop RGE+ no $\t q$ R.C. +simple $m_h$ R.C.)        
& $m_{1/2}$  & 250.0 $\pm$ 3.7 & 250 \\
 &            & (249.5 $\pm$ 4.7) &    \\
 &  $A_0$ &   -104.2 $\pm 379$ & -100\\
 &        &   (-100.6 $\pm 136$) &     \\
 &  $\tb(m_Z)$ & 9.9 $^{+9.4}_{-4.7}$ &     10 \\
        &        &  (9.96 $\pm 4.11$)  &      \\
\hline\hline
\end{tabular}
\end{center}
\end{table}
%
\newpage
\section{Conclusion}
We have examined some specific bottom-up
reconstruction strategies at the LHC, both for general and 
universality-constrained MSSM parameters, 
starting from a plausible set of incomplete measurements of a few
MSSM sparticles. Using sparticle mass measurements  
mainly from cascade decays of gluino and squarks,
and the lightest Higgs boson mass, we have constructed different 
algorithms, based on rather simple semi-analytical inverse relations
between the MSSM basic parameters and mass spectrum, incorporating
radiative corrections, when known, at a realistic level.   
We have determined constraints on the relevant basic MSSM parameters from
the expected mass accuracies, under different theoretical assumptions on 
the degree of universality of some of the parameters. We have also exhibited
analytically the possible discrete ambiguities in the reconstruction of some 
of the basic parameters, when using only mass input in a general unconstrained 
MSSM without extra knowledge or assumptions 
on the relative hierarchy of the relevant parameters. 
This is the case in particular for the gaugino/Higgsino sector parameters.      
These constraints have been also compared at different stages in a 
sector-by-sector analysis, with those obtained from more conventional 
top-down approaches of fits to data with minimization procedures. 
The results are overall consistent, which is an a posteriori check that
our rather naive semi-analytic approach, with many approximations, 
does essentially capture the sensitivity on parameters. \\

Regarding the SPS1a reconstruction example studied here, 
more quantitatively we have
shown that a rather limited data set, consisting of merely the measurement
of sparticle
masses involved in gluino cascade decay to a few percent accuracy,
 may still provide reasonably good constraints on some of the
relevant MSSM parameters. This is in particular the case  for the gaugino mass
parameters and the squark and slepton soft mass terms, even for an 
unconstrained MSSM without universality assumptions. 
If a precise measurement of the lightest Higgs mass is available, additional  
constraints (though moderate) on $\tb, X_t\equiv A_t-\mu/\tb$ are obtained,  
but only if Higgs-sfermion mass term universality is assumed. 
Interesting constraints for a non-universal MSSM are
however more challenging to obtain in general 
for the Higgs sector parameters (as well as
for $\tb$), unless more precise measurements would be available
(or data from another sector, like heavy Higgses and/or scalar 
top masses typically), 
as is also known from other analysis\cite{LHCstudy}. We stress again that 
considering only the gluino/squark cascade (plus the lightest
Higgs) sparticle identification in our analysis is not motivated by a strong
prejudice against other potential SUSY-discovery processes at the LHC.    
Indeed the cascade in Eq.~(\ref{casc})
may be considered already quite specific from a general MSSM viewpoint, 
but it gives us a well-defined and rather minimal input set for 
testing our approach, 
and comparing it with other analyses for the very much studied SPS1a 
benchmark. Even if our results are probably not  
very new to the experts, they illustrate that a step-by-step semi-analytic
approach can help to
exhibit better the sensitivity of basic MSSM parameters with respect to given 
sparticle mass or other data, which may be more difficult to grasp from 
global top-down fits. In many cases, the non-linear and non-symmetric 
behaviour of error propagations is exhibited by our analytic bottom-up approach 
in a more explicit way (as illustrated typically for $\tb$, comparing
Tables \ref{tabcomb} and \ref{tabfit}), and similarly for the possible 
discrete ambiguities e.g. in the gaugino/Higgsino sector.  
As compared to other recent analysis of MSSM constraints at LHC 
(e.g. \cite{LHCstudy,dzerwas,mssmbayes,ben_bayes}), our results 
are difficult to compare quantitatively in very detail, since the data used are often different (with generally more input sparticle masses 
assumed in most other analysis). 
We find however a rough consistency on the expected sensitivity of
the basic MSSM parameters, as discussed above.  \\
  
The relatively simple algorithms described here
are rather flexible, and may be easily
interfaced with more elaborated simulation tools, that would allow
in particular a more realistic statistical treatment of the different
sources of experimental and theoretical uncertainties. Moreover, some of our
analytical relations may be used at least 
as ``Bayesian priors'' guideline to other analysis, 
in a way complementary to the Markov chain techniques. 
We plan indeed a more refined statistical analysis in the future\cite{prepa},
by possibly combining our approach with Bayesian and Markov chain techniques 
of ref. \cite{mssmbayes,ben_bayes,dzerwas}. 

Our approach may thus provide a useful 
complementarity to more elaborated simulations, as well as possible
cross-checks. 
Moreover it is not at all restricted to the 
LHC phenomenology: some of the algorithms described here, for instance
in the neutralino and squark/slepton sectors, may be
readily used for ILC data upon straightforward changes in sparticle 
mass accuracies. 
However, a similar approach for the ILC, following e.g. the studies made   
in refs. \cite{zerwasetal,otherinv},  
deserves specific analysis beyond the scope of the present paper, due to the 
different sparticle spectrum expected to be reached at the ILC, which will imply
slightly different inversion algorithms.  

\vspace{1cm}

{\bf Acknowledgements}\\ 
This work is partially supported by ANR contract ``PHYS@COL$\&$COS'' 
and GDR 2305 ``Supersym\'etrie''. 
We are grateful to Gilbert Moultaka for stimulating
conversations at a preliminary stage of this work. We also thank 
Dirk Zerwas, 
Ulrich Ellwanger and Sabine Kraml for interesting comments or discussions.   

\newpage
\appendix 
\section{Solving the gaugino/Higgsino parameters for different input}
In this appendix, we give for completeness explicit analytic solutions relating
neutralino masses and gaugino/higgsino parameters, 
depending on different input and output choice and depending on general MSSM or
more constrained universality assumptions, relevant to sections 3.1-3.4. 
\subsection{Two neutralino mass input}
From Eqs.~(\ref{b4}), (\ref{b5}), one can solve $M_1, M_2$ for input
masses $m_{\t N_1}, m_{\t N_2}$, plus $\mu$, $\tb$ input:
after expressing $M_1$ e.g. from Eq.~(\ref{b4}) (see also \cite{inv1}): 
\be 
M_1= -\frac{P_{12} \left[\mu^2 +m^2_Z +P_{12} +(M_2-S_{12}) S_{12}\right] 
+M_2\,\mu\, m^2_Z 
s^2_W \sin 2\beta} 
      {S_{12} P_{12} -M_2(\mu^2 + P_{12}) + c^2_W m^2_Z\mu \sin 2\beta}\;,
\label{M1sol}
\ee
(with $S_{12}\equiv m_{\t N_1}+m_{\t N_2}$, $P_{12}\equiv m_{\t N_1}m_{\t N_2}$), 
$M_2$ is given by a standard second order equation with solution:

\be
M_2 =\frac{-b_1 \pm\sqrt{b^2_1-4 a_1 c_1}}{2a_1}
\label{M2sol}
\ee
where $+\sqrt{\dots}$ corresponds to $M_1 < M_2$ and $-\sqrt{\cdots}$ to $M_2 <M_1$ respectively, and 
\beq 
      a_1 & = & P_{12}\left[\mu^4 + P_{12} (P_{12} + m^2_Z s^2_W)  
      +\mu^2 (2 P_{12} - S_{12}^2 + 
       m^2_Z s^2_W) + \mu\, m^2_Z s^2_W \sin 2\beta\, S_{12}\right] \nn \\
      b_1 &= & P_{12}\left[S_{12}(-\mu^4 + \mu^2 (-m^2_Z - 2 P_{12} + S_{12}^2)  
       -P_{12}(P_{12} + m^2_Z (s^2_W-c^2_W)) \right.\nn \\ 
       & & \left. -\mu\, m^2_Z (P_{12} + 
      (S_{12}^2-P_{12}) (s^2_W-c^2_W) + 2 \mu^2 c^2_W) \sin 2\beta\right] \nn \\
      c_1 &= & P_{12}(P_{12}(\mu^2 + m^2_Z + 
      P_{12})(\mu^2 + c^2_W m^2_Z + P_{12}) - 
       (\mu^2 + c^2_W m^2_Z) P_{12} S_{12}^2 \nn \\
       & &+ \mu\, m^2_Z \,\sin 2\beta 
       (S_{12} (P_{12} + c^2_W(\mu^2 + m^2_Z + 2P_{12} - S_{12}^2)) 
       +c^2_W m^2_Z\, \mu\,\sin 2\beta))\;.
\label{abcm1m2}
\eeq

The occurrence of two $M_2$ solutions in Eq.~(\ref{M2sol})
is due to the structure of Eqs.~(\ref{b4}), (\ref{b5}), but this ambiguity
cannot be resolved from the sole knowledge of two neutralino mass input, since
all the available relations from Eqs.~(\ref{4inv}) are completely symmetrical 
under $m_{\t N_1} \leftrightarrow m_{\t N_2}$ permutations.
Typically in the simple limit of neglecting all $D$-terms i.e. $m_Z\to 0$ in 
the neutralino mass matrix 
Eq.~(\ref{mino}), the four neutralino mass eigenvalues are trivially given as 
\be
m_{\t N_i} (i=1,..,4) = M_1, M_2, |\mu|, |\mu|\;. 
\ee 
with all possible permutations i.e no ordering implied. 
The exact mass eigenvalues with $m_Z\ne 0$
are no longer invariant under e.g. $M_1 \leftrightarrow M_2$ permutations, but 
concerning the inverted relations, even if all physical
masses were known and thus ordered according to e.g $m_{\t N_1} < \cdots < m_{\t N_4}$,  
there is no way to determine from the sole knowledge of a
given mass $m_{\t N_i}$ if it corresponds to a Bino, Wino or Higgsino.
This needs information on diagonalization matrix elements, that
can be partly accessed if measuring the couplings of neutralinos to other
particles.
  (However, for known $\mu$ input, measuring a third neutralino mass could in 
 principle solve the $M_1, M_2$ ambiguity, since the two different solutions 
gives different $m_{\t N_3}, 
m_{\t N_4}$ (which are most simply obtained by solving
Eq.~(\ref{mu3N}) as a second order equation for e.g. $m_{\t N_4}$). But in
practice this needs an accurate knowledge of both $\mu$ and 
the three neutralino masses).

In Fig.~\ref{figm1m2ex} the two $M_1, M_2$ solutions are illustrated as functions of $\mu$ (see respectively $M_2(1,2)$ and $M_1(1,2)$), with
the two neutralino mass input and $\tb$ fixed to SPS1a values for reference. 
Note that in Eq.~(\ref{M2sol}) there can be values of $\mu$, $\tb$ for which
$a_1 \to 0$: for example for SPS1a values of $\tb$, 
$a_1(\mu,\tb)$ has four zeros for real values of $\mu$: $\mu \sim
-165, -110, 118, 157$ GeV. However these poles with respect to $\mu$ are artifacts 
since for $a_1=0$ strictly, the (unique) solution reduces to $M_2 = -c_1/b_1$
(moreover the possible zeros 
of $b_1(\mu)$ do not coincide with those of $a_1(\mu)$). 
Nevertheless sufficiently close to these $\mu$ ``pole'' values, one of the two
$M_2$ solutions 
 can become arbitrarily large, as can be seen on 
Fig.~\ref{figm1m2ex}~\footnote{In fact only two of the $a_1(\mu)$ zeroes 
lead to this pole behaviour of $M_2(\mu)$, as seen on Fig.~\ref{figm1m2ex}
(e.g. at $\mu \sim -165$ and 
$ 157$ GeV for $M_2 > M_1$ and the two other zeros for $M_2 < M_1$), 
whereas the other zeros of $a_1$ give 
$b_1(\mu) >0$ such that the correct behaviour
 $M_2 \to -c_1/b_1$ for $a_1\to 0$ is recovered.}.  
More physically, this means that such large values of $M_2$, with 
their corresponding $\mu$ pole values, are still consistent with 
the same neutralino masses $m_{\t N_1}, m_{\t N_2}$. 
      
As such, Eq.~(\ref{M2sol}) is not very illuminating: it may be useful 
to expand to first order in the $D$-term,
i.e. to ${\cal O}(m^2_Z)$, which gives a very simple expression:
\be
      M_2= m_{\t N_2} +c^2_W m^2_Z 
      \frac{(m_{\t N_2}+\mu \sin 2\beta)}{\mu^2-m^2_{\t N_2}}
\label{M2appr}
\ee
for the case $M_1 < M_2$ (and the same expression with $m_{\t N_2}\to m_{\t N_1}$
for the case $M_2 < M_1$). (For $M_1$ one better keeps the exact and
relatively simple expression 
Eq.~(\ref{M1sol})).    
Here the pole at $\mu= m_{\t N_i}$ in Eq.~(\ref{M2appr}) is clearly an 
 artifact of first order in $m^2_Z$ expansion, and simply reflects that
 this approximation is not valid for $\mu \sim m_{\t N_i}$.  
 Eq.~(\ref{M2appr}) is certainly a good approximation 
 for $\mu \gg m_{\t N_1}, m_{\t N_2}$
 and $m_{\t N_i} \gg m_Z$. Moreover it can still be
 a rather good approximation even for moderate $\mu$ and neutralino mass values. 
For instance for SPS1a values in Table \ref{susoutput}, and neglecting
self-energy radiative corrections to neutralino masses 
one has $\mu\sim 357$ GeV,
$m_{\t N_2}\sim 176.2$ GeV and  
$M^{exact}_2(SPS1a)
\sim 191.6$ while  $M^{approx}_2(SPS1a)\sim 192.6$ which is only $.5\%$ difference,
i.e. less than the magnitude of radiative corrections to neutralino masses, 
which are about $\sim 4$ GeV for $m_{\t N_2}$. 
We have used however the exact expressions (\ref{M2sol}) throughout our 
analysis in section 3.
%
\begin{figure}[h!]
\begin{center}
\epsfig{figure=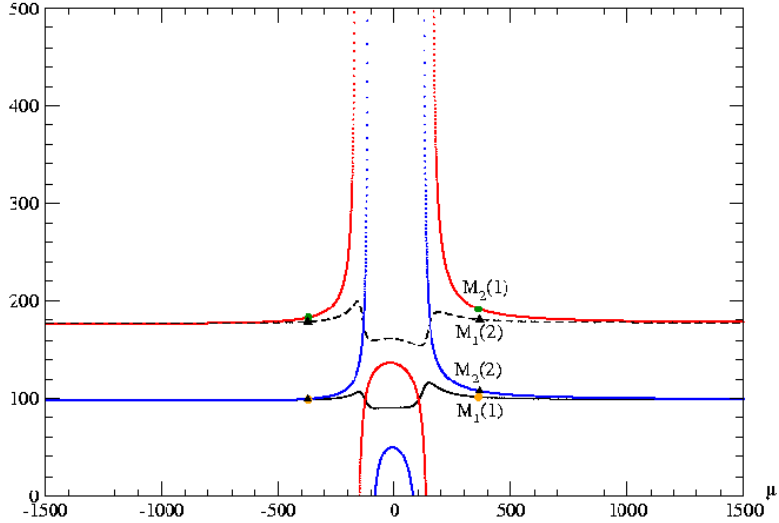,angle=-90,width=12cm}
\caption{The two $M_1, M_2$ solutions (respectively $M_2(1,2)$ and $M_1(1,2)$) 
from Eqs.~(\ref{M1sol}), (\ref{M2sol}) for two neutralino mass $m_{\t N_1}$, $m_{\t N_2}$  input, as functions of $\mu$. The dots and triangles corresponds to the solutions
once a third neutralino mass $m_{\t N_4}$) input is used, completely determining $|\mu|$
(for fixed $\tb$). The neutralino mass input values
and  $\tb$ are fixed to their SPS1a values for illustration.} 
\label{figm1m2ex}
  \end{center}
 \end{figure}
\subsection{Three neutralino mass input: general case}
Knowing three neutralino masses allow to use one more equation from
(\ref{4inv}) to solve e.g. for $M_1$, $M_2$ and $\mu$ in terms of the
three neutralino masses and $\tb$ input values. 
(Alternatively, we could also solve this
system to determine rather $\tb$ for fixed $\mu$, but since it  
only depends on $\sin 2\beta$, any $\tb$ dependence drops out for large $\tb$.
Thus we can anticipate without calculations that there
can be no interesting $\tb$ upper bounds, irrespectively of the number of
neutralino masses measured, given their expected LHC accuracies).       
The extra relation to determine $\mu^2$ is Eq.~(\ref{mu3N}) obtained 
from the trace and second invariant in Eq.~(\ref{4inv}). Note however that
the sign of $\mu$ is still not determined. 
Combining Eq.~(\ref{mu3N}) with Eqs.~(\ref{b4}), (\ref{b5})
gives a high (sixth) order polynomial equation in $\mu$ (or equivalently
in $M_1$ or $M_2$) which cannot be solved fully analytically. It is however
easy to solve iteratively using e.g. Eq.~(\ref{mu3N}) on the solutions 
(\ref{M2sol}), (\ref{M1sol}) (upon having chosen a definite order for
the relative values of $M_1$ and $M_2$). This iterative solution converges
very quickly. The solutions for $M_1$, $M_2$ and $\mu$ obtained for $\tb$ input
fixed to SPS1a value are illustrated by the different dots and triangles in
Fig.~\ref{figm1m2ex}.  

Alternatively it is useful to consider, as in the case
of two neutralino masses above, approximate solutions obtained by expanding to
first order in $m^2_Z$. This gives very simple expressions:    
\beq     
       M_1 & = & m_{\t N_1} +s^2_W m^2_Z\frac{(m_{\t N_1}+  
       m_{\t N_3}\sin 2\beta)}{m^2_{\t N_3}-m^2_{\t N_1}} \nn \\       
       M_2 &= & m_{\t N_2} +c^2_W m^2_Z\frac{(m_{\t N_2}+  
       m_{\t N_3}\sin 2\beta)}{m^2_{\t N_3}-m^2_{\t N_2}}  \nn \\
      |\mu| &= & m_{\t N_3} -\frac{m^2_Z}{2} (1+\sin 2\beta)\:
      \frac{[m_{\t N_3}-m_{\t N_2}+c^2_W (m_{\t N_2}-m_{\t N_1})]}
       {(m_{\t N_3}-m_{\t N_1})(m_{\t N_3}-m_{\t N_2})}  
       \label{3Napprox}
\eeq 
corresponding to the first case $M_1 < M_2$ (i.e. a Bino LSP), and similar expressions
with $m_{\t N_1}\leftrightarrow m_{\t N_2}$ for the case $M_2 <M_1$. Actually,
considering three neutralino mass input and $M_1, M_2, \mu$ output adds 
more discrete ambiguities than in the two neutralino mass case: accordingly
in a most general MSSM, without further knowledge on e.g. 
neutralino couplings, or further theoretical
assumptions, one should in principle consider
all possible ordering among $M_1$, $M_2$, $\mu$ values, i.e.
six possible cases (where the LSP 
mass $m_{\t N_1}$ can be either Bino, Wino or Higgsino). It is 
straightforward to derive
such other solutions by appropriate permutations of the three $m_{\t N_i}$,
within Eqs.~(\ref{3Napprox}).   
For all cases Eqs.~(\ref{3Napprox}) are very good approximations, at least 
as far as $m_{\t N_i}$ are not
small compared to $m_Z$, e.g. for the SPS1a case typically
the differences with exact results is of order or below the percent level. 
\subsection{gaugino mass universality: two or three neutralino mass input}
Assuming gaugino mass universality one can solve again Eqs.~(\ref{b4}), (\ref{b5}) but changing
input/output: now the EWSB scale values of $M_1$ and $M_2$ are given
from $M_3$ by Eq.~(\ref{guniv}), so that one can determine $|\mu|$ and $\tb$ assuming two neutralino mass input from
a linear system, which gives
\be     
      \mu^2  = -\frac{b_2}{a_2}
\label{mu2N}
\ee     
with
\beq
      a_2 & = & P_{12}\,( M_1 + M_2 - S_{12} + \frac{P_{12}-M_1 M_2}
      {M_{12}}\, )   \nn \\
      b_2 &= & P_{12}\,\left[ P_{12}\,(M_1 + M_2 -S_{12}) + 
     S_{12} \left( (M_1 - m_{\t N_1}) (M_2 - m_{\t N_1}) - (M_1 + M_2 - m_{\t N_1})
     m_{\t N_2} \right. \right. \nn \\ & & \left. \left. 
     +m^2_{\t N_2}- m^2_Z\right)      
      + m^2_Z \left(M_{12} + 
        \frac{P_{12} \left(P_{12} + (M_1 -S_{12})(S_{12}-M_2) +m^2_Z\right)}
         {m^2_Z M_{12}}     \right) \right]
\label{a2b22N}
\eeq
and 
\beq      
      \sin 2\beta &=& -\frac{\left[ P^2_{12} - M_1 M_2 \mu^2 + 
         P_{12}\, ( (M_1 +M_2)S_{12} -M_1 M_2 -S^2_{12} 
          +\mu^2+ 
         m^2_Z )\right]}{\mu\, m^2_Z M_{12}} 
\label{tb2N}
\eeq
with $M_{12} \equiv c^2_W M_1 + s^2_W M_2$ and other notations as in Eqs.~(\ref{M1sol})-(\ref{abcm1m2}). Note that we assume $\mu$ to be real.  
The case of three neutralino mass input is treated by combining 
Eq.~(\ref{mu3N})-Eq.~(\ref{tb2N}).  
\section{Bottom-up renormalization group evolution in SuSpect code}
In this section we illustrate in some details an important ingredient
of our bottom-up reconstruction procedure which is the 
renormalization
group evolution (RGE)\cite{rge,RGE2} of MSSM parameters from low to high energy. 
We take this opportunity to present the results from an  
available option of the SuSpect code, illustrating some 
properties of this bottom-up RG evolution which are 
quite general and independent of the present analysis and
LHC cascade decay phenomenology. (This option has recently been adapted to
the more suitable ``Les Houches accord'' input file conventions\cite{slha}.)\\
The fact that the RGE for the MSSM parameters are ``invertible", i.e
can be evolved from high to low scale 
and backward, is a rather obvious feature
of any such coupled differential equations. However, what makes it less 
straightforward in the standard approach to MSSM calculation with RGE, is that
there are actually three (at least) different energy scales in the game, with
corresponding boundary conditions: 
\begin{itemize}
\item the high (generally GUT) scale,
at which one defines e.g. the mSUGRA parameters with eventually
some universality relations;     
\item the (low) electroweak symmetry breaking scale, at which the
soft breaking and other relevant MSSM parameters are evolved to in a
standard top-down evolution;
\item finally the scale $m_Z$, or eventually some other low energy 
scale, which enter as other boundary conditions e.g; via the precise
measurements of the gauge couplings. 
\end{itemize}
The interplay between these different scales in MSSM
spectrum calculation codes (such as SuSpect \cite{suspect} and similar 
codes\cite{isasugra,softsusy,spheno})
needs, among other things, iterative algorithms for the evolution
between the three different
scale, with a consistent implementation of any possibly known radiative
corrections to the parameters (gauge couplings, top, bottom masses, etc)
entering as boundary conditions to the RG evolution. (We refer to the 
SuSpect manual \cite{suspect} for details on such RGE algorithms.) 

Such a bottom-up evolution option has been
available in a beta-version for some time in SuSpect but was not 
publicly available
nor illustrated until now.  
In Table \ref{rgebup} we illustrate for the SPS1a benchmark point
this bottom-up RG evolution of all parameters. We show in particular some
important features on the error propagation in such a procedure, and
which parameters are more sensitive to this dispersion.   
The input parameters (in the second column) 
were obtained in a first stage from a standard (top-down) run from the
mSUGRA SPS1a input parameters in Eq.~(\ref{defsps1a}) 
(with $m_{top}=175$ GeV). The third column gives 
the corresponding output values resulting from 
a RG evolution up to a GUT scale 
from a bottom-up SuSpect run under the most general MSSM assumptions,
i.e. without any a priori on possible universality relations at the
GUT scale. One can already notice from this that the agreement with the
original mSUGRA parameter values is excellent: 
the discrepancies are of order ${\cal O}(10^{-3})$ that are
in fact consistent with the accuracy chosen
(i.e. the intrinsic error of the numerical evolution of the RGE as
performed with a Runge-Kutta algorithm, as well as the intrinsic error due
to necessary iterative procedures\cite{suspect}).
We point out however that the results shown here correspond
to the choice of $m_{H_u}$, $m_{H_d}$
input, while the correspond results for $m_A$, $\mu$ input choice are a little
bit worse for $m_{H_u}(Q_{GUT})$ and $m_{H_d}(Q_{GUT})$, which is attributed to
a certain precision loss in our algorithm
when passing from one to the other input set, that involves
calculations and iterations including two-loop radiative corrections.
(This is in particular due to the well-known $Tr [Y M^2]$ terms
present in the RGE\cite{rge,RGE2}, which involve a combination of soft
scalar terms, including $m^2_{H_u}$ and $m^2_{H_d}$: this combination vanishes
by definition in mSUGRA, and remains zero at all scales
in a top-down RG evolution. However, when 
using bottom-up RGE the approximate 
$m^2_{H_u}, m^2_{H_d}$ values as obtained from $m_A, \mu$ 
from EWSB conditions Eqs.~(\ref{eq:ewsb}) does not satisfy $Tr [Y M^2]=0$
exactly, and this induces a not completely negligible departure
when evolving from low to GUT scales.)    

Next, the fourth
and subsequent columns give the deviations in the output parameter values
corresponding to 
$\pm 1\%$ deviations of some relevant parameters input values:
$M_3$, $m_{H_u}$, $m_{Q3_L}$, that we chose
on purpose as they give the most important sensitivity to other
parameters deviations. One can see
that the deviations induced in other parameters remain generally 
reasonable, at the percent level, notably 
for the gaugino masses, and most of the squark and sleptons (except for
the one sfermion mass which is varied in each case). In contrast, the deviations
induced on the scalar mass $m_{H_u}$ can be huge, even for a moderate
percent deviation, i.e. there is a large amplification or dispersion
of error. This is of course explained
from the detailed RGE dependence of $m_{H_u}$ on other soft terms, resulting in a very strong sensitivity
on certain other parameters. The same is true to some
extent for $m_{t_R}$. The strong sensitivity  of 
$m_{H_u}$ e.g. on the top mass value and other parameters through its RGE
is a well-know feature, but more precisely our illustration here
indicate that it will be very difficult to have a precise determination
of $m_{H_u}$ at a high GUT scale, even in the very optimistic case where 
all other MSSM parameters would be know quite precisely. Consequently it 
will be very difficult to check for eventual universality of the
soft scalar mass terms with the squark and slepton soft terms. 
One can however turn this argument the other way round, and deduce from this illustration that,  while $m_{H_u}$
plays a crucial role in the (radiative) electroweak symmetry breaking,
the sparticle masses at low scale determining most of the collider 
phenomenology are not very much dependent on its precise
value at GUT scale. (This is somewhat similar here to the ``focus point"
properties observed for large $m_0$ values in other mSUGRA parameter
choices). More generally, many features as those illustrated in Table 
\ref{rgebup} may be important to keep in mind for any realistic 
bottom-up procedure. The above bottom-up RGE procedure had been used
in several stages of our analysis, as indicated in the main text.             
\begin{table}[h!]
\begin{center}
\caption{\label{rgebup} Bottom-up RG evolution of SPS1a parameters from
SuSpect 2.41 with 
illustration of error propagations. Input parameters (2nd column) 
were obtained from a standard (top-down) SuSpect run from SPS1a input parameters
in Eq.~(\ref{defsps1a}) 
(with $m_{top}=175$ GeV). The third column gives corresponding output values 
once evolved back to a GUT scale 
from a bottom-up SuSpect run with general MSSM option. Fourth
and subsequent columns give deviations in the output parameter values
corresponding to 
$\pm 1\%$ deviations of some relevant parameters input values as indicated.
(Variation range is given explicitly when non-symmetrical)} 
\vspace{.2cm}
\begin{tabular}{|c||c||c||c|c|c|}
\hline
par. & input(GeV) & GUT output  & $\Delta M_3=\mp 1\%$ & $\Delta m_{H_u}=
\mp 1\%$ & $ \Delta m_{Q3_L}=\mp 1\%$ \\ \hline\hline
$Q_{EWSB}$&  465.5 & $\simeq 2.47\: 10^{16}$ &  0.1 $\%$ &   0.1$\%$ & $0.1 \%$     \\ 
\hline
 $M_1$ &      101.5    & 250.004  & negl. & negl.   & negl.       \\
 $M_2$ &      191.6    &  249.998 & " "   & " "  & " "     \\
 $M_3$ &     586.6   &   249.999  & $\pm 2.2$ & " " &  " " \\
\hline
$m^2_{H_d}$ &$(179.9)^2$ & $(100.004)^2$ & $(100.6)^2$--&$(100.7)^2$--&
$(101.2)^2$--\\
            &           &   &  $(99.4)^2$ & $(99.2)^2$ & $(98.7)^2$\\
\hline
$m^2_{H_u}$ & $-(358.1)^2$ & $(100.017)^2$ & $(132.6)^2$--& $(64.9)^2$-- &
$(63.7)^2$--\\
            &           &   & $(48.4)^2$ &  $(124.4)^2$& $(126.4)^2$\\
\hline
($m^{pole}_A$) &   398.8  &           &          &      &             \\ 
($\mu$)        &   356.9  &   353     &          &      &              \\
\hline
$m_{e_L}$ &     195.5  & 100.004       & 100.2--99.8&  100.8--99.2& 101.5--98.5 \\
$m_{\tau_L}$&      194.7 &  100.004    & 100.2--99.8&  100.8--99.2& 101.5--98.5\\
$m_{e_R}$ &    136   &  99.998          & 100--99.9&   98.4--101.6 &96.8--103.1\\
$m_{\tau_R}$&      133.5 &  99.998    & 100--99.9&   98.4--101.6 & 96.8--103.1  \\
 $m_{Q1_L}$&     545.8   & 100.001     & 121--72&  99.7--100.3&  99.1--100.8  \\
$m_{Q3_L}$&      497  &  100.005        &131--52 & 94.6--104.6&  55.2--130.4    \\
$m_{u_R}$&     527.8   & 99.997       &121--72 &  101--99 &  101.8--98.1\\
$m_{t_R}$&         421.5  &   100.006   & 140--14& 90.6--107.5&  81.9--115.3   \\
$m_{d_R}$&    525.7   &  99.997         & 121--72& 99.4--100.6& 98.7--101.3    \\
$m_{b_R}$&        522.4 &  99.997      & 122--72&  99.4--100.6& 98.5--101.5    \\
\hline
$-A_t$  &  494.5   & 100.009                   & $111--89$& " " & " "\\
 $-A_b$ &   795.2   & 100.002                 & $106--94$&  " " & " "\\
 $-A_\tau$ &   251.7 & 100.002                & $100--99.9$ & " " &" " \\
 $-A_u$ &  677.3   & 100.005                  & $108--92$& negl. & negl. \\
 $-A_d$ &  859.4  & 100.001                   & $105--95$ & " " & " "  \\
 $-A_e$ &   253.4  & 100.002                  & $100--99.9$& " " & " " \\
\hline 
$\tan\beta$ &   9.74 &   &            &   &              \\
\hline\hline
\end{tabular}
\end{center}
\end{table}

%
\end{document}